\documentclass{aa} 
\usepackage{graphicx}
\usepackage{amsmath}
\usepackage{txfonts}
\usepackage{lscape}
\usepackage{longtable}
\usepackage{rotating}
\usepackage{appendix}
\usepackage{url}
\usepackage{float}
\usepackage[multiple]{footmisc}
\usepackage{threeparttable}

\begin{document}
\def\HII{H\,{\sc{ii}}}
\def\HI{H\,{\sc{i}}}
\def\farcs{\hbox{$.\!\!^{\prime\prime}$}}
\def\Ks{$K_{\rm{s}}$}
\def\sun{\hbox{$_\odot$}}
\def\degr{\hbox{$^\circ$}}

\def\h{\hbox{$^{\reset@font\r@mn{h}}$}}
\def\m{\hbox{$^{\reset@font\r@mn{m}}$}}
\def\s{\hbox{$^{\reset@font\r@mn{s}}$}}

\def\msol{\hbox{\kern 0.20em $M_\odot$}}
\def\lsol{\hbox{\kern 0.20em $L_\odot$}}

\def\kms{\hbox{\kern 0.20em km\kern 0.20em s$^{-1}$}}
\def\pc{\hbox{\kern 0.20em pc$^{2}$}}


   \title{Bipolar \HII\ regions - Morphology and star formation in their vicinity}   
   \subtitle{I - G319.88$+$00.79 and G010.32$-$00.15}
\author{L. Deharveng\inst{1}
 \and A. Zavagno\inst{1}
 \and M.R. Samal\inst{1}
 \and L.D. Anderson\inst{2}
 \and G. LeLeu\inst{1}
 \and D. Brevot\inst{1}
 \and A. Duarte-Cabral\inst{3}
 \and S. Molinari\inst{4}
 \and M. Pestalozzi\inst{4}
 \and J.B. Foster\inst{5}
 \and J.M. Rathborne\inst{6}
 \and J.M. Jackson\inst{7}}

\offprints{lise.deharveng@oamp.fr }
\institute{Aix Marseille Universit\'e, CNRS, LAM (Laboratoire d'Astrophysique de Marseille) UMR 7326, 
13388, Marseille, France 
	\and Department of Physics and Astronomy, West Virginia University, Morgantown, WV 26506, USA
	\and School of Physics and Astronomy, University of Exeter, Stocker Road, Exeter EX4 4QL, UK
	\and INAF-Istituto Fisica Spazio Interplanetario, Via Fosso del Cavaliere 100, I-00133 Roma, Italy
    \and Yale Center for Astronomy and Astrophysics, Yale University, New Haven, CT 06520
    \and CSIRO Astronomy and Space Science, PO Box 76 Epping NSW 1710 Australia 
    \and Institute for Astrophysical Research, Boston University, Boston, MA 02215, USA }
   \date{Received 18/03/2014; Accepted }

  \abstract
   {}
   {Our goal is to identify bipolar \HII\ regions and to understand their morphology, their evolution, and the role they play in the formation of new generations of stars.}
   {We use the {\it Spitzer}-GLIMPSE, -MIPSGAL, and {\it Herschel}-Hi-GAL surveys to identify bipolar \HII\ regions, looking for (ionized) lobes extending perpendicular to dense filamentary structures. We search for their exciting star(s) and estimate  their distances  using near-IR data from the 2MASS or UKIDSS surveys. Dense molecular clumps are detected using {\it Herschel}-SPIRE data, and we estimate their temperature, column density, mass, and density. MALT90 observations allow us to ascertain their association with the central \HII\ region (association based on similar velocities). We identify Class~0/I young stellar objects (YSOs) using their Spitzer and  {\it Herschel}-PACS emissions. These methods will be applied to the entire sample of candidate bipolar \HII\ regions to be presented in a forthcoming paper.}
   {This paper focuses on two bipolar \HII\ regions, one that is especially interesting in terms of its morphology, G319.88$+$00.79, and one in terms of its star formation, G010.32$-$00.15. Their exciting clusters are identified and their photometric distances estimated to be  2.6~kpc and 1.75~kpc, respectively; thus G010.32$-$00.15 (known as W31 north) lies much closer than previously assumed.  We suggest that these regions formed in dense and flat structures that contain filaments. They have a central ionized region and ionized lobes extending perpendicular to the parental cloud. The remains of the parental cloud appear as dense (more than 10$^4$~cm$^{-3}$) and cold (14--17~K) condensations. The dust in  the photodissociation regions (in regions adjacent to the ionized gas) is warm (19--25~K). Dense massive clumps are present around the central ionized region. G010.32-00.14 is especially remarkable because five clumps of several hundred solar masses surround the central \HII\ region; their peak column density is a few 10$^{23}$~cm$^{-2}$, and the mean density in their central regions reaches several 10$^5$~cm$^{-3}$. Four of them contain at least one massive YSO (including an ultracompact \HII\ region and a high-luminosity Class~I YSO); these clumps also contain extended green objects (EGOs) and Class~II methanol masers. This morphology suggests that the formation of a second generation of massive stars has been triggered by the central bipolar \HII\ region. It occurs in the compressed material of the parental cloud.}  
   {}
\titlerunning{Bipolar nebulae}
\authorrunning{L.~Deharveng et al.}
\maketitle
%

\section{Introduction}

Hi-GAL, the {\it Herschel} infrared GALactic Plane survey (Molinari et al. \cite{mol10a}), gives an overall view of the distribution of the different phases of the Galactic interstellar medium (ISM). According to Molinari et al. (\cite{mol10b}), ``The outstanding feature emerging from the first images is the impressive and ubiquitous ISM filamentary nature.'' Now that all of the Galactic plane has been surveyed by Hi-GAL we still have the same view of a very filamentary structure, especially at {\it Herschel}-SPIRE wavelengths. Several papers present a preliminary analysis of filaments (for example, Arzoumanian et al. \cite{arz11}, \cite{arz13}; Hill et al. \cite{hil11}; Peretto et al. \cite{per13}) and suggest that the filaments are the main birth site of prestellar cores (for example, Molinari et al. \cite{mol10b}; Andr\'e et al. \cite{and10}). 

The {\it Spitzer}-GLIMPSE survey has shown a ``bubbling Galactic plane'' (Churchwell et al. \cite{chu06}, \cite{chu07}; Simpson et al. \cite{sim12}). These bubbles observed at 8.0~$\mu$m enclose classical \HII\ regions (Deharveng et al. \cite{deh10}; Anderson et al. \cite{and11}). The exact morphology of most of these bubbles is still unknown; are they 3D spherical structures or 2D rings? The vicinity of 43 bubbles has been observed by Beaumont \& Williams (\cite{bea10}) in the $^{12}$CO~(3-2) transition. These observations suggest that these bubbles are not spherical structures, but rings, and that \HII\ regions form in flat molecular clouds (with a thickness of a few parsecs). If massive stars form in filamentary 1D or 2D structures, thus dense filaments or sheets, so in a medium where density gradients are present, many bipolar nebulae should form and be observed. The Green Bank Telescope \HII\ Region Discovery Survey (Bania et al. \cite{ban10}; resolution 82$\arcsec$) shows the opposite situation, the rarity of bipolar nebulae, hence the conclusion of Anderson et al. (\cite{and11}), based on the comparison of the radio and {\it Spitzer} 8.0~$\mu$m  emissions and on the radio recombination linewidths, that the majority of bubble sources are 3D structures. Statistics on bipolar nebulae should resolve this controversy. 

Bipolar \HII\ regions are also interesting because they have a simple morphology, allowing one to locate the ionized and neutral components in space in their associated complex. And, as predicted by Fukuda \& Hanawa (\cite{fuk00}; Sect.~3), star formation can be triggered by the expansion of such nebulae adjacent to dense filaments. In the following, we use {\it Spitzer} and {\it Herschel} observations of the Galactic plane to search for these nebulae.\\

Using the {\it Spitzer} GLIMPSE survey, and the {\it Herschel} Hi-GAL survey we have identified 16 candidate bipolar \HII\ regions in a zone of the Galactic plane between $\pm$60$\degr$ in longitude and $\pm$1$\degr$ in latitude. In this first paper we describe two of them, G319.88$+$00.79 and G010.32$-$00.15, that we consider the most outstanding in terms of morphology and star formation,  respectively. In this paper, we give in Section 2 a quick description of the observations we use for this study, mainly the {\it Spitzer}-GLIMPSE and -MIPSGAL surveys, the {\it Herschel} Hi-GAL survey, and the MALT90 survey. The few existing models of bipolar nebulae are discussed in Section~3, and we discuss how these nebulae can be identified and how star formation in their vicinity can be observed. The general methods used to estimate the physical parameters of various structures (molecular clumps, cores, etc.) are described in Section~4, where we also explain  how we detect young stellar objects (YSOs), especially candidates Class~0/I YSOs. Sections~5 and 6 are devoted to the study of the G319.88$+$00.79 and G010.32$-$00.15 bipolar nebulae. We discuss our findings in Sect.~7 and give conclusions in Sect.~8. A second paper will contain the catalogue of bipolar nebulae, where a study of each of them uses the same observations and the same methods as described in this first paper, and our general conclusions concerning  star formation in their surroundings.\\

Following Foster et al. (\cite{fos11}) we refer to molecular ``clumps'' when speaking about molecular structures with typical masses in the range 50\msol\ -- 500\msol, typical sizes in the range 0.3~pc -- 3~pc, and typical mean densities in the range 10$^3$~cm$^{-3}$ -- 10$^4$~cm$^{-3}$. These structures may contain smaller ($\leq$0.1~pc) and denser ($\geq$10$^4$~cm$^{-3}$) substructures called ``cores''.

In the following we use the term ``filament'' to describe a cylindrical structure (1D), and  ``sheet'' to describe a flat structure (2D). It has a small thickness in one direction and a larger section perpendicular to this direction. A sheet observed   edge-on appears as a filament (and so some structures called filaments may be sheets).

\section{Observations}

We use several surveys for this study.  We summarize their wavelengths and resolutions in Table~\ref{obs}. The Hi-GAL survey (Molinari et al. \cite{mol10a}) covers a two-degree-wide strip of the Galactic plane, in five bands centred at 70~$\mu$m, 160~$\mu$m, 250~$\mu$m, 350~$\mu$m, and 500~$\mu$m. The data reduction pipeline is described by Traficante et al (\cite{tra11}). The Hi-GAL maps, which trace the dust continuum emission, are complemented by: 

$\bullet$ DSS2-red or SuperCOSMOS H$\alpha$ maps (Parker et al. \cite{par05}) and radio-continuum maps (NVSS survey, Condon et al. \cite{con98}; SUMSS survey, Bock et al. \cite{boc99}) to describe the emission of the ionized gas (respectively H$\alpha$ emisssion or free-free emission). We use  in particular the MAGPIS survey (Multi-Array Galactic Plane Imaging Survey; Helfand et al. \cite{hel06}) and the CORNISH survey (Hoare et al. \cite{hoa12}; Purcell et al. \cite{pur13}) to detect ultracompact (UC) \HII\ regions. Radio recombination lines are also used to  give  the velocity of the ionized gas.

$\bullet$  2MASS (Skrutskie et al. \cite{skr06}),  UKIDSS  (from the Galactic plane Survey; Lawrence et al. \cite{law07}), or VISTA (Minniti et al. \cite{min10})  maps and catalogues to describe the stellar content of the selected regions.

$\bullet$ {\it Spitzer}-GLIMPSE  maps at 3.6~$\mu$m, 4.5~$\mu$m, 5.8~$\mu$m, 8.0~$\mu$m (Benjamin et al. \cite{ben03}), and {\it Spitzer}-MIPSGAL maps at 24~$\mu$m (Carey et al. \cite{car09}) to detect candidate YSOs, and describe the distribution of specific dust grains. WISE (Wide-Field Infrared Survey Explorer; Wright et al. \cite{wri10}) maps are used if the {\it Spitzer} maps are missing or saturated. The {\it Spitzer}-GLIMPSE and -MIPSGAL catalogues are used, as well as the Robitaille et al. (\cite{rob08}) catalogue of intrinsically red sources.  

$\bullet$ Maser observations of various molecular species (especially methanol) and observations of extended green objects (EGOs; Cyganowski et al. \cite{cyg08}) to indicate the presence of YSOs with outflows (Breen et al. \cite{bre10}, and references therein).

$\bullet$ MALT90 (Millimeter Astronomy Legacy Team 90 GHz) observations of the molecular clumps present in the vicinity of the \HII\ regions. A general description of the MALT90  survey  is given by Jackson et al. (\cite{jac13}). About 2000 dust clumps observed at 870~$\mu$m (ATLASGAL survey, Schuller et al. \cite{sch09}) have been mapped with the Mopra telescope at a frequency $\sim$90~GHz and a beam size of 38$\arcsec$. Each map has a size of 3$\arcmin \times$3$\arcmin$, with a pixel size of 9$\arcsec$. Sixteen molecular lines are mapped simultaneously with a spectral resolution of 0.11~km~s$^{-1}$. More details can be found in Foster et al. (\cite{fos11}, \cite{fos13}). Here we use these observations to obtain the velocity fields of the molecular clumps, and ascertain their association with the ionized regions.

%
\begin{table}[h!]
\centering
\caption{List of the main surveys used for this study}                               
\begin{tabular}{lll}                
\hline\hline                          
IR surveys          & Wavelength         &  FWHM           \\
                 &($\mu$m)            & ($\arcsec$)     \\
\hline 
2MASS $ J, H, K_s$  & 1.25, 1.65, 2.16  &  $\geq$2.5   \\
UKIDSS $ J, H, K$ &  1.25, 1.63, 2.20  &   $\leq$1   \\
VISTA  $ J, H, K_s$ & 1.25, 1.65, 2.15 &  $\leq$1   \\
{\it Spitzer}-IRAC & 3.6, 4.5, 5.8, 8.0 & $\leq$2  \\
WISE   & 3.4, 4.6, 11, 22  &  6.1, 6.4, 6.5, 12.0  \\
{\it Spitzer}-MIPS & 24  & 6  \\
Hi-GAL & 70, 160, 250, 350, 500 &  8.5, 12.5, 18, 25, 36 \\
\hline
Radio surveys          & Wavelength         &  FWHM           \\
                       &(cm)                & ($\arcsec$)     \\
\hline 
NVSS & 21 & 45 \\
SUMSS & 35 & 45$\times$45cosec$\delta$ \\        
MAGPIS   & 20    &  6   \\
CORNISH  &  6    &  1.5 \\
\hline
\label{obs}  
\end{tabular}
\end{table}

\section{Simulations of bipolar nebulae and triggered star formation}

\begin{figure*}[tb]
\centering
\includegraphics[width=18cm]{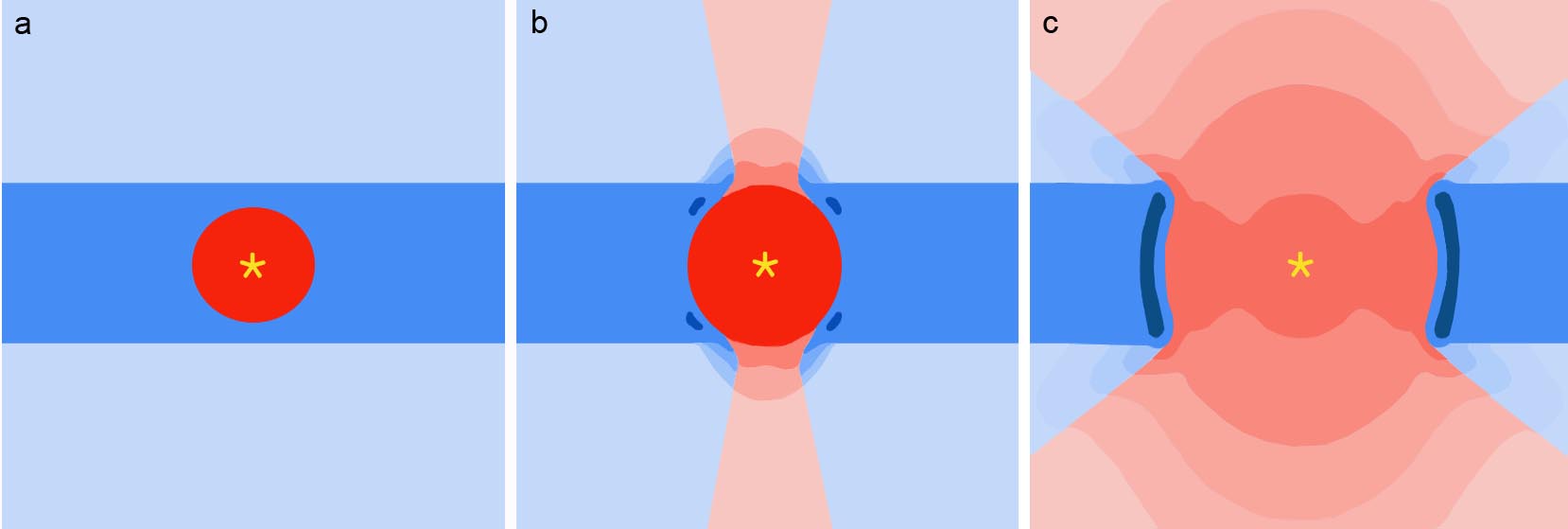}
  \caption{Formation of a bipolar \HII\ region, according to the simulation of Bodenheimer et al. (1979; their figure 4; the thickness of the parental plane is 1.3~pc, its density 300~H$_2$~cm$^{-3}$). The ionized material appears in red, the neutral material in blue; the density of the material is indicated by the saturation level. Schema a shows the dense molecular plane surrounded by low density material, the initial Str{\"{o}}mgren sphere around its central exciting star (T$_*$=4$\times$10$^4$~K). In schema b, at 3$\times$10$^4$~yrs, the expanding \HII\ region reaches the border of the parental cloud, and the low density material is quickly ionized; a bipolar nebula forms. In schema c, at 1$\times$10$^5$~yrs, the high density ionized material flows away from the central region; high density molecular material accumulates at the waist of the bipolar nebula, forming a torus of compressed material.}
\label{schema}
\end{figure*}

Very few studies describe the formation and evolution of a bipolar \HII\ region\footnote{In contrast, many studies consider the formation of bipolar planetary nebulae, but the situation is very different since these result from the ejection of a small amount of material by an evolved low-mass star.}. Bodenheimer et al. (\cite{bod79}) present a 2D hydrodynamic simulation following the evolution of an \HII\ region excited by a star lying in the symmetry plane of a flat homogeneous molecular cloud surrounded by a low-density medium. (The parental cloud is in pressure equilibrium with the surrounding medium.) Figure~\ref{schema} illustrates the formation of a bipolar \HII\ region according to this simulation. Initially (schema a), a spherical \HII\ region grows inside the cloud. 
A bipolar nebula forms when the ionization front breaks through the two opposite faces of the cloud simultaneously (schema b). Then, the dense ionized material pours out of the dense cloud with velocities of up to 30~km~s$^{-1}$, forming a double cone structure (schema c); with time the angle of the cone widens, and the density decreases in the central parts of the \HII\ region; neutral material accumulates at the waist of the bipolar nebula. These \HII\ regions appear bipolar when the angle between the line of sight and the cloud's symmetry plane is small (when the flat parental cloud is almost seen edge-on). \\

\begin{figure*}[tb]
\centering
\includegraphics[width=18cm]{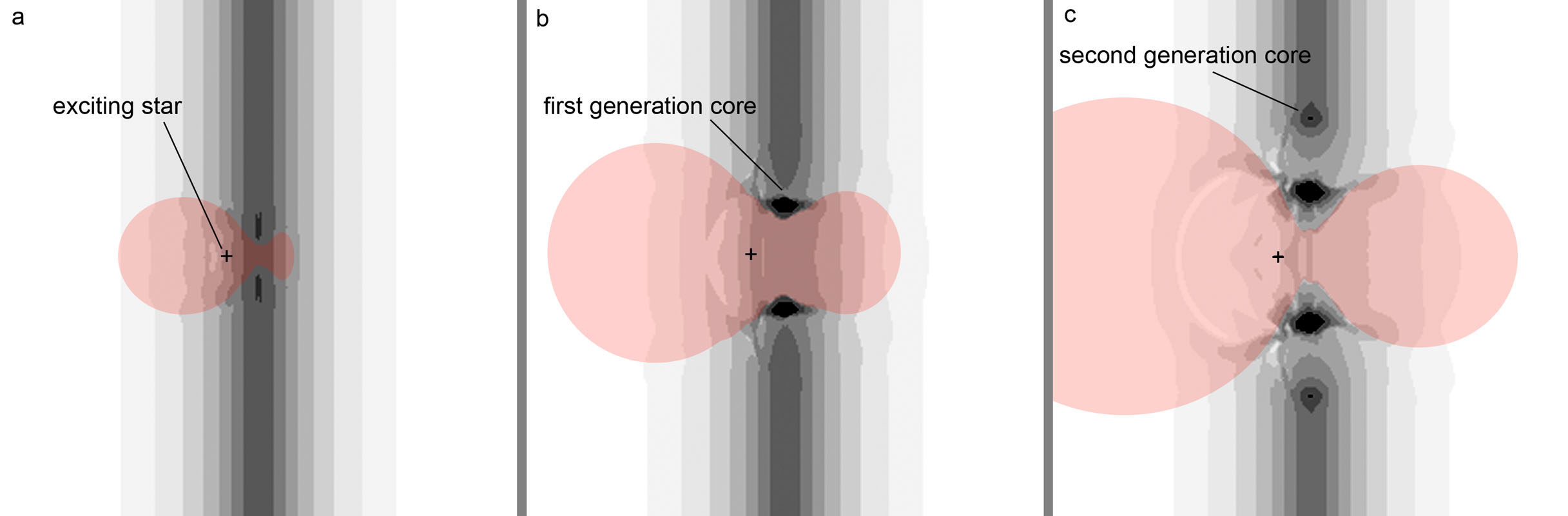}
  \caption{Star formation triggered by the expansion of an \HII\ region close to a filament, according to the simulation of Fukuda \& Hanawa (2000; their figure 1; model C1{\it l}': neutral sound speed 0.3~km~s$^{-1}$, maximum density on the axis of the filament 2$\times$10$^6$~cm$^{-3}$, exciting star at 0.025~pc of the filament's axis). The neutral filament appears in grey, the ionized gas in pink; we show what we believe to be the extent of the ionized gas in the plane containing the axis of the filament and the exciting star (black cross) of the \HII\ region. Schema a shows the expanding \HII\ region (age 8.5$\times$10$^4$~yr) pinching the molecular filament. In schema b, first-generation cores have formed by compression of the neutral material by the ionized gas (age 2.1$\times$10$^5$~yr). In schema c, second-generation cores have formed due to gravity (age 3.0$\times$10$^5$~yr). The first- and second-generation cores are separated by 0.1~pc.}
\label{fukuda}
\end{figure*}

Fukuda \& Hanawa (\cite{fuk00}) study sequential star formation in a filament, triggered by the expansion of a nearby \HII\ region. Their 3D simulations take both (magneto) dynamical and gravitational effects that may induce star formation into account. In these simulations (Fig.~\ref{fukuda}) a dense filamentary cloud is compressed by a nearby  expanding \HII\ region and fragments to form cores sequentially. The \HII\ region is initially off the filament; as it expands and grows in size it interacts with the filament. The filament is compressed and pinched, and it possibly separates into two parts\footnote{Fukuda \& Hanawa (\cite{fuk00}) only represent in their figures the distribution and kinematics of the neutral material. No limits are given for the ionized region. The limits drawn in Fig.~\ref{fukuda} correspond to our best guess.} ; the \HII\ region appears bipolar to an observer if the line of sight is roughly perpendicular to the plane containing the filament and the exciting star. The dynamical compression of the filament triggers the formation of a first-generation of cores at the waist of the bipolar nebula. With time the density in the cores increases (from compression by the \HII\ region and self-gravity), they become gravitationally bound and they collapse. Star formation probably occurs, but this is not followed in the Fukuda \& Hanawa paper. Later, the formation of the first-generation cores changes the nearby gravitational field and induces the formation of second-generation cores. The time of the formation of the first-generation cores depends upon the distance of the exciting star to the axis of the filament, of the  energy of the expansion (through its velocity and the radius of the \HII\ region), and of the magnetic field. Depending on the distance between the exciting star of the \HII\ region and the filament's axis, the number of first-generation cores ranges from 2 to 4. 

Fukuda \&  Hanawa did not simulate what happens if an \HII\ region forms in or near a dense sheet of material (in a flat structure instead of a filament).

\subsection{What do we expect to see?}

The signatures of bipolar \HII\ regions that we expect to see are:

$\bullet$ A dense filament, observed as a cold elongated high column density feature. This structure will be observed in emission at {\it Herschel}-SPIRE wavelengths. It can appear in absorption, as an elongated infrared dark cloud (IRDC) at 8.0~$\mu$m or 24~$\mu$m. This ``filament'' could also be a sheet seen edge-on. 

$\bullet$ An \HII\ region, bright in its central part and displaying two ionized lobes perpendicular to the filament. It will appear as a radio-continuum source, centred on the filament and slightly elongated in two directions perpendicular to it. The two lobes will be especially well traced at 8.0~$\mu$m, a wavelength band dominated by the emission of PAHs located in the photodissociation region (PDR) surrounding the ionized gas and excited by the UV radiation leaking from the \HII\ region (PAHs are destroyed inside the ionized region, Povich et al. \cite{pov07}; Pavlyuchenkov et al. \cite{pav13}). At 24~$\mu$m we will see an extended central source, similar to the radio-continuum source (Deharveng et al. \cite{deh10}; Anderson et al. \cite{and11}); also 24~$\mu$m emission will be observed in the PDRs surrounding the two ionized lobes. The 24~$\mu$m emission observed in the direction of the ionized region comes from very small dust grains (size of a few nanometres) located inside the ionized gas and out of thermal equilibrium after absorption of ionizing photons (Pavlyuchenkov et al. \cite{pav13}). At 70~$\mu$m a small fraction of the emission comes from the ionized region, but the bulk of the emission comes from the PDRs surrounding the ionized region (emission of silicate grains; size of a few hundred nanometres; Pavlyuchenkov et al. \cite{pav13}). 

$\bullet$ The exciting star(s) or cluster. If the extinction is not too high in its direction, it should be observed in the near~IR, near the symmetry axis of the filament, at the centre of the \HII\ region.

$\bullet$ Dense material and possibly dense cores at the waist of the bipolar nebula, containing YSOs or not, depending on whether star formation is presently at work or not. If present, Class 0/I YSOs will possibly appear as point-like sources at {\it Herschel} wavelengths, especially at 70~$\mu$m; Class I/II YSOs will appear as {\it Spitzer}-mid IR sources, prestellar cores will be without any associated mid-IR sources. 

$\bullet$ More cores located along the filament, but not adjacent to the \HII\ region, if second-generation cores have already formed (according to the simulations of Fukuda \& Hanawa~\cite{fuk00}). If present, they will appear as {\it Herschel}-SPIRE compact sources. 

Of course, all the components of the complex, the filament, the clumps, and the \HII\ region must have similar velocities to ascertain their association.

In the simulation of Bodenheimer et al. (\cite{bod79}) the medium surrounding the molecular condensation has a very low density. Thus, it is very quickly ionized, and the two lobes are density bounded and not bounded by ionization. If this medium is denser (but however less dense than the parental cloud), we can expect the two lobes to be ionization-bounded and surrounded by an ionization front (IF) of D type\footnote{An IF is of D type when the expansion velocity of the \HII\ region is comparable to the sound speed in the ionized gas, of the order of 10~km~s$^{-1}$. This IF is preceded on the neutral side by a shock front; with time neutral material accumulates between the two fronts.}. Then they should appear as closed lobes surrounded by PAHs emission, and also possibly surrounded by neutral material collected during the expansion of the ionized gas. Thus the presence of a layer of neutral material surrounding the ionized lobes can also be a feature of  bipolar nebulae.

We search for these signatures and discuss them for each candidate bipolar \HII\ region.\\

\begin{figure*}[tb]
\centering
\includegraphics[width=16cm]{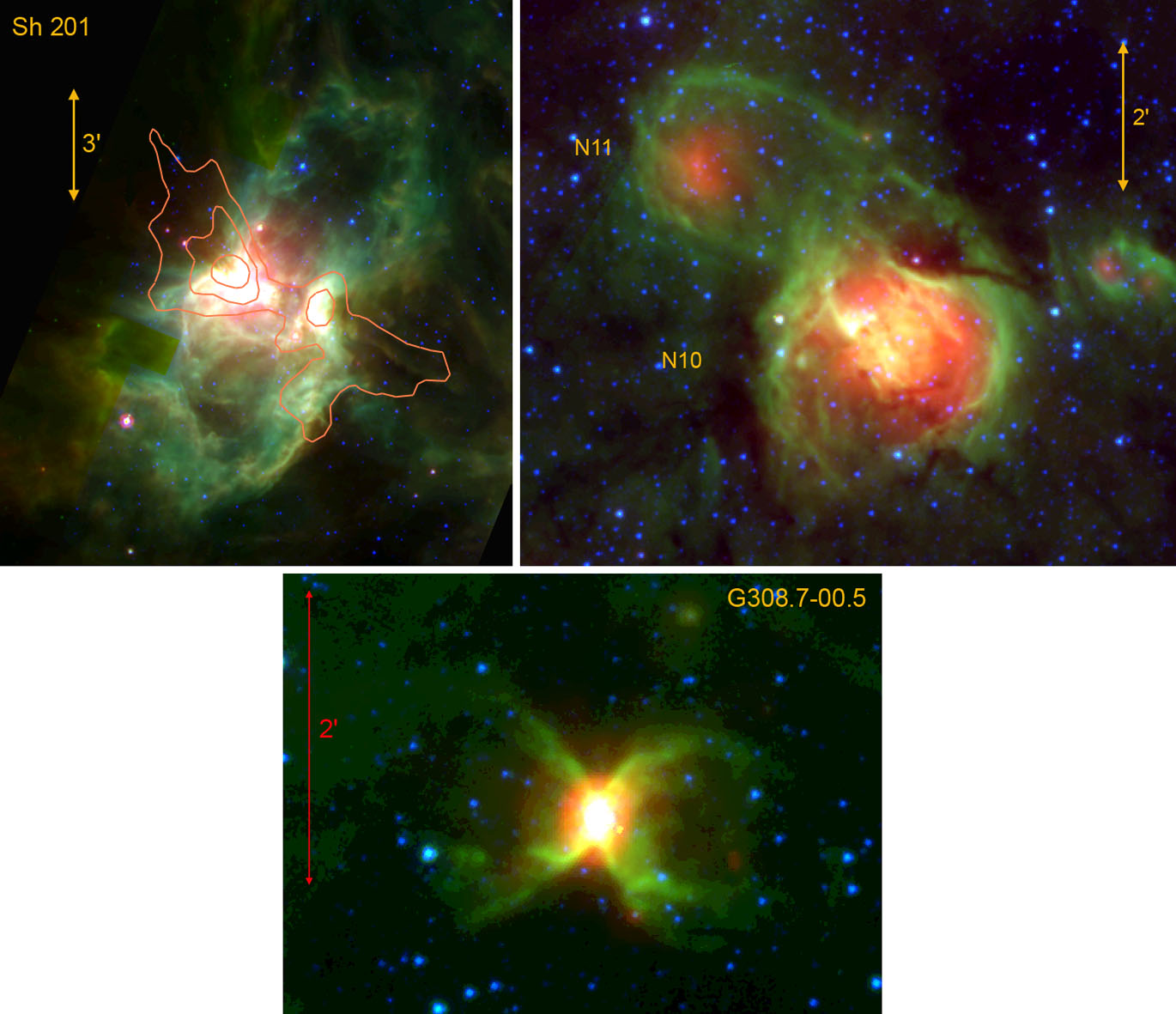}
  \caption{Composite colour image of \HII\ regions, as seen by {\it Spitzer}. Red is the MIPSGAL image at 24~$\mu$m showing the emission of the hot dust, green is the GLIMPSE image at 8.0~$\mu$m dominated by the PAHs emission from the PDRs, and blue is the GLIMPSE image at 4.5~$\mu$m showing the stellar sources. Sh~201 is a textbook example of a bipolar \HII\ region; the red column density contours trace the parental molecular filament (levels of 2, 5, and 15 $\times$10$^{22}$~cm$^{-2}$). N10 and N11 are adjacent bubbles around distinct \HII\ regions. This is a case of mis-identified bipolar nebula. G308.7-00.5 is a nebula associated with an evolved star.}  
\label{example}
\end{figure*}

Several objects can be mis-identified as bipolar \HII\ regions. They are:

$\bullet$ Nebulae associated with evolved stars, such as G308.7-00.5 (Fig.~\ref{example}). These ``well defined nebulae are formed during a brief post-main-sequence phase when a massive star becomes a Wolf-Rayet star or a Luminous Blue Variable  star'' (Gvaramadze et al. \cite{gva10}). Some of them show a bipolar structure when observed by {\it Spitzer}. 

$\bullet$ Adjacent bubbles surrounding distinct \HII\ regions, such as N10 and N11 (Fig.~\ref{example}). According to the Churchwell et al. catalogue (\cite{chu06}) `` BP indicates a bipolar bubble or a double bubble whose lobes are in contact.'' N10 and N11 fall within the last case. Figure~\ref{example} shows that the 8.0~$\mu$m bubbles enclose two separate \HII\ regions: we clearly see two separate zones of extended 24~$\mu$m emission at the centre of each bubbles. In bipolar \HII\ regions like Sh~201 (see below), the extended 24~$\mu$m emission comes from the central region, the region where the lobes originate. Furthermore, the {\it Herschel}-SPIRE maps show no cold filament between N10 and N11.

The \HII\ region Sh~201 is a textbook example of a bipolar \HII\ region (Fig.~\ref{example}). It has been studied using deep {\it Herschel} observations (see  Appendix C of Deharveng et al. \cite{deh12}). The complex displays a cold filament observed at SPIRE wavelengths, an \HII\ region centred on this filament, and two lobes, well traced at 8.0~$\mu$m (Fig.~\ref{example}), perpendicular to the filament. The ionized region is bright in its central part. It shows a narrow waist, limited on each side by two massive clumps. YSOs of Class~0 and Class~I are observed in the direction of these clumps (Deharveng et al. \cite{deh12}).\\

In all the figures of this paper the regions are presented in Galactic coordinates.

\section{Reduction methods}

\subsection{Determination of the physical parameters of the filaments and clumps}

For all estimates in this paper, we assume a gas-to-dust ratio of 100. We assume the following values for the dust opacity $\kappa_{\nu}$, 35.2, 14.4, 7.3, and 3.6~cm$^2$~g$^{-1}$, respectively, at 160, 250, 350, and 500~$\mu$m\footnote{These values correspond to the formula 
\begin{equation}
\kappa_{\nu}=10\times\left(\frac{\nu}{1000~GHz}\right)^{\beta}\,,
\end{equation}
which assumes a spectral index $\beta$ of 2 for the dust emissivity and gives a dust opacity at 300~$\mu$m of 10.0~cm$^2$~g$^{-1}$ (Andr\'e et al. \cite{and10}). These values have been discussed by Sadavoy et al. (\cite{sad13}); they show that assuming $\beta$=2 with the Herschel bands alone (using the 160~$\mu$m -- 500~$\mu$m range) affects the measured best-fit temperature by $\leq$2 K if $\beta$=  1.5 or $\beta$ = 2.5, and that ``the Herschel-only bands provide a decent first look at the clump masses for an assumed value of $\beta$ = 2''.}\footnote{The dust opacities determined by Ossenkopf \& Henning (\cite{oss94}; dust with thin ice mantle, density 10$^6$~cm$^{-3}$) are also often used; they correspond to a $\beta\sim$1.8 and have values of 40, 18, 10, and 5~cm$^2$~g$^{-1}$ respectively at 160~$\mu$m, 250~$\mu$m, 350~$\mu$m, and 500~$\mu$m.}. The opacities are uncertain, as is 
the gas-to-dust ratio. The dust opacity probably differs depending on the content of the clumps, for example whether they have internal sources (Paradis et al. \cite{par14}).  This remains the main source of uncertainty of the temperature, column density, and mass determinations. (The column density and the mass are inversely proportional to $\kappa_{\nu}$.)\\

We have obtained dust temperature maps for all the regions, using {\it Herschel} data between 160~$\mu$m and 500~$\mu$m. We assume that the dust emission is optically thin at these wavelengths. We do not use the 70~$\mu$m data; the 70~$\mu$m maps show that a fraction of this emission comes from ionized regions, as is the case for a larger portion of the 24~$\mu$m emission; i.e. the 24~$\mu$m and 70~$\mu$m  emissions probe a different gas/dust component than emission observed at longer wavelengths (Sect. 3.1). Furthermore, some of the clumps are not optically thin at 70~$\mu$m; for example, most of the clumps in the field of G010.32$-$00.15 (Table~4). The temperatures that we determine are those of cold dust grains in thermal equilibrium, located in the PDRs of \HII\ regions and in the surrounding medium. The steps in the temperature determination are: 1) the 160~$\mu$m, 250~$\mu$m, and 350~$\mu$m maps have been smoothed to the resolution of the 500~$\mu$m map; 2) all the maps have been regridded so they have the same pixel size of 11.5$\arcsec$ as the 500~$\mu$m map (at the same location); 3) the SED of each pixel has been fitted, using a  modified blackbody model; thus a first temperature map. We applied colour corrections to the temperature map of the G319.88+00.79 field (using the first temperature map, and following the PACS and SPIRE documentation).  The colour-corrected temperatures are higher than those obtained without corrections, with a  mean difference of 0.58~K (in the range 0.3~K -- 0.7~K). However, since the colour corrections were still uncertain at the time we created the temperature maps, and because they are not very large (less than 6\% on the fluxes,) we decided to present in the following the temperature maps obtained without colour-corrections.

No background subtraction has been made, so the temperature obtained is a mean temperature for the dust along the line of sight, weighted by the emission. If assigned to a specific structure, it assumes that the emission from  the structure is dominant, which is possibly not the case. These temperature maps will be used to study the temperatures of different structures present in the field. \\

The molecular hydrogen colum density is given by
\begin{equation}
N(\mathrm {H_2}) = 100\,\, \frac{F_{\nu}\,\,10^{20}}{\kappa_\nu\,\,B_\nu(T_{\mathrm {dust}})\,\,2.8\,\,m_{\mathrm H}}
\end{equation}
where the brightness $F_{\nu}$ is expressed in MJy~sr$^{-1}$ (the units of the Hi-GAL maps), the Planck function $B_{\nu}$ in J~s$^{-1}$~m$^{-2}$~Hz$^{-1}$, the factor of 2.8 is the mean molecular weight, $m_{\mathrm H}$ is the hydrogen atom mass.

To keep the best possible spatial resolution, we use temperature and column density maps regridded the resolution of the
250~$\mu$m data. The original temperature maps have the resolution of the 500~$\mu$m maps; however, since the dust temperature varies only smoothly across the fields, we expect that this extrapolated temperature does not differ too strongly from the one that  would be obtained if all the maps had the resolution of the 250~$\mu$m observations. Here again, no background emission has been subtracted, and thus, the column density corresponds to the whole line of sight. \\

Two methods can be used to estimate the mass of a structure. In the first method, we integrate the column density map within an aperture. To compare the parameters of the clumps observed in the vicinity of the bipolar nebulae, we use apertures  defined by the intensity at half the peak value.  We prefered to use this column density method for extended structures. In the second method, after assuming a mean temperature for the structure (either by fitting the SED of the structure or by integrating the temperature map), we determine the mass of the structure following the method of Hildebrand (\cite{hil83}). The total (gas+dust) mass of a feature is related to its integrated flux density $S_{\nu}$ by
\begin{equation}
M_{\mathrm{(gas+dust)}} = 100\,\, \frac{S_\nu\,\,D^2}{\kappa_\nu\,\,B_\nu(T_{\mathrm{dust}})}
\end{equation}
where $D$ is the distance to the source and $B_{\nu} (T_{\mathrm{dust}})$ is the Planck function for a temperature $T_{\mathrm{dust}}$. Here again we use the 250~$\mu$m fluxes (to have the best resolution). This method is  preferably used for point sources. We have used both methods to estimate the mass of the three clumps in the G319.88+00.79 field  (Table~\ref{G319condensations}). The masses obtained  differ by less than 20\% (see Table~2).

If a clump can be modelled as an elliptical Gaussian of uniform temperature, its total mass is twice the mass measured using an aperture following the level at half the peak's value. Also, the derived mass is not dependent on the beam size, which therefore allows us to compare the masses of clumps at different distances and angular sizes.

We also estimate the mean density in the central regions of the clumps, using their mass and beam deconvolved size obtained with an aperture following the level at half the maximum value, assuming a spherical morphology and a homogeneous medium.

A background subtraction is done in only one instance: to estimate the parameters of the bright clumps discussed in Sects.~5.3 and 6.3 more accurately (details are given in the text and in the notes to Tables~\ref{G319condensations} and \ref{G10condensations}).

\subsection{The velocity field of the molecular clumps}

The MALT90 observations are used to obtain the velocity field of each molecular clump. In both regions we used the brightest molecular transitions, H$^{12}$CO$^+$ (1-0), N$_2$H$^+$ (1-0), and  HNC (1-0).  We performed Gaussian fittings of the main lines to retrieve the distribution and velocity field of the molecular gas. We also used the H$^{13}$CO$^{+}$ (1-0) to confirm the localized existence of H$^{12}$CO$^{+}$ self-absorption, and to validate the use of a single Gaussian fit of the double profile of H$^{12}$CO$^{+}$ as providing a good approximation of the systemic velocities.

We present the velocity fields resulting from this Gaussian fitting, and use those to ascertain the association of the observed molecular clumps with the HII regions.  Details about the reduction procedure can be found in Zavagno et al. (in preparation).

\subsection{Identification of young stellar objects}

In the field of each region we try to detect Class~0 or Class~I YSOs and prestellar cores, if any. We do not consider Class~II YSOs because, depending on the mass of their central source, they can possibly be as old as the exciting stars of the bipolar nebulae. Only Class~0/I YSOs, with an age $\leq$10$^5$~yr (Andr\'e et al. \cite{and00} and references therein) compared to an age of 10$^6$~yr or more for extended \HII\ regions, are most probably second-generation forming stars. Different indicators have been used to identify candidate Class~I or Class~0 YSOs: 

$\bullet$ The {\it Spitzer} [3.6]$-$[4.5] versus [5.8]$-$[8.0] diagram (Allen et al. \cite{all04}).  We have used the {\it Spitzer} GLIMPSE Source Catalogue (which combines the three surveys GLIMPSE I, II, and 3D)  to draw these diagrams, measuring a few missing magnitudes when necessary\footnote{Some interesting sources seen in the {\it Spitzer} images are missing in the catalogues. Most often such sources are difficult to measure because they are found against a bright background. Most such sources are observed in the direction of PDRs, and therefore are interesting when discussing star formation triggered by \HII\ regions.}. The limits in these diagrams of the location of Class~I and Class~II sources are only indicative, since the position of the sources depends of the external extinction, which is often unknown, and relies on a good background estimate, which is often difficult to obtain. 

$\bullet$ Emission of  point sources at 24~$\mu$m. We used the 24~$\mu$m magnitudes from the Robitaille et al. (\cite{rob08}) catalogue of intrinsically red objects anf from the MIPSGAL catalogue, when available; otherwise, we used these magnitudes to calibrate our own measurements.  The spectral index, given by
\begin{equation}
\alpha=\frac{dlog(\lambda F_\lambda)}{dlog(\lambda)},
\end{equation}
has been estimated between $K$ and 24~$\mu$m, when possible, to confirm or estimate the nature of the point sources. According to Lada (\cite{lad87}) and Greene et al. (\cite{gre94}), Class~I YSOs have $\alpha\geq$0.3 in the range $K$--20~$\mu$m; following Greene et al. sources with $-$0.3$\leq\alpha\leq$0.3 are "flat-spectrum" sources (of uncertain evolutionary status between Class~I and Class~II), and sources with $\alpha\leq$-0.3 are Class~II or Class~III YSOs. This index is however strongly affected by an external extinction\footnote{An uncorrected (visual) external extinction $\sim$50~mag (corresponding to a column density $\sim$5$\times$10$^{22}$~cm$^{-2}$) decreases the value of $\alpha$ by $\sim$1.04.} that is generally unknown because we do not know the position along the line of sight of sources observed in the direction of clumps. Thus we prefer to consider the colour [8.0]$-$[24] of the point sources, since it is almost unaffected by extinction; $\alpha\geq$0.3 corresponds to a colour [8.0]$-$[24]$\geq$3.9~mag, $\alpha\leq$-0.3  to [8.0]$-$[24]$\leq$3.2.

$\bullet$ Emission of compact sources at {\it Herschel} wavelengths - mainly the detection of 70~$\mu$m point sources, the counterparts of Class~0/I YSOs or prestellar cores. Owing to the resolution of {\it Herschel} data and the distance of our regions, it is difficult to distinguish between point sources and compact sources in our fields. Most of our 70~$\mu$m point sources have no detectable counterpart at longer wavelengths. What can we deduce from their 70~$\mu$m fluxes?

Dunham et al. (\cite{duh08}) find a tight correlation between the 70~$\mu$m flux and the internal luminosity of low-mass protostars (that of their central objects), and it is valid up to 50~$\lsol$, which corresponds to a flux of 16~Jy at 70~$\mu$m for a source at a distance of 1~kpc. Most of our measured sources are brighter at 70~$\mu$m (see Tables~\ref{G319_YSOs} and \ref{G10_YSOs}). Ragan et al. (\cite{rag12}) have shown (their figure 8) that the 70~$\mu$m luminosity of more massive sources correlates well with their total luminosity. There is good agreement between the correlations established for low and high-luminosity sources. We use this correlation to roughly estimate  the luminosity of sources detected only at 70~$\mu$m.

Stutz et al. (\cite{stu13}) have shown that a ratio $log(\lambda$F$_{\lambda}$(70)/$\lambda$F$_{\lambda}$(24))$\geq$1.65  defines what they call ``PACS Bright Red sources'' (PBRs), which are extreme Class~0 sources with high infall rates. We estimate this ratio and identify such objects when possible.\\

As discussed by Robitaille et al. (\cite{rob08}, and references therein), evolved stars (AGB stars) can be mistaken for YSOs. We have used the criteria recommended by Robitaille et al. (\cite{rob08}) to identify candidate xAGB stars (extreme AGB stars; AGBs with very high mass-loss rates): [4.5]$\leq$7.8~mag. When possible we measured the [8.0]$-$[24] colour, because another criterion to identify AGB stars is [8.0]$-$[24]$\leq$2.5~mag (Robitaille et al. \cite{rob08}). However, as stressed by Robitaille et al., the separation between AGBs and YSOs is only approximate.\\

The magnitudes of {\it Spitzer} and {\it Herschel} point sources are measured, if necessary, using aperture photometry for isolated sources superimposed on a faint background or using PSF fitting with DAOPHOT (Stetson \cite{ste87}) in more difficult cases (sources superimposed on a bright variable background; details are given in Deharveng et al. \cite{deh12}).\\

\section{G319.88$+$00.79}

\begin{figure*}[tb]
\sidecaption
\includegraphics[width=12cm]{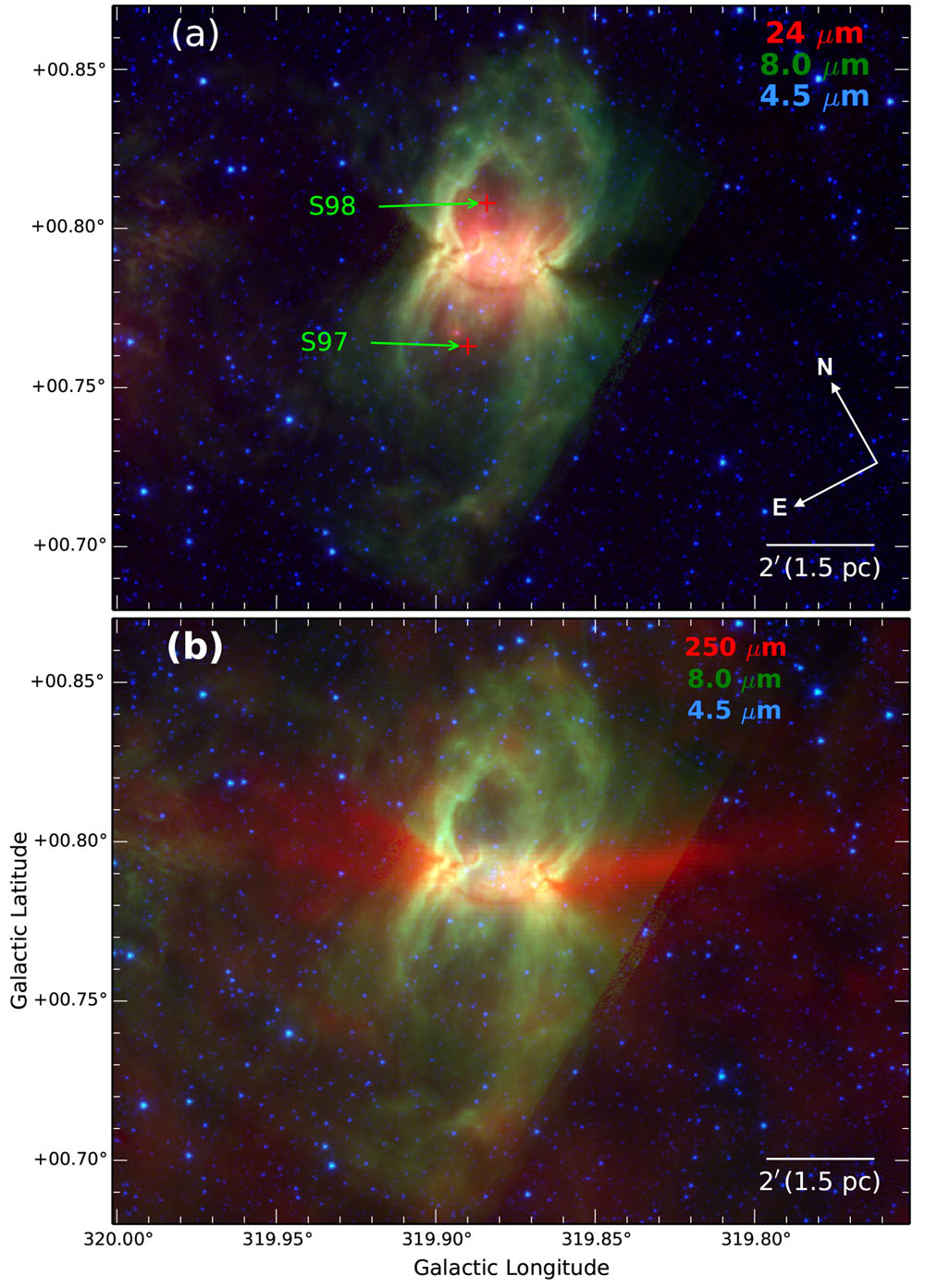}
  \caption{Bipolar nebula G319.88$+$00.79, composed of the S97 \& S98 bubbles (Churchwell et al. \cite{chu06}; the centres of the bubbles are indicated). {\it (a)} Composite colour image based on {\it Spitzer} observations. Red, green, and blue are for the 24~$\mu$m, 8.0~$\mu$m, and 4.5~$\mu$m emissions, respectively. {\it (b)} Composite colour image, with red, green, and blue for the 250~$\mu$m, 8.0~$\mu$m, and 4.5~$\mu$m emissions, respectively (logarithmic units). The cold parental filament (or sheet) appears in red, and the PDR surrounding the ionized lobes in green }
\label{G319a}
\end{figure*}

\begin{figure}[tb]
\centering
\includegraphics[width=85mm]{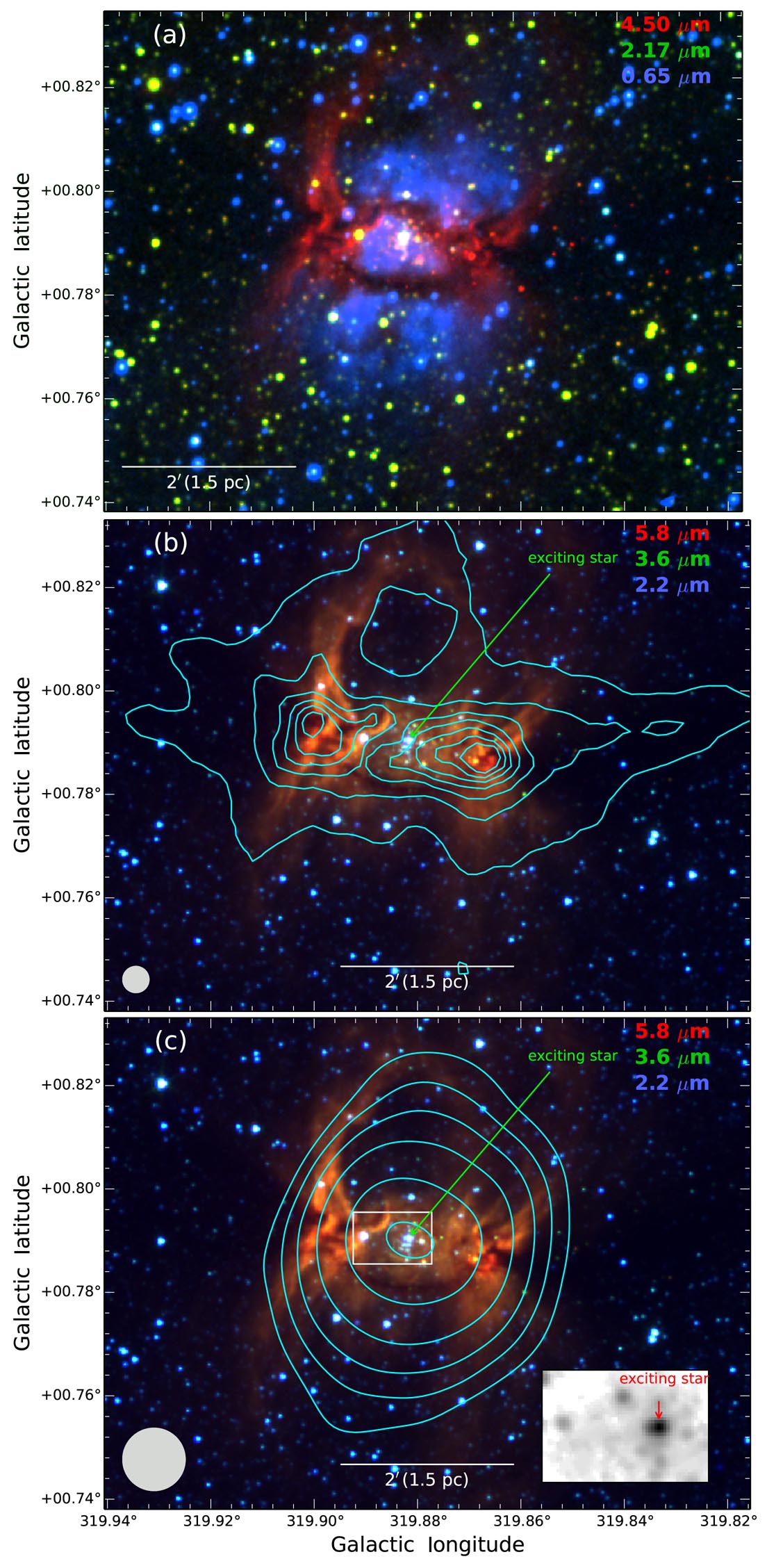}
  \caption{Centre of G319.88+00.79. {\it (a)} Colour image with -- in red, green, and blue, respectively -- the 4.5~$\mu$m emission, the $K$ 2MASS emission, and the SuperCOSMOS H$\alpha$ emission of the ionized gas (linear units). {\it (b)} Colour image with 5.8~$\mu$m, 3.6~$\mu$m, and $K$ 2MASS, in red, green, and blue, respectively (linear units). The 250~$\mu$m blue  contours are projected on the colour image (levels of 1000 to 7000~MJy/sr by steps of 1000~MJy/sr). The exciting star is identified. The beam of the 250~$\mu$m image is at the lower left. {\it (c)} the radio continuum blue contours from the SUMSS survey at 843~MHz (emission of the ionized gas; levels of 0.010, 0.025, 0.050, 0.100, 0.200, and 0.300~Jy/beam; the SUMSS beam is at the lower left) are projected on the previous colour image. {\it Inset} 2MASS $J$ band image of the exciting cluster (logarithmic units); the exciting star that appears elongated is probably double. }
\label{G319c}
\end{figure}

This nebula is a textbook example of a bipolar \HII\ region. It is composed of the two {\it Spitzer} bubbles S97 and S98 (Churchwell et al. \cite{chu06}).  Figure~\ref{G319a}~{\it (a)} (displaying the central 24~$\mu$m emission region) and Fig.~\ref{G319c}~{\it (c)} (displaying the central radio-continuum emission region) show that the two bubbles only enclose one \HII\ region, bright in its central part. Figure~\ref{G319a}~{\it (b)} shows that a bright cold dust filament is present at the waist of the nebula, which is conspicuous on all the {\it Herschel} SPIRE images (and even more conspicuous in the column density map, Fig.~\ref{dusttemperature1}~{\it (d)}). The two lobes, perpendicular to the filament, are observed at 8.0~$\mu$m, 24~$\mu$m, and 70~$\mu$m. All the region is symmetric with respect to the filament. However, one ionized lobe is enclosed well by the S98 bubble, whereas the other one (S97 bubble) is larger and open. Thus the \HII\ region is ionization-bounded on one side and possibly density-bounded on the other side.  The SPIRE images shows that the lobes (especially the upper one) are surrounded by a low emission layer. This emission possibly comes from neutral material collected during the expansion of the ionized gas.\\ 

\begin{figure}[tb]
\centering
\includegraphics[width=85mm]{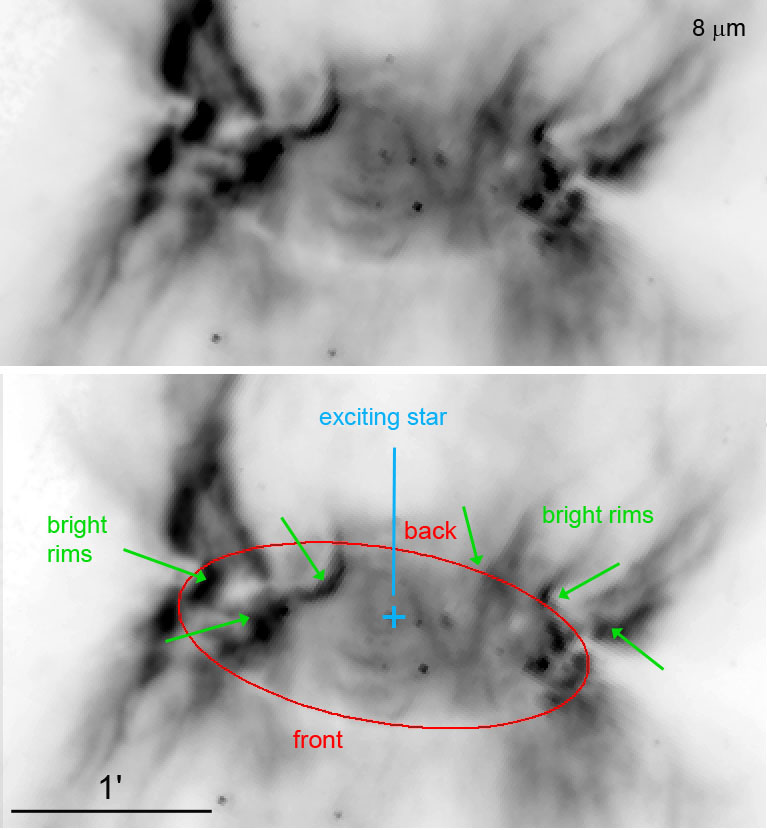}
  \caption{Centre of G319.88+00.79 at 8.0~$\mu$m showing the absorption feature (the ellipse), the bright rims present at the waist of the nebula (green arrows), and the position of the exciting star.}
\label{G319cc}
\end{figure}

G319.88+0.79 is an optical \HII\ region. Figures~\ref{G319c} and \ref{G319cc} show the central region at the waist of the bipolar nebula. We very clearly see a thin absorption feature, shaped like an incomplete elliptic ring encircling the waist of the nebula. It is observed in H$\alpha$ (SuperCOSMOS image, Fig.~\ref{G319c}~{\it (a)}), in the near-IR (2MASS) and also at {\it Spitzer}  8.0~$\mu$m (Fig.~\ref{G319cc}) and 24~$\mu$m. The extinction is high on the lower side of the ring, probably corresponding to the front side (along the line of sight). This absorption is due to the dust present in the dense molecular material encircling the waist of the nebula. Two bright clumps, called C1 and C2 (Sect.~5.3; Fig.~\ref{dusttemperature1}~{\it (d)}), are observed on each side at the waist of the nebula. We see, especially well at 5.8~$\mu$m or 8.0~$\mu$m (Fig.~\ref{G319cc}),  several bright rims, probably bordering small size high density structures embedded inside the two bright clumps observed at 250~$\mu$m (Fig.~\ref{G319c}~{\it (b)}).  A stellar cluster lies between these two condensations, at the centre of the elliptic absorption ring. It has been detected on the 2MASS images by Dutra et al. (\cite{dut03}). Its centre lies at $\alpha$(2000)$=15^{\rm h} 03^{\rm m} 33^{\rm s}$, $\delta$(2000)$=-57\degr$40.1$\arcmin$ ($l$=319\fdg882, $b$=$+$0\fdg790). We discuss it in Sect.~5.2. 

The overall morphology of the G319.88$+$00.79 complex is discussed in Sect.~7.2.

\subsection{What more do we know about this region?}

 It is a thermal radio source. The H109$\alpha$ and H110$\alpha$ radio recombinaison lines indicate a velocity of $-$38~km~s$^{-1}$ for the ionized gas (Caswell \& Haynes \cite{cas87}; the position observed,  $l$=319\fdg874  $b$=$+$0\fdg770, does not correspond to the radio peak, but the large beam $\sim$4$\farcm$4 covers most of the radio emission); in the same direction V(H$_2$CO)=$-41$~km~s$^{-1}$ (formaldehyde absorption). They conclude that G319.88$+$0.79 lies at the near kinematic distance of 2.6~kpc (using the Brand \cite{bra86} rotation curve). The near distance is confirmed by the fact that this \HII\ region has an H$\alpha$ counterpart; the distance is discussed in Sect.~5.2. The radio-continuum flux near 5~GHz is 1.6~Jy (Caswell \& Haynes, \cite{cas87}), and at 4.85~GHz it is 1.7~Jy (Kuchar \& Clark \cite{kuc97}). Using Equation~(1) in Simpson and Rubin (\cite{sim90})  and assuming an electron temperature of 5700~K as determined by Caswell \& Haynes, we obtain an ionizing photon flux of 1.20$\times$10$^{48}$~s$^{-1}$; according to Martins et al. (\cite{mar05}), this points to an O8.5V exciting star, if it is single. The SUMSS map of the region shows a peak of radio emission in the  direction of the central cluster, and a slight elongation in the direction of the two lobes (Fig.~\ref{G319c}~{\it (c)}).

The IRAS source IRAS14597-5728 (coordinates $l$=319\fdg8805, $b$=$+$0\fdg7911) lies 5$\arcsec$ away from the exciting star. Four IRDCs are listed by Peretto \& Fuller (\cite{per09}) in the direction or vicinity of G319.88$+$00.79 (Fig.~C.1 in Appendix~C). They are discussed in Sect.~7.1.

\subsection{The exciting cluster - The distance of the complex}

A cluster is present in the central region (Fig.~\ref{G319c}), visible in the near-IR on the 2MASS images and also in the {\it Spitzer} bands at 3.6~$\mu$m and 4.5~$\mu$m. The brightest star lies at $l$=319\fdg8818, $b$=$+$0\fdg7903; its 2MASS magnitudes are $J=10.476$, $H=9.610$, and $K=9.100$. We have seen (Sect.~5.1) that the radio flux of the \HII\ region indicates an exciting star of spectral type O8.5V or more massive if ionizing photons are absorbed by dust. Assuming that this star is an O8.5V star, its near-IR photometry indicates that it is affected by a visual extinction of $\sim$9.2~mag and lies at a distance of 2.1~kpc (using the photometry of O stars by Martins \& Plez \cite{mar06} and the extinction law of Rieke \& Lebofsky \cite{rie85}). This confirms a near distance for this region; however, this distance is uncertain for two reasons: 1) ionizing photons can be absorbed by dust inside the ionized region (the 24~$\mu$m emission comes from almost the same region than the radio continuum emission); if 70\% of the ionizing photons are absorbed by dust we need an O7V exciting star. Thus a distance of 2.45~kpc. 2) Figure~\ref{G319c} shows ({\it Inset}) that the exciting star is double, composed  of two stars separated by some 2$\farcs$4; the 2MASS magnitudes are thus overestimated. If the two stars were of equal intensity in $K$, their magnitude would be $K=9.85$; this leads to a distance of 2.85~kpc. Thus in the following we keep the near kinematic distance of 2.6~kpc ($\pm$0.4~kpc)\footnote{The median kinematic distance uncertainty estimated for a large sample of \HII\ regions by Anderson et al. (\cite{and12}) is 0.5~kpc. The distance uncertainty we estimate for this region is slightly smaller; the adopted distance range contains the two determinations of the photometric distance.}.\\

\subsection{Molecular condensations}

\begin{figure*}[tb]
\centering
\includegraphics[width=15cm]{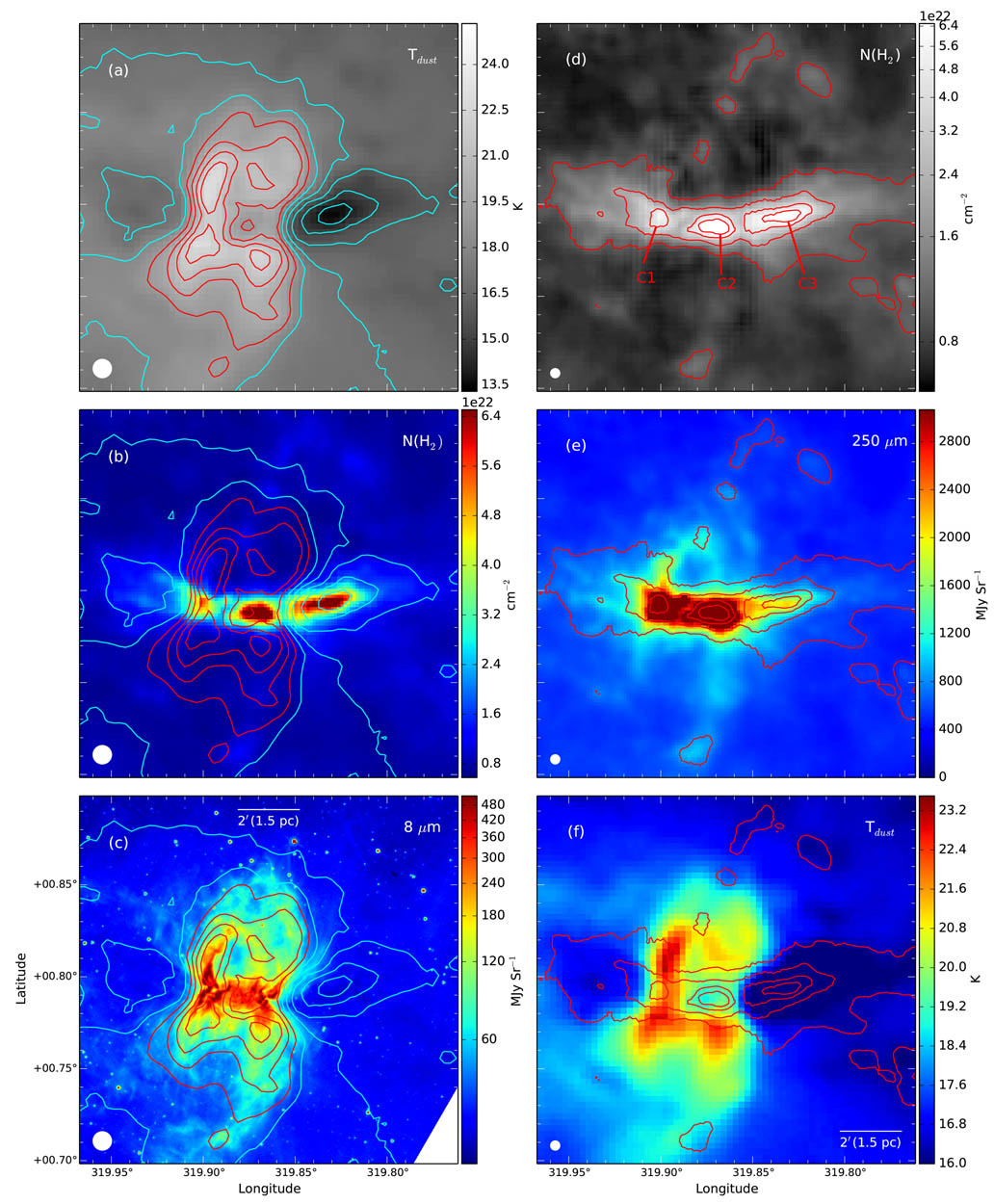}
  \caption{Temperature and column density maps of G319.88$+$00.79. The left side of the figure concerns the dust temperature. {\it (a)} The underlying grey image is the temperature map; the red contours correspond to temperatures of 19, 20, 21, and 22~K; the blue ones to temperatures of 14, 15, 16, 17, and 18~K. The 36$\arcsec$ beam of the temperature map (also that of the 500~$\mu$m map) is at the lower left. {\it(b)} The underlying image is the column density map in false colours, showing the cold parental filament. {\it (c)} The underlying image is the 8.0~$\mu$m emission map in false colours, showing  the bipolar lobes. The right side of the figure shows the column density. {\it (d)} The underlying grey image is the column density map; the red contours correspond to column densities of 1, 2, 4, and 6 times 10$^{22}$~cm$^{-2}$. The 18$\arcsec$ beam of the column density map (also that of the 250~$\mu$m map) is at the lower left. The three clumps discussed in the text are identified. {\it (e)} The underlying image is the 250~$\mu$m emission map in false colours. {\it (f)} The underlying image is the temperature map in false colours. } 
\label{dusttemperature1}
\end{figure*}

 Figure~\ref{dusttemperature1}~{\it (a)} shows the dust temperature map of the region. The parental filament (or sheet) is cold with a minimum temperature of 13.4~K in the direction of a region located along the filament. The dust in the PDRs surrounding the two ionized lobes is warmer with a maximum temperature of 23.2~K.
 
When looking at the {\it Herschel} SPIRE maps, we see two bright clumps, C1 and C2. They are adjacent to the ionized region, and their temperature is characteristic of PDRs (Anderson et al. \cite{and11}), respectively 22.3~K and 19.7~K (temperatures obtained by fitting the SEDs of these clumps, between 160~$\mu$m and 500~$\mu$m, with background subtraction). However, the column density map (Fig.~\ref{dusttemperature1}~{\it(d)}) shows the presence of another clump, C3, whose column density is as high as that of C2. Clump C3 is found at the coldest parts of the parental filament. It shows the dramatic influence of the dust temperature in the column density and mass estimates, hence the importance of the {\it Herschel} observations. The column density reaches values $\geq$8$\times$10$^{22}$~cm$^{-2}$  in the direction of C2 and C3.  Clump C2, which is relatively warm (19.7~K), is clearly under the influence of the adjacent \HII\ region, whereas clump C3 (mean temperature 14.4~K) is a zone of the parental filament (or sheet) that lies farther away from the \HII\ region; it has possibly not been  influenced by it (as shown by its low temperature). \\

\begin{table*}[h!]
\begin{threeparttable}[b]
\caption{Properties of the clumps in the field of G319.88$+$00.79.}
\begin{tabular}{rrrrrrrr}
\hline\hline\\
Name & $l$         & $b$          & N(H$_2$)\tnote{1}      & T$_{dust}$\tnote{2}  & M\tnote{3}          & R$_{\rm eq}$\tnote{4} & n(H$_2$)\tnote{5} \\
     & ($\degr$) & ($\degr$)  & (cm$^{-2}$)   & (K)         & ($\msol$)  & ($\arcsec$) & (cm$^{-3}$) \\
\hline \\
C~1 & 319.9012 & $+$00.7935     & 5.3$\times$10$^{22}$ & 22.3$\pm$1.9        & 303 (246$^{-49}_{\,+72}$)   & 25.2 (0.32~pc)   & 5.4$\times$10$^4$ \\ \\
C~2 & 319.8668 & $+$00.7868     & 8.3$\times$10$^{22}$ & 19.7$\pm$1.4        & 497 (458$^{-85}_{\,+120}$)   & 24.9 (0.31~pc)   & 8.6$\times$10$^4$ \\ \\
C~3 & 319.8324 & $+$00.7925     & 7.5$\times$10$^{22}$ & 14.4$\pm$1.2        & 766 (623$^{-170}_{\,+280}$)   & 119.7$\times$24.3 (1.5~pc$\times$0.3~pc) &  1.6--8.0$\times$10$^4$ \\ \\                
\hline
\label{G319condensations}
\end{tabular}
\begin{tablenotes}
    \scriptsize
\item[1]  Peak column density  corrected for a background value $\sim$4.6$\times$10$^{21}$~cm$^{-2}$ (the lowest value of column density in a field of radius 10$\arcmin$ centred on the exciting star; this low column density value is located at $l$=320\fdg04 $b$=$+$0\fdg78, outside any feature associated with the bipolar nebula and its parental cloud).  
\item[2]  Dust temperature derived from the SED, based on the integrated fluxes of the clumps in the range 160~$\mu$m--500~$\mu$m.
\item[3]  Mass obtained by integrating the column density in an aperture following the level at the N(H$_2$) peak's half intensity. In brackets is the mass obtained by integrating the 250~$\mu$m emission in an aperture following the level at the 250~$\mu$m peak's half intensity and using the temperature in column 5.
\item[4]  Equivalent radius of the aperture following the level at the N(H$_2$) peak's half intensity (beam deconvolved).
\item[5]  Mean density of the central region, enclosed by the aperture following the level at the N(H$_2$) peak's half intensity.
\end{tablenotes}
\end{threeparttable}
\end{table*}

The parameters of the three clumps are given in Table~\ref{G319condensations}. The Galactic coordinates given in columns 2 and 3 are those of the peaks of column density. The peak's column densities are given in column 4. The clumps C1, C2, and C3 are the only structures of the field with a column density $\geq$5$\times$10$^{22}$~cm$^{-2}$.  The dust temperatures, in Col.~5, are derived from the SEDs and based on the integrated fluxes of the clumps at 160~$\mu$m, 250~$\mu$m, 350~$\mu$m, and 500~$\mu$m, corrected for the background emission. The uncertainty on T$_{dust}$ given in Table~\ref{G319condensations} corresponds to an uncertainty of $\sim$15\% on the fluxes\footnote{Note that the uncertainty on T$_{dust}$ is possibly slightly higher.  Using the dust opacity values of Ossenkopf \& Henning (\cite{oss94}) leads to temperatures that are higher by about 10\%. The temperatures obtained without background correction are fainter by about 4\%}.  To estimate the mass given in Col.~6 we have integrated the column density in an aperture corresponding to half the peak's value. (These apertures are displayed in Fig.~B.1, Appendix~B.) We are thus dealing with the central regions of the clumps. The mass given in brackets has been estimated using the 250~$\mu$m flux integrated in an aperture following the level at the 250~$\mu$m peak's half-intensity and assuming the mean temperatures given in Col.~5. They differ by less than 20\% of the masses estimated from the column density map. The uncertainty on the mass given in Col.~6 corresponds to the uncertainty on the temperature given in Col.~5\footnote{The uncertainty in the mass is increased by the uncertainty on the distance. For example, a distance uncertainty of 0.4~kpc leads to a mass uncertainty of $\sim$75~\msol.  A larger source of uncertainty in the mass comes from the uncertain opacities. As an example, using the opacities of Ossenkopf \& Henning (\cite{oss94}) results in an increase in the dust temperature (24.6~K instead  of 22.3~K for C1) and therefore in a large decrease in the mass (150$\msol$ instead of 246$\msol$ for C1).}. The equivalent beam deconvolved radius at half-intensity is given in Col.~9. A mean density is estimated in Col.~10. (Using the mass and size of the central region and assuming an uniform density in this structure. This density is  uncertain because we do not know the length of the clump along the line of sight and assume that the clump is spherical.) Clump C3 is far from spherical, and its size at half column density is 119.7$\arcsec$ (along the filament) $\times$ 24.3$\arcsec$ (1.5~pc $\times$ 0.3~pc). If we assume that it is cylindrical (like a 1D filament) with a length of 119.7$\arcsec$ and a radius of 12.15$\arcsec$, then its mean density is $\sim$8.0$\times$10$^4$~cm$^{-3}$. If its length along the line of sight is $\sim$119.7$\arcsec$ (more like a 2D structure), then its mean density is $\sim$1.6$\times$10$^4$~cm$^{-3}$.\\

\subsection{Velocity field of the molecular clumps}

Two fields have been observed by MALT90 in this region: G319.901$+$00.792 and G319.872$+$00.788. They overlap and cover the C1 and C2 clumps, and the second field covers part of C3. (C3, less bright than C1 and C2 at 870~$\mu$m, has not been targeted by MALT90.)

\begin{figure}[h!]
\centering
\includegraphics[width=85mm]{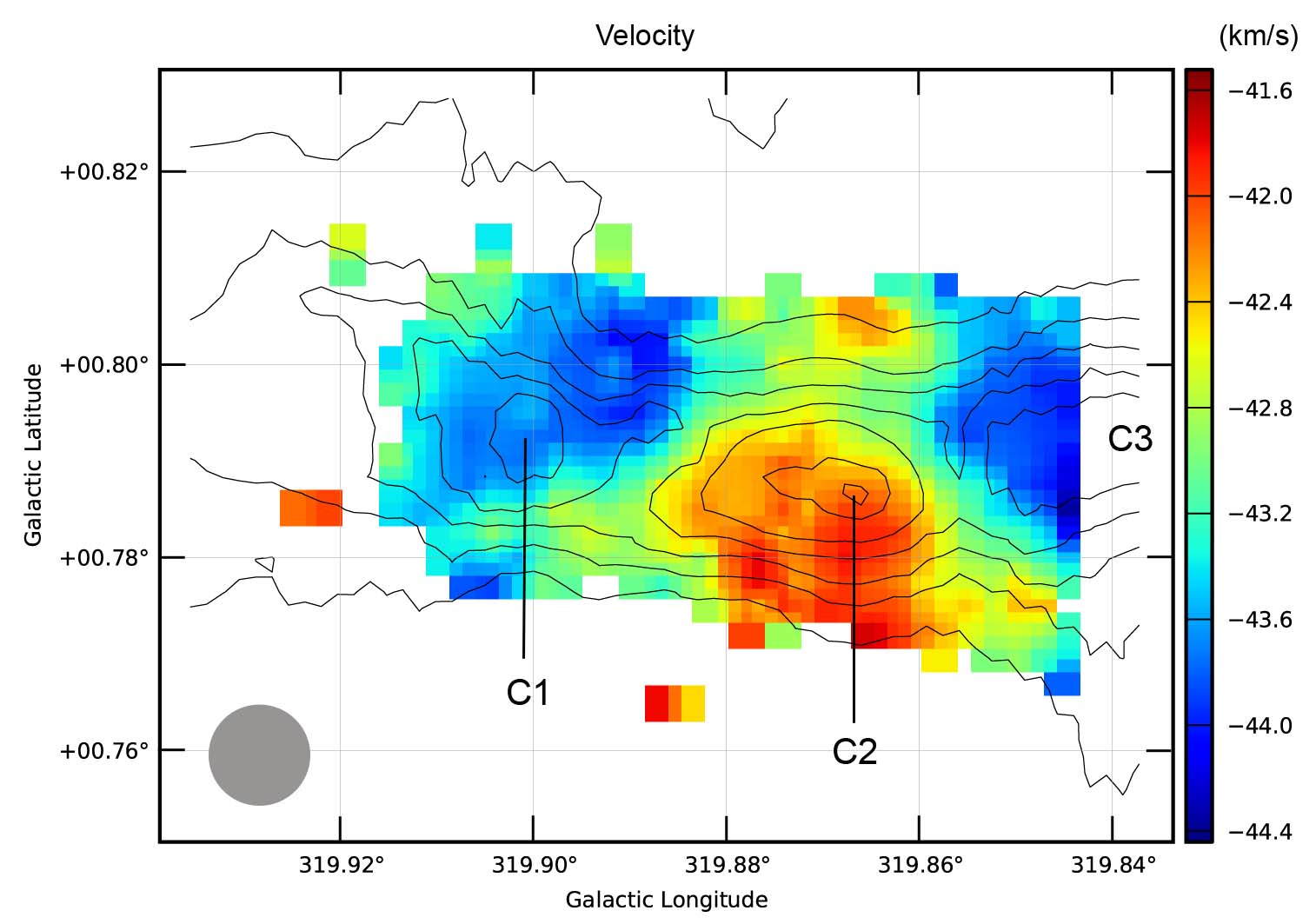}
  \caption{Velocity field of the clumps located at the waist of G319.88$+$00.79. These results are obtained by MALT90 with the HCO$^+$ (1-0) transition. What is shown here is the central velocity from doing a Gaussian fit to the line; only pixels with a ratio S/N$\geq$5 are shown here. The contours are for the column density (levels of 1, 1.5, 2, 3, 4.5, 7, and 8.5$\times$10$^{22}$~cm$^{-2}$). The 38$\arcsec$ beam of the MALT90 observations is at the lower left. }
\label{malt1}
\end{figure}

Figure~\ref{malt1} presents the radial velocity field of the C1 and C2 clumps, obtained with MALT90 observations in the H$^{12}$CO$^+$ (1-0) emission line. Similar results (not shown here) are obtained with the N$_2$H$^+$ molecular line. The velocity of the clumps lies in the range $-$41.5 to $-$44.5~km~s$^{-1}$, which is similar enough to that of the ionized gas ($-$38~km~s$^{-1}$), indicating that the molecular clumps and the \HII\ region are associated\footnote{The velocity of the ionized gas has been obtained with a rather low resolution ($\sim$4$\arcmin$; Caswell \& Haynes \cite{cas87}), which prevents us from estimating the relative position of the clumps with respect to the ionized region.}. Clump C1 and the filament (only a part of C3) have a mean velocity in the range $-$43.5 to $-$44~km~s$^{-1}$ and a line width in the range 3 to 4~km~s$^{-1}$. C2 has a mean velocity in the range $-$42 to $-$42.5~km~s$^{-1}$. Towards C2, the clump with the highest column density, the HCO$^+$ line shows a double-peaked profile, which we believe is due to self-absorption. Nevertheless, because the velocity field shown is that obtained by doing a single Gaussian fit to a double-peaked profile, the resulting velocity lies close to the position of the self-absorption, so close to the expected systemic velocities of the clump (within 1~km/s).

\subsection{Young stellar objects}

\begin{figure}[tb]
\centering
\includegraphics[width=85mm]{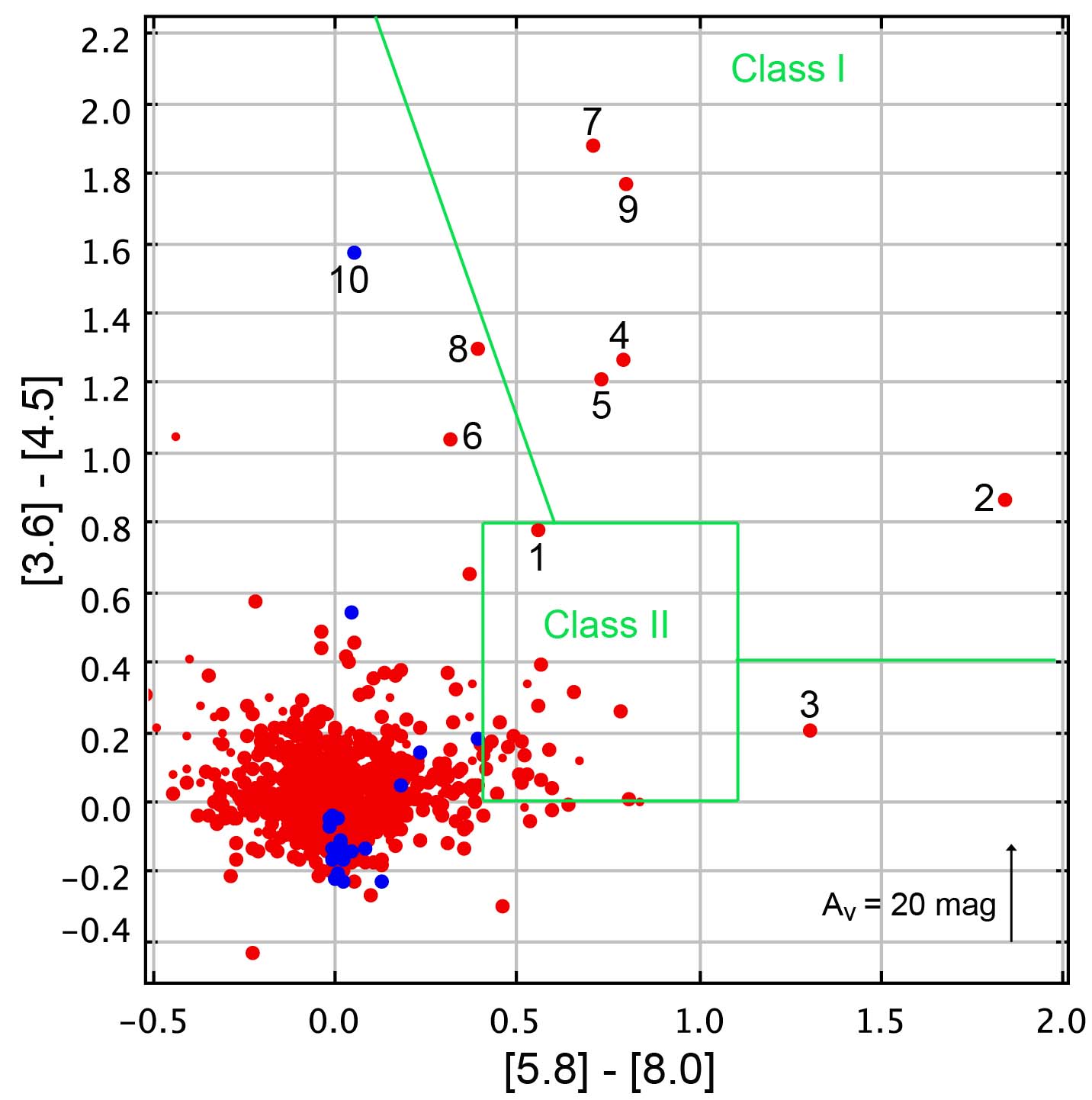}
  \caption{Colour-colour diagram, [3.6]$-$[4.5] versus [5.8]$-$[8.0], for all {\it Spitzer}-GLIMPSE sources with measurements in the four bands, and situated at less than 7$\arcmin$ from the exciting star of G319.88$+$00.79. The big red dots correspond to sources with measurements more accurate than 0.2~mag in all four bands; the small red dots are for the less accurate sources. The blue dots are for candidate xAGB stars, following Robitaille et al. (\cite{rob08}).  The location of YSOs dominated by a disk (Class~II YSOs) and by an envelope (Class~I YSOs) are indicated, following Allen et al. (\cite{all04}; see also Megeath et al. \cite{meg04}). The extinction law is that of Indebetouw et al. (\cite{ind05}). The sources discussed in the text, mainly candidate Class~I YSOs, are identified.}
\label{champ319f}
\end{figure}

\begin{figure}[tb]
\centering
\includegraphics[width=85mm]{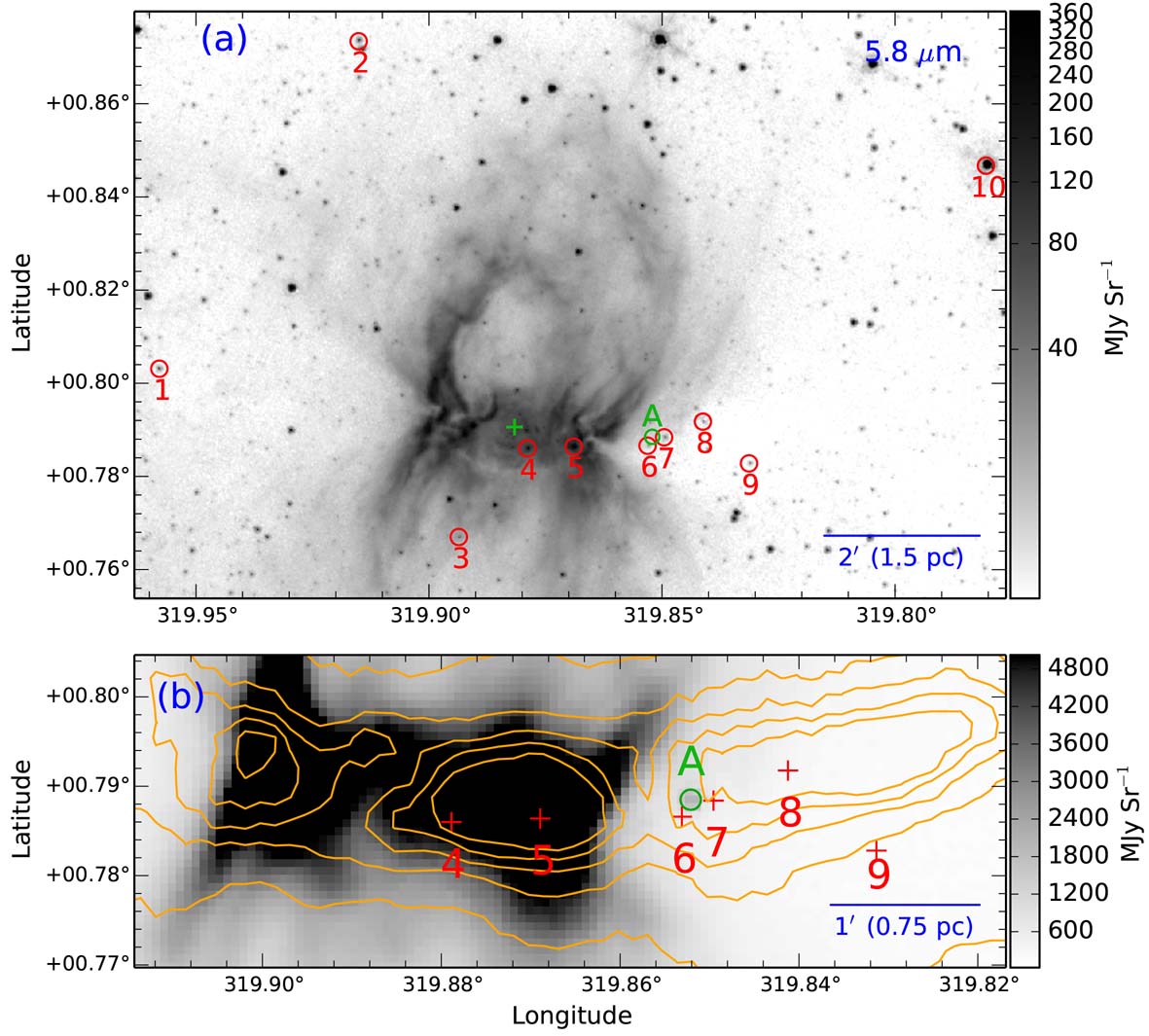}
  \caption{Identification of the sources discussed in the text in the field of G319.88$+$00.79. {\it (a)} All sources located at less than 7$\arcmin$ of the central exciting star (green plus). Source A is a compact 70~$\mu$m source. The underlying grey image is the {\it Spitzer} 5.8~$\mu$m image. {\it (b)} Sources located at the waist of the bipolar nebula and along the parental filament. The orange contours are for the column density (contours at 1, 2, 3, 4, and 5 $\times$ 10$^{22}$~cm$^{-2}$). The underlying grey image is the 70~$\mu$m image showing source A.}
\label{champ319g}
\end{figure}

\begin{table*}
\caption{Sources in the field of G319.88$+$00.79, mainly candidate Class~I YSOs or objects discussed in the text.}  
\resizebox{18cm}{!}{
\begin{tabular}{rrrrrrrrrrrrrr}
\hline\hline
Name & l         & b         & $J$   & $H$   & $K$   & [3.6]  & [4.5]  & [5.8]  & [8.0]  & [24]    & S(70$\mu$m) &  $\alpha$ & [8]$-$[24]  \\
     & ($\degr$) & ($\degr$) & (mag) & (mag) & (mag) & (mag)  & (mag)  & (mag)  & (mag)  & (mag)   & (Jy)      &  & (mag) \\ \\
\hline
1 & 319.95786 & +00.80312 & 14.334$\pm$0.060 & 12.813$\pm$0.035 & 11.980$\pm$99.999 & 10.879$\pm$0.053 & 10.109$\pm$0.056 & 9.551$\pm$0.037 & 8.992$\pm$0.033         & 6.63 ROB & $\leq$0.2$^*$ & -0.83  & 2.36 \\
2 & 319.91510 & +00.87341 & 15.521$\pm$0.478 & 14.278$\pm$0.358 & 13.042$\pm$0.137 & 11.532$\pm$0.086 & 10.679$\pm$0.044 & 9.063$\pm$ 0.030 & 7.225$\pm$0.020         & 1.43 ROB & 2.97$\pm$0.19 CuTEx & 1.58 & 5.79\\
3 & 319.89360 & +00.76700 & 12.029$\pm$0.024 & 11.430$\pm$0.027 & 11.134$\pm$0.021 & 10.901$\pm$0.047 & 10.706$\pm$0.068 & 10.274$\pm$0.102 & 8.967$\pm$0.129         & (0.129 ROB) & $\leq$0.3$^*$ &  &  \\
4 & 319.87887 & +00.78600 & & &                                                    & 10.961$\pm$0.156 & 9.701$\pm$0.199 & 8.378$\pm$0.065 & 7.584$\pm$0.079            & 3.76$\pm$0.3$^*$ & 4$\pm$2$^*$ & & 3.82 \\  
5 & 319.86893 & +00.78639 & 17.314$\pm$99.999 & 16.758$\pm$99.999 & 13.791$\pm$0.048 & 10.457$\pm$0.167 & 9.249$\pm$0.071 & 7.694$\pm$0.092 & 6.96$\pm$0.10$^*$                & 3.54$\pm$0.3$^*$ &  & 1.06 & 3.42 \\
6 & 319.85309 & +00.78660 & & &                                                    & 12.578$\pm$0.058 & 11.545$\pm$0.079 & 10.732$\pm$0.081 & 10.416$\pm$0.139          & $\geq$8.0$^*$ & $\leq$0.2$^*$  & & $\leq$2.4 \\
7 & 319.84955 & +00.78840 & & &                                                    & 12.859$\pm$0.096 & 10.988$\pm$0.069 & 10.006$\pm$0.055 & 9.301$\pm$0.047         &  5.71 ROB & $\leq$0.2$^*$  & & 3.59 \\  
8 & 319.84122 & +00.79177 & & &                                                    & 14.011$\pm$0.092 & 12.724$\pm$0.115 & 11.891$\pm$0.131 & 11.502$\pm$0.109        & 8.00$\pm$0.50$^*$ & $\leq$0.2$^*$ &  & 3.50 \\
9  & 319.83133 & +00.78281 &  &  &  & 13.286$\pm$0.064 & 11.528$\pm$0.060 & 10.391$\pm$0.058 & 9.590$\pm$0.028                                                        & 5.13$\pm$0.15$^*$ & $\leq$0.2$^*$ & & 4.46 \\
10 & 319.78050 & +00.84672 & 8.057$\pm$0.024 & 6.459$\pm$0.017 & 5.749$\pm$0.016 & 7.207$\pm$0.278 & 5.641$\pm$0.066 & 5.286$\pm$0.033 & 5.231$\pm$0.021              & 4.92$\pm$0.15$^*$ & $\leq$0.2$^*$ & -2.57 & 0.31 \\
A & 319.85210$^*$ & +00.78850$^*$ & & & & & & & & 5.98$\pm$0.15$^*$ & 3.88$\pm$0.24 CuTEx &  & \\
\hline
\label{G319_YSOs}
\end{tabular}\\
}
\tablefoot{
Our own measurements are indicated by an $^*$. ROB indicates 24~$\mu$m magnitudes from Robitaille et al. (\cite{rob08}). CuTEx indicates values taken from the Hi-GAL catalogue, to be published (Molinari et al. 2015, in prep.). They have been obtained using the package CuTEx (Molinari et al. 2011); for an example of using CuTEx, see Elia et al. (2013). 
}
\end{table*}

We used the colour-colour diagram [3.6]$-$[4.5] versus [5.8]$-$[8.0] displayed Fig.~\ref{champ319f} to detect candidate Class~I YSOs or some other noteworthy sources. The field considered here is centred on the central exciting star, and it has a radius of 7$\arcmin$ to encompass the two lobes of the bipolar \HII\ region. The astrometry and photometry of these sources are given in Table~\ref{G319_YSOs}, extracted from the {\it Spitzer} GLIMPSE Source Catalogue and completed by 24~$\mu$m and 70~$\mu$m measurements. The spectral index $\alpha$ and the colour [8]$-$[24] are also given in this table.  Furthermore, a candidate Class~0 YSO is detected at 70~$\mu$m.  All the sources discussed in the text are identified in Fig.~\ref{champ319g}. We first discuss two distinctive sources: 

$\bullet$ Source \#3 appears as a stellar source on the {\it Spitzer}-GLIMPSE images, but as a region of extended emission on the MIPSGAL 24~$\mu$m image. Source 3 is one of the ``intrinsically red'' sources discussed by Robitaille et al. (\cite{rob08}), and  according to these authors, it is very bright at 24~$\mu$m ([24]=0.129~mag). It is probably the emission of the extended region that has been measured. The 2MASS photometry of the central source points to an early B star (possibly  B3V), affected by 5.8~mag of visual extinction, if at the same distance than the bipolar nebula (an assumption).

$\bullet$ Source \#10 lies outside the top lobe of the bipolar nebula. It is most probably an evolved star, according to its 2MASS photometry, its brightness in the {\it Spitzer}-IRAC bands, and its low [5.8]$-$[8.0] and [8.0]-[24] colours. Also its spectral index is strongly negative ($\alpha\leq-2.5$), and it has no detectable 70~$\mu$m counterpart. This evolved star is probably not associated with the bipolar nebula. \\

Source \#1 is located along the parental filament (on the left). It is probably a Class~II YSO, according to its spectral index of $-$0.83 and colour [8]$-$[24]=2.36. It has no detectable 70~$\mu$m counterpart. Source \#2 is a Class~I YSO with a spectral index of 1.58 and colour [8]$-$[24]=5.79; it is also detected and point-like at 70~$\mu$m and at 160~$\mu$m (flux of 0.77~Jy according to CuTEX). Source \#2 is located in a direction of rather low column density, 6.7$\times$10$^{21}$~cm$^{-2}$, thus a maximum external extinction A$_{V}\sim$7.2~mag leading to a corrected $\alpha\sim$1.4. This confirms that source \#2 is a Class~I YSO.) But the association of YSO \#2 with the bipolar nebula is highly uncertain because it lies outside the northern lobe. 

All the other candidate Class~I YSOs, sources \#4 to \#9, lie along the parental filament. According to their colour [8]$-$[24] YSOs \#4 and \#9 are Class~I, YSOs \#5, \#7, and \#8 are flat-spectrum sources, and YSOs \#6 is probably a highly reddened Class~II.  None of the YSOs \#6 to \#9 has a detectable 70~$\mu$m counterpart: it indicates that they  are low-luminosity sources ($\leq$5$\lsol$). YSO \#4 lies near the centre of the nebula. It has a 70~$\mu$m counterpart, but its flux is very uncertain, indicating a luminosity $\sim$80$\lsol$. YSO \#5 lies in the direction of the very centre of clump C2. Due to its position close to the PDR and close to bright rims emitting at 70~$\mu$m, we are unable to detect any counterpart. YSOs \#6, \#7, \#8, and \#9 lie on the border or centre of C3, in the direction of the cold parts of the parental filament. It is also in this direction that a 24~$\mu$m and 70~$\mu$m compact source is detected without any counterpart at shorter wavelengths: we call it A. Its SED peaks at 70~$\mu$m and decreases at longer wavelengths. (Its 160~$\mu$m flux is very uncertain, and it is not detectable at SPIRE wavelengths.) YSO A is probably the least evolved protostar. Its 70~$\mu$m flux $\sim$3.8~Jy indicates a central object with a luminosity $\sim$80~$\lsol$. For this source, $log(\lambda$F$_{\lambda}$(70)/$\lambda$F$_{\lambda}$(24))=1.66$\pm$0.90, so it is possibly an extreme Class~0 YSO with a high infall rate (a PBR according to Stutz et al. \cite{stu13}, Sect.~4.3). 

To conclude, except for YSO \#2, which is possibly not associated with the nebula, all candidate Class~0/I/flat-spectrum  sources lie either at the waist of the nebula (YSOs \#4 and \#5) or along the parental filament.

\begin{figure*}[tb]
\sidecaption
\includegraphics[width=12cm]{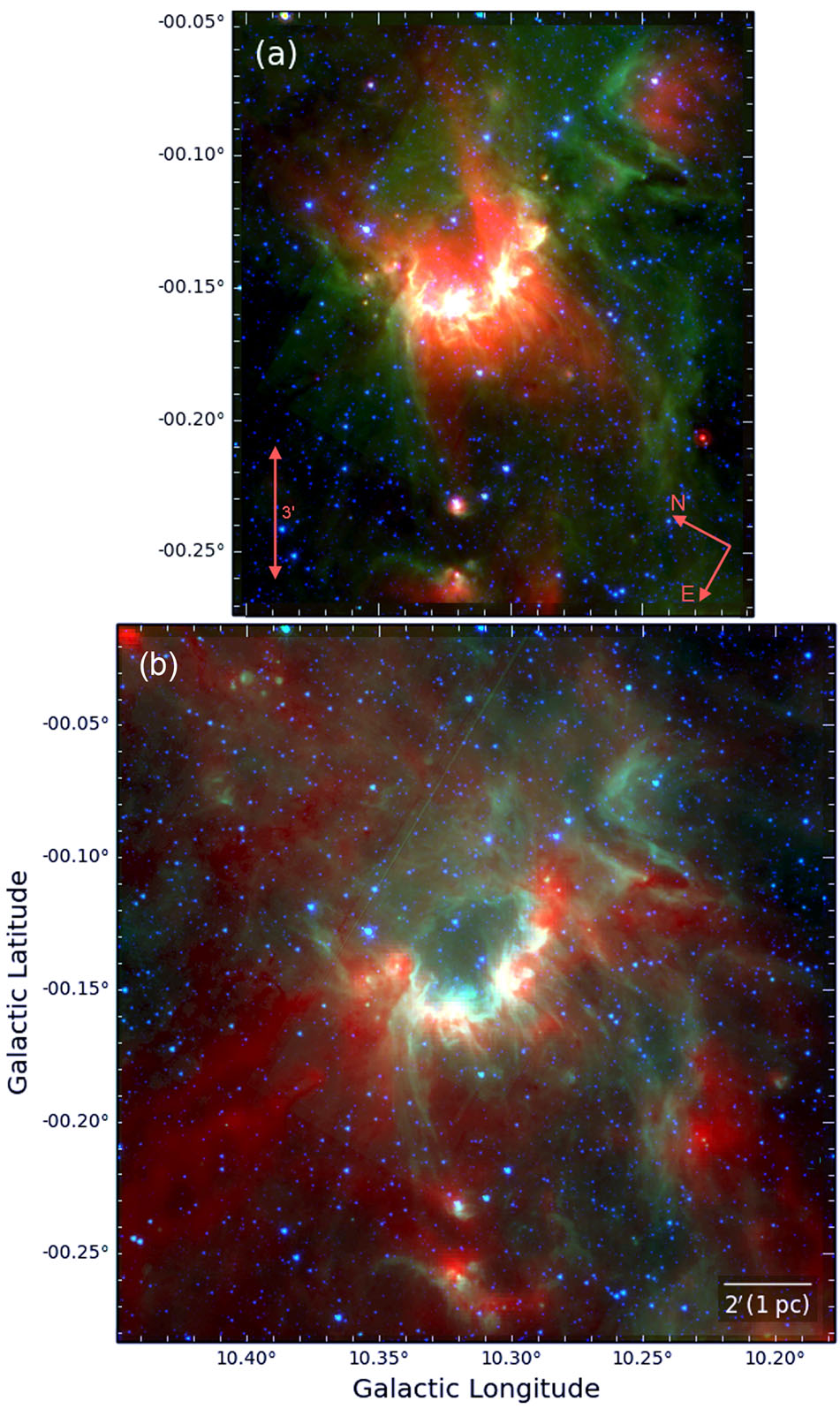}
  \caption{Bipolar nebula G010.32$-$00.15. {\it (a)} Composite colour image with red, green, and blue for the {\it Spitzer} 24~$\mu$m, 8.0~$\mu$m, and 4.5~$\mu$m emissions,  respectively. (Note that the 24~$\mu$m emission is saturated in the central region). {\it (b)} Composite colour image with red, green, and blue for the 250~$\mu$m cold dust emission, 8.0~$\mu$m emission, and 3.6~$\mu$m stellar emission,  respectively (the 250~$\mu$m emission is saturated in the C3 condensation, see Sect.~6.3).}
\label{champ10a}
\end{figure*}

\section{G010.32$-$00.15}

G010.32$-$00.15 is part of the well known bright star-forming region W31, studied by Beuther et al.~(\cite{beu11}) and presented as a superluminous star-forming complex. W31 contains numerous \HII\ regions. Two of them dominate the infrared and radio emission: G010.32$-$00.15 and G010.16$-$00.35 (see, for example, Kuchar \& Clark~\cite{kuc97}).  G010.32$-$00.15 is clearly a bipolar \HII\ region. Figure~\ref{champ10a}~{\it (a)} gives a {\it Spitzer} view of it; the 24~$\mu$m emission (red in the colour image) shows two lobes, the upper one open, the lower one possibly closed. The narrow waist is traced well by the bright 8.0~$\mu$m emission. This emission forms a closed ellipse (visible in Fig.~\ref{champ10a}~{\it (b)}), with a bright lower side  and a faint upper one. At this point it is difficult to know which one lies in the foreground and which in the background.

\begin{figure}[h!]
\centering
\includegraphics[width=75mm]{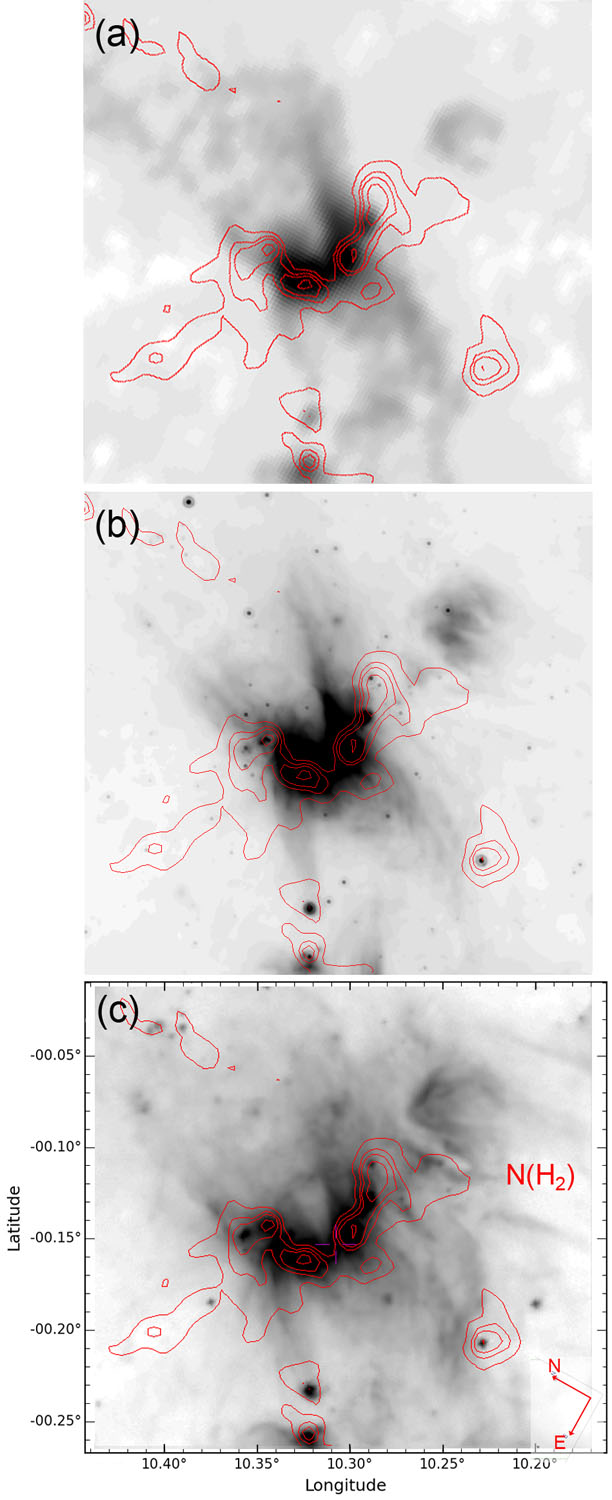}
  \caption{G010.32$-$00.15 at different wavelengths, showing the bipolar nature of the \HII\ region. The red contours correspond to the column density showing the dust clumps encircling the waist of the nebula (levels of 1.5, 1.0, 0.75, and 0.5$\times$10$^{23}$~cm$^{-2}$). The underlying grey images are {\it (a)} the VLA radio-continuum image at 21 cm, {\it (b)} the MIPSGAL map at 24~$\mu$m, and {\it (c)} the PACS map at 70~$\mu$m (all in logarithmic units to enhance the diffuse emission regions, especially the two ionized lobes).}
\label{champ10aa}
\end{figure}

G10.32$-$00.15 is a radio-continuum source. The radio map at 21 cm obtained by Kim \& Koo (\cite{kim01}) and displayed in Fig.~\ref{champ10aa}~{\it (a)} clearly shows the bipolar structure of the ionized region. The radio emission resembles the 24~$\mu$m emission (Fig.~\ref{champ10aa}~{\it (b)}) which is bright in the central region (where the 24~$\mu$m emission is saturated), and which presents two diffuse elongated lobes. As discussed in Sect.~3, a fraction of the 24~$\mu$m emission comes from dust located inside the ionized region. The lobes seen at 70~$\mu$m (Fig.~\ref{champ10aa}~{\it (c)}) surround the ionized lobes on the outside; this 70~$\mu$m emission is  mainly due to dust located in the PDR surrounding the ionized region. The bipolar morphology of G010.32$-$00.15 has been discussed by Kim \& Koo (\cite{kim01}, \cite{kim02}).\\

Molecular material is associated with this region. Kim \& Koo (\cite{kim02}) mapped the region in $^{13}$CO~(1-0) and CS~(2-1) with a resolution of the order of 1$\arcmin$. Their $^{13}$CO map shows that: i) the molecular material associated with G010.32$-$00.15 and that associated with G010.16$-$00.35 share similar velocities, mainly in the range 10 to 13 km~s$^{-1}$ (their figure 5); ii) ``the distribution of molecular gas is severely elongated in the direction orthogonal to the bipolar axis of G10.3$-$0.1.'' 
and  their figure 9 illustrates this last point. Beuther et al. (\cite{beu11}) discuss C$^{18}$O and $^{13}$CO observations obtained with an angular resolution $\sim$27.5$\arcsec$ and cold dust emission obtained at 870~$\mu$m (ATLASGAL survey) with a resolution $\sim$19$\arcsec$. Their results agree with those of Kim \& Koo for the velocity field. The high resolution of their 870~$\mu$m observations allowed them to detect four very bright cold dust clumps in the direction of G010.32$-$00.15  (their figure 1). These clumps are discussed in Sect.~6.3.\\

Figure~\ref{champ10a}~{\it (b)} presents a general view of the complex with the 250~$\mu$m emission of the cold dust in red, the 8.0~$\mu$m emission tracing the ionization front limiting the ionized material in green, and the stellar emission at 3.6~$\mu$m in blue. It shows the parental filament east of the nebula, and also the bright clumps present at the waist of the nebula.\\

Seven IRDCs are listed by Peretto \& Fuller (\cite{per09}) in the vicinity of G010.32$-$00.15 (Fig.~C.2 in Appendix~C). They are discussed in Sect.~7.1.

The overall morphology of the G010.32$-$00.15 complex is discussed in Sect.~7.2.

\subsection{Distance of G010.32$-$00.15 - Exciting cluster}

The distance of this region is very uncertain because different indicators give very different distances covering the range 2~kpc to 19~kpc. This very peculiar situation is explained in Appendix~A.1. In the following we try to identify the exciting star (or cluster) of G010.32$-$00.15 and determine its photometric distance.\\

What can we learn from the radio-continuum emission of the \HII\ region? The radio flux density of the \HII\ region is $\sim$14.7~Jy at 5~GHz (Du et al. \cite{du11}; Kuchar \& Clark \cite{kuc97}). Assuming an electron temperature $\sim$7000~K (Reifenstein et al. \cite{rei70}), we estimate an  ionizing photon flux (N$_{\rm Lyc}$ in s$^{-1}$) such that log(N$_{\rm Lyc}$)=48.65 or 49.72 for distances of 1.75~kpc or 6~kpc, respectively (justified in Appendix~A.1), hence the spectral type of the exciting star, if single, respectively O7V or O3V according to Martins et al. (\cite{mar05}). For a far kinematic distance of 15~kpc the exciting cluster should contain more than seven O3V stars. (Note that we do not take the possible absorption of ionizing photons by dust into account.)\\

\begin{figure}[h!]
\centering
\includegraphics[width=80mm]{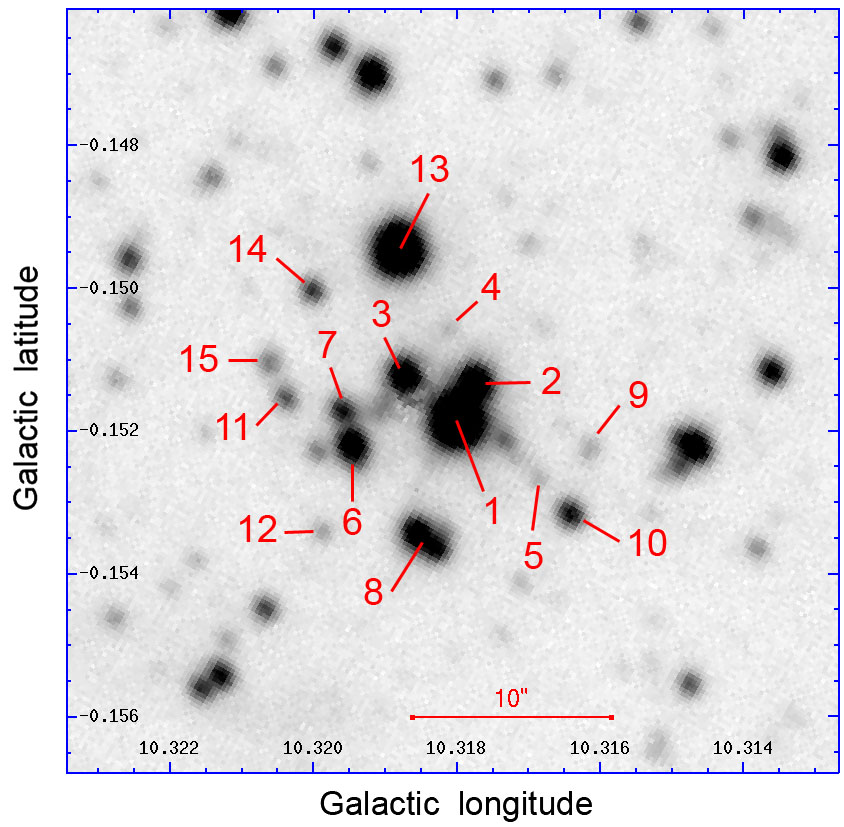}
  \caption{Exciting cluster of G010.32$-$00.15. We have identified the stars located at less than 10$\arcsec$ of the exciting star (\#1). (Their number increases with their distance to the exciting star.) The underlying grey image is the UKIDSS $J$ image (logarithmic units).}
\label{champ10bb}
\end{figure}

\begin{figure}[h!]
\centering
\includegraphics[width=85mm]{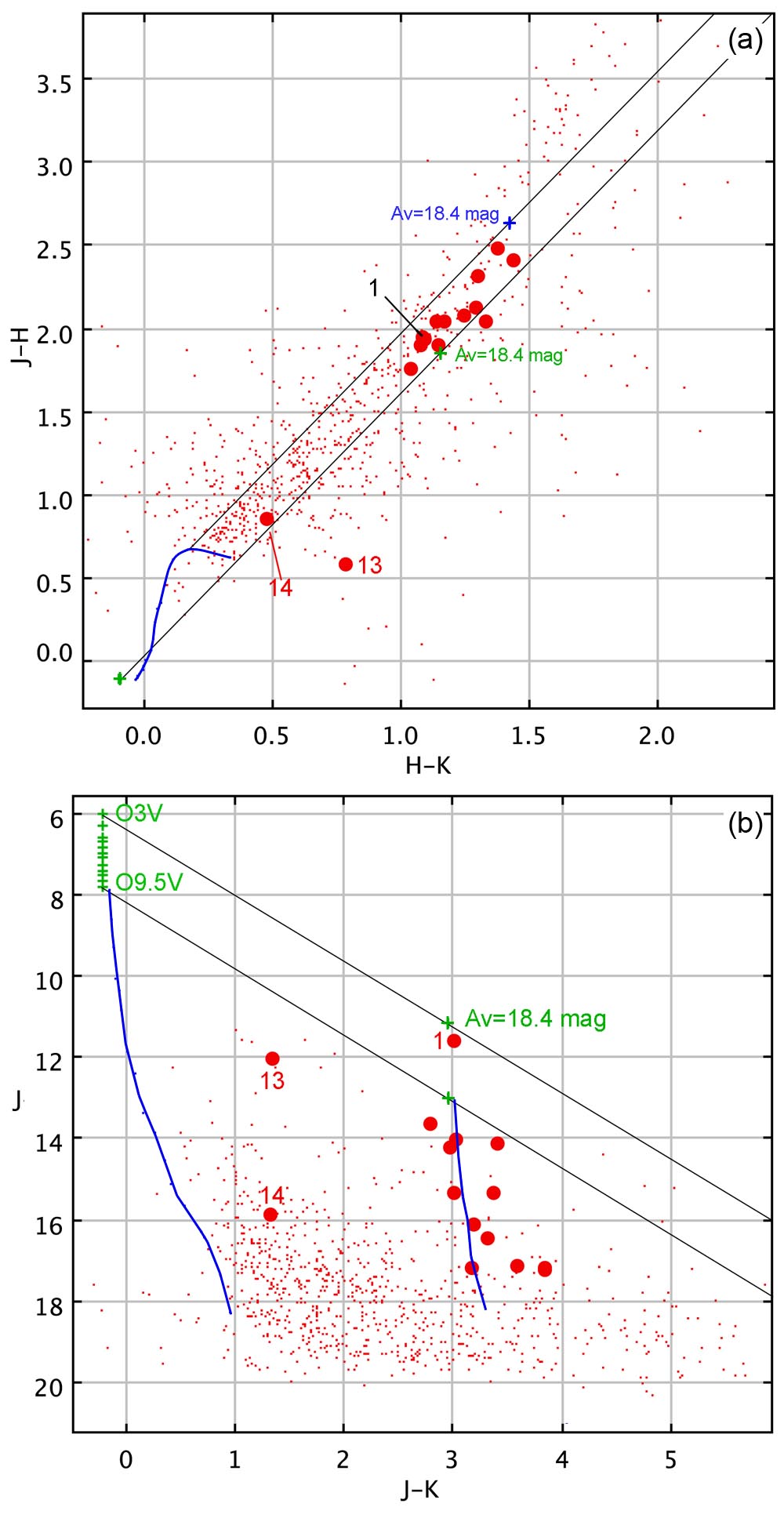}
  \caption{Near IR photometry of the central exciting cluster of G010.32$-$00.15. {\it (a)} $J-H$ versus $H-K$ diagram of the cluster's stars discussed in the text (photometry from the UKIDSS catalogue). The small red dots correspond to stars located at less than 90$\arcsec$ of the exciting star (\#1). The big red dots correspond to the stars identified in Fig.~\ref{champ10bb}, located at less than 10$\arcsec$ from the exciting star. The green plus correspond to O stars, affected by 0 and 18.4~mag of visual extinction. The blue line is the main sequence for B stars and stars of later spectral types. Main sequence stars are expected to lie in the region contained between the two black reddening lines (derived from main sequence O and M0 stars). {\it (b)} $J$ versus $J-K$ diagram; the red symbols have the same meaning as in Fig.~\ref{champ10c}~{\it (a)}. The two black lines limit the zone occupied by reddened O stars at a distance  of 1.75~kpc. The main sequence is also drawn for this distance, and for a visual extinction of 0~mag or 18.4~mag (extinction law of Rieke \& Lebofsky \cite{rie85}).}
\label{champ10c}
\end{figure}

The deep near-IR UKIDSS images show that a cluster is present at the centre of the G010.32$-$00.15 \HII\ region (facing the C2 clump; Sect.~6.3). A star, our \#1, dominates the cluster at all near-IR wavelengths (and up to 4.5~$\mu$m). Figure~\ref{champ10bb} displays the cluster in the $J$ band. We have identified the 14 stars located at less than 10$\arcsec$ of the exciting star, and brighter than 17.5~mag in $J$ (for accurate photometry). Table~\ref{G10cluster} (Appendix~A) gives the position of these stars, their $JHK$ magnitudes according to the UKIDSS catalogue, and their distance $d$ to star~\#1. Only the data for the central star (\#1), saturated on the UKIDSS images, comes from the 2MASS catalogue or from Bik et al. (\cite{bik05}). Figure~\ref{champ10c} presents the $(J-H)$ versus $(H-K)$ and $J$ versus $(J-K)$ diagrams for all the stars located at less than 90$\arcsec$ from the central star and for those at the centre of the cluster. The main sequence for a distance of 1.75~kpc, and a visual extinction of 18.4~mag is also drawn. (These  distance and extinction are justified below.) Bik et al.~(\cite{bik05}), from $K$-band spectroscopy, derive a spectral type O5V-O6V for star \#1. Using the synthetic photometry of O stars from Martins \& Plez (\cite{mar06}) and assuming a spectral type O5V for star~\#1, we derive a visual extinction of 18.8~mag and a distance of 1.6~kpc from the 2MASS magnitudes of this star (respectively 18.00~mag and 1.94~kpc from Bik et al. photometry). This explains the adopted mean values used in Fig.~\ref{champ10c}. This figure shows that,  except for two stars (\#13 and \#14), star \#1 and surrounding stars are most probably main sequence stars, lying at the same distance thus forming a cluster; the visual extinction of the stars in the cluster is in the range 16.6~mag -- 22.3~mag. Stars \#13 and \#14 are probably foreground stars.

Star \#1 dominates the cluster and is probably the  main exciting star of the \HII\ region. Independent of its spectral type, its near-IR magnitudes can constrain the maximum possible distance of the cluster. Let us assume that star \#1 is a O3V star; using 2MASS photometry, its visual extinction is 18.8~mag (independent of the distance), and its observed $K$ magnitude indicates a distance of 2.1~kpc (using the Bik et al.'s magnitudes, the maximum distance is 2.5~kpc). Since star \#1 cannot be brighter than an O3V star, its distance cannot be greater than 2.5~kpc. How can we reconcile a maximum distance of 2.5~kpc for G010.32$-$00.15 with the H$_2$CO and \HI\ absorption spectra that point to greater distances? A tentative explanation is given in Appendix~A.1.\\ 

To conclude, we have identified the exciting star of G010.32-00.14. It lies at the centre of a cluster, itself at the centre of the ionized region. This star is an O5V--O6V star with a visual extinction of 18.4$\pm$0.4~mag. Its spectrophotometric distance is 1.75$\pm$0.25~kpc. All this agrees with the radio-continuum flux (assuming that some ionizing photons are absorbed by dust) and with the near kinematic distance. We adopt this distance of 1.75~kpc in the rest of this paper\footnote{At this near distance, W31 is no longer one of the most luminous giant molecular cloud/high-mass star formation complexes in the Galaxy, as claimed by Beuther et al. (\cite{beu11}).}.

\begin{figure*}[tb]
\centering
\includegraphics[width=15cm]{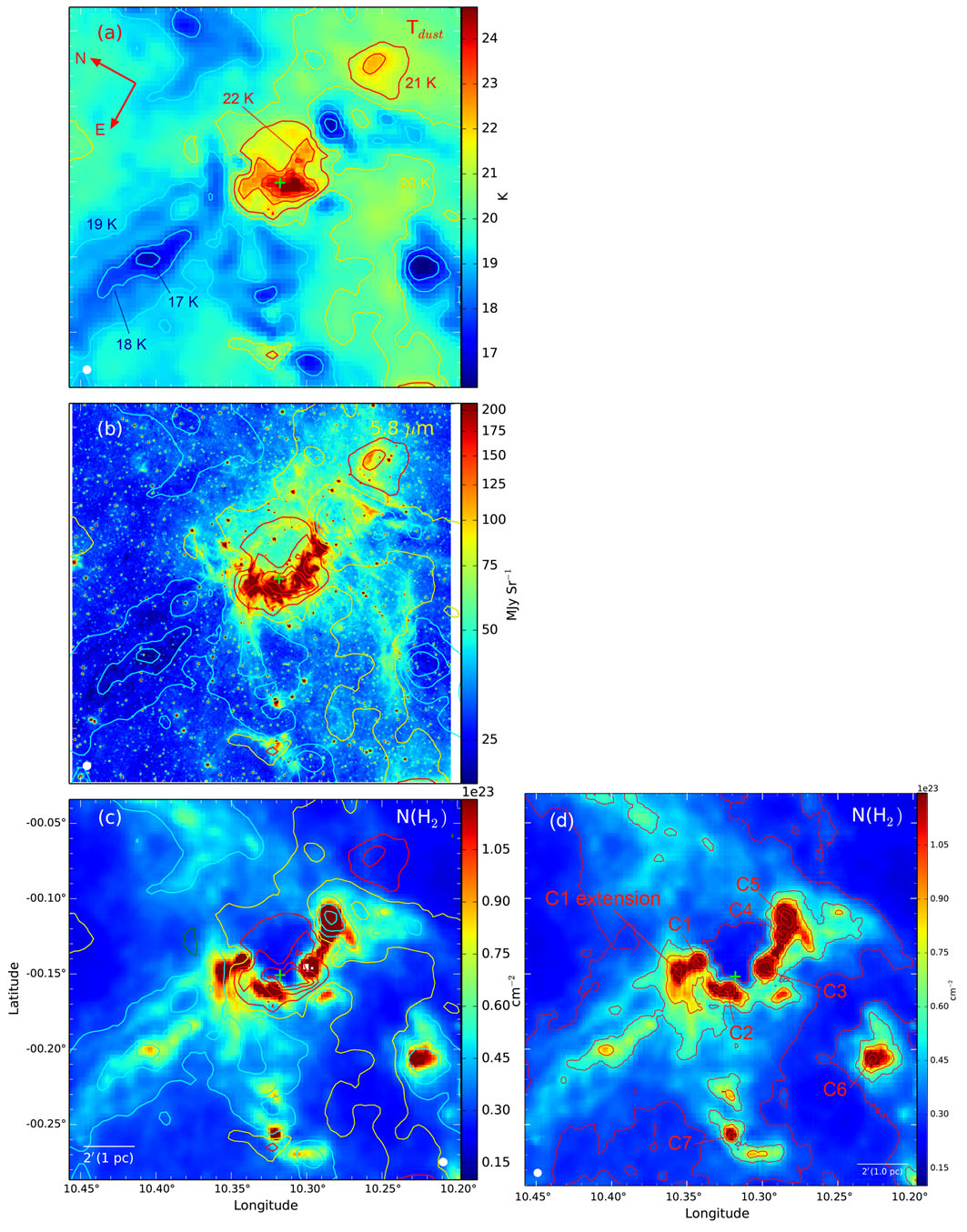}
  \caption{Dust temperature and column density maps of G010.32$-$00.15. The three left images deal with the temperature; the green cross shows the position of the exciting star. {\it (a)} The underlying grey image is the temperature map; the red contours correspond to temperature of 21, 22, 23, and 24~K, the blue ones to temperatures of 19, 18, and 17~K, the yellow one to 20~K. {\it (b)} The same contours are superimposed on the {\it Spitzer} 8.0~$\mu$m image, which enhances the  waist  of the nebula. {\it (c)} The same contours are superimposed on the column density map. {\it (d)} Column density map (contour levels of 0.3, 0.5, 0.75, 1.0, 1.5, and 2.0 $\times$10$^{23}$~cm$^{-2}$). The clumps discussed in the text are identified.} 
\label{champ10e}
\end{figure*}

\subsection{Dust temperature and column density maps} 

Figure~\ref{champ10e}~{\it (a)} presents the dust temperature map of this region. It shows that the warmer dust is found in the central region, close to the central ionized region, with the highest temperature (25.4~K) obtained in the direction of the PDR close to the central exciting cluster (at the edges of  clumps C2 and C3, on their sides turned towards the exciting cluster and between them; the clumps are identified in Fig.~\ref{champ10e}~{\it (d)}). Low temperatures are found in the eastern filament (as low as 16.6~K), and in the direction of two clumps (as low as 16.2~K in clump C6 and 16.6~K in C5).\\

Figure~\ref{champ10e}~{\it (d)} presents the column density map of this region. The column density is especially high over the region: higher than 2$\times$10$^{23}$~cm$^{-2}$ in the centre of the six clumps, C1 to C6  (Fig.~\ref{champ10e} and Table~\ref{G10condensations}). All these clumps (except C6) lie at the waist of the bipolar nebula. Clump C3 is saturated at 250~$\mu$m in its central region, and in unsaturated pixels the column density reaches 3.8$\times$10$^{23}$~cm$^{-2}$. A hole in the column density is present in the central region (in the direction of the exciting cluster) where the column density is of the order of 2$\times$10$^{22}$~cm$^{-2}$. The ionized gas lies in this central region, so this column density is probably that of the foreground or background extended emission. (We adopt this value for subtracting the background to estimate the parameters of the clumps in Sect.~6.3.) \\ 

In this region we again see the influence of the temperature on the determination of the column density. Clumps C5 and C6 are not among the brightest at 250~$\mu$m, but they are cold, and as a consequence their column density is high.\\

\subsection{Molecular clumps}

\begin{figure}[tb]
\centering
\includegraphics[width=85mm]{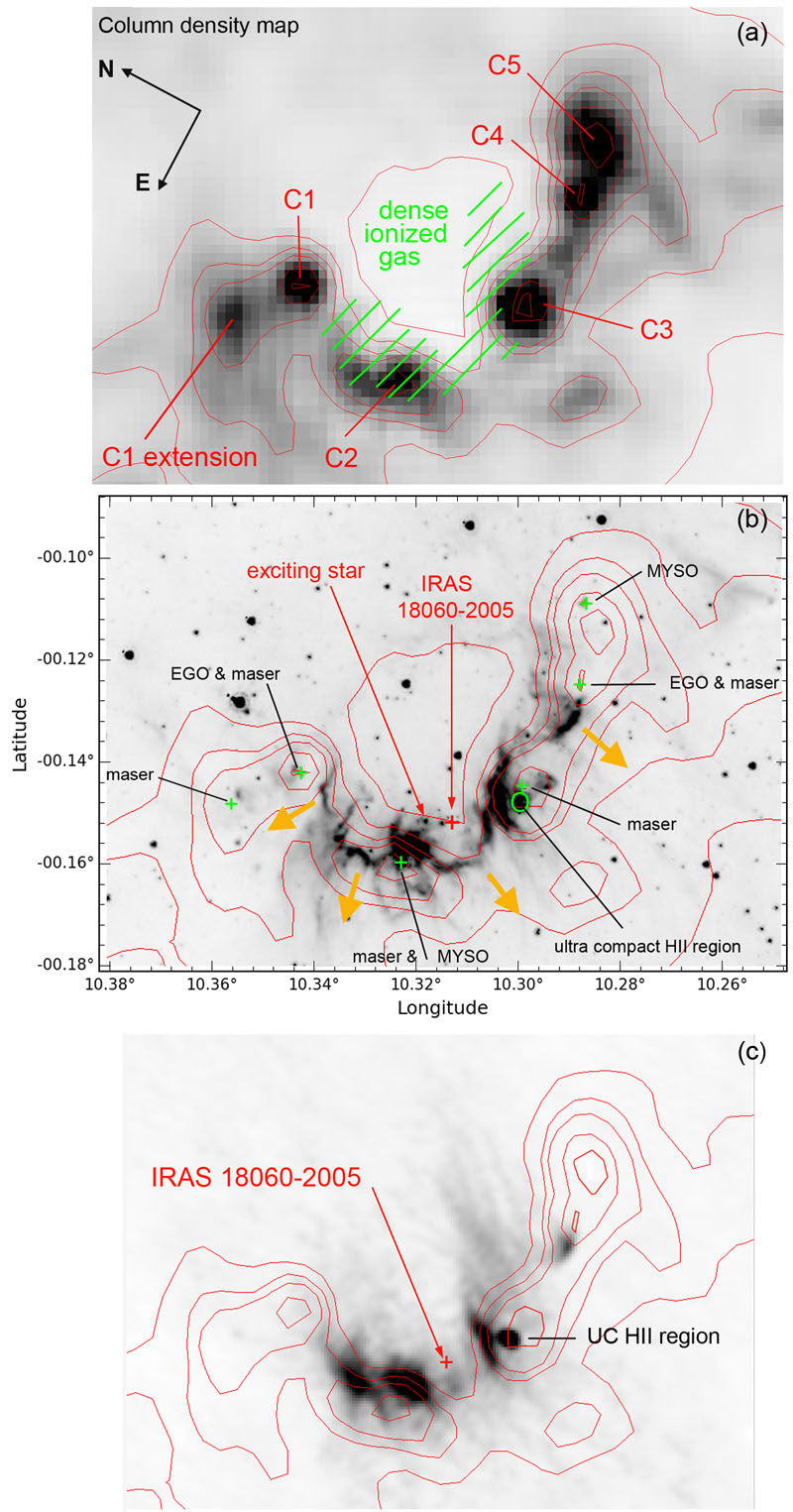}
  \caption{Identification of sources discussed in the text, in the vicinity of G010.32$-$00.15. {\it (a)} Column density map showing the five clumps present at the waist of the bipolar nebula (red contour levels at 0.3, 0.5, 0.75, 1.0, 1.5, and 2.0 $\times$10$^{23}$~cm$^{-2}$). {\it (b)} {\it Spitzer} 5.8~$\mu$m image; the plus symbols show the positions of the methanol masers, of the EGOs, the MYSOs, the UC \HII\ region, and IRAS18060-2005; the exciting star is also identified. The orange arrows show the pressure exerted by the ionized gas trying to escape from the central ionized region through the molecular shell, between the dense clumps. {\it (c)} MAGPIS image at 21~cm showing the UC \HII\ region and the dense ionized layers bordering the IF (especially the borders of C2 and C3 turned towards the exciting cluster). } 
\label{champ10f}
\end{figure}

The field of Fig.~\ref{champ10e} contains eight structures with a peak column density higher than 1.5$\times$10$^{23}$~cm$^{-2}$. The five brightest are located at the waist of the nebula; we call them C1 to C5. C1 and C3 are almost circular, and C1 is the most compact. All are adjacent to the ionization front and bordered by a bright rim at 8.0~$\mu$m (on the side turned towards the central exciting cluster). The parameters of the high column density clumps can be found in Table~\ref{G10condensations}. The first line of Cols.~2 and 3 gives the Galactic coordinates of their column density peaks. The second line of Cols.~2 and 3 gives the coordinates of the ATLASGAL emission peaks at 870~$\mu$m\footnote{The positions of the column density peaks and of the ATLASGAL 870~$\mu$m emission peaks given by Beuther et al. differ by less than 4$\arcsec$. This is compatible with the uncertainties on positions expected for these two sets of observations obtained with beams $\sim$19$\arcsec$.}, while Col.~4 gives their velocity, as  obtained from the NH$_3$ emission (Wienen et al. \cite{wie12}) or from the C$^{18}$O emission (Beuther et al. \cite{beu11}). Column 5 gives their peak column density. Column 6 gives -- only for information -- the mean dust temperature estimated from the temperature map (Fig.~\ref{champ10e}).  The mass, given in Col.~7, has been obtained by integration of the column density in an aperture following the level corresponding to half of the peak value (after correction for the background). This mass is the mass of the central region of the clump.  Column 8 gives the size of the central region, in arcsec (and parsec in brackets), estimated as the equivalent radius of the apertures used (hence the beam-deconvolved size at half peak value). Column 9 gives an estimate of the mean density of the central region, assuming that it is homogeneous.

The central regions of the C2 to C4 clumps are rather similar in terms of size, with a radius in the range 0.10~pc -- 0.12~pc, a peak column density a few 10$^{23}$~cm$^{-2}$, a mass in the range 220~$\msol$ -- 330~$\msol$, and a mean density of several 10$^5$~cm$^{-3}$. They are dense.  The central region of C1 is not very different, but it is smaller and denser, so it deserves to be called a core.  Clumps C5 and C6 are also dense and are slightly more extended and more massive than the C1 to C4 clumps. Clump C7 which is not associated with the region (on the basis of its velocity, see below), lies at an unknown distance, so we cannot estimate its mass and density. Two velocity components are observed in the direction of the C1 extension (Sect.~6.4); thus, two structures are present along this line of sight, and we do not know which fraction of the molecular material is associated with the bipolar nebula.

The C1 to C5 clumps form a structure that limits, at least on the southern side, the brightest part of the ionized region. Their velocities are similar to the velocity of the ionized gas, as shown by Table~\ref{G10condensations} and discussed in Sect.~6.4. Figure~\ref{champ10f}~({\it b}) shows that the IF is distorted between the clumps. We suggest that this is due to the high-pressure ionized gas trying to escape from the central region via low-density paths between the clumps. All clumps display signatures of massive-star formation at work (see Sect.~6.5); each of the C1 to C4 clumps contains a 6.67~GHz methanol maser in its very centre, at less than 3$\arcsec$ (0.025~pc) from the column density peak (Green et al. \cite{gre10}). Two of these masers in C1 and C4 lie close to an extended green object (EGO;  Cyganowski et al. \cite{cyg09}), another one in C3 lies close to an UC \HII\ region (Walsh et al. \cite{wal98}), and the last one in C2 lies close to a massive YSO  (YSO~\#4). Another methanol maser lies in the direction of the C1 extension. All these objects are identified in Fig.~\ref{champ10f} and are discussed in Sect.~6.5.\\

\begin{table*}[tb]
\begin{threeparttable}[b]
\caption{Molecular clumps in the field of G010.32$-$00.15 (C7 and a fraction of the C1 extension are not associated with the bipolar \HII\ region).}
\begin{tabular}{llllrrrlr}
\hline\hline\\
Name & $l$\tnote{1} & $b$\tnote{1} & V$_{\rm LSR}$                      & N(H$_2$)\tnote{1}       & T$_d$\tnote{2} &   Mass\tnote{3}    & R$_{\rm eq}$\tnote{4} & n(H$_2$) \\
     &   &   & (km~s$^{-1}$)                             & (10$^{23}$~cm$^{-2}$)    & (K)   & ($\msol$) & ($\arcsec$) (pc)  & (10$^5$~cm$^{-3}$) \\ 
\hline\\
C1 & 10.3430 & -00.1418 &                         & 3.27 & 19.6  & 184 & 5.7 (0.05)   & 55 \\
   & 10.3419 & -00.1422  & 12.14\tnote{6}, 10.1\tnote{7}   &                &       &           &              &   \\
C2 & 10.3232 & -00.1600 &                         & 2.09 & 22.0  & 223 & 14.4 (0.12)  & 4.2 \\
   & 10.3224 & -00.1604  & 12.02\tnote{6}, 11.9\tnote{7}   &                &       &           &              &   \\
C3\tnote{5} & 10.2998    & -00.1460         &        & $\geq$2.66 & 21.8  & 330 & 12.6 (0.11)  & 9.2  \\
   & 10.3001 & -00.1469  & 12.83\tnote{6}, 11.2\tnote{7}   &                &       &           &              &   \\
C4 & 10.2868 & -00.1249 &                         & 2.80 & 19.4  & 216 & 11.7 (0.10)  & 7.6  \\
   & 10.2876 & -00.1243  & 11.7\tnote{7}   &                &       &           &              &   \\ 
C5 & 10.2837 & -00.1137 &                         & 2.50 & 17.1  & 506 & 21.0 (0.18)  & 3.1 \\
   & 10.2836 & -00.11.36  & 10.6\tnote{7}   &                &       &           &              &   \\ 
C6 & 10.2282 & -00.2067 &                         & 1.76 & 16.3  & 250 & 18.7 (0.16)  & 2.2 \\
   & 10.23   & -00.21    & 12.14\tnote{6}  &                &       &           &              &   \\
\hline
C7 & 10.3215 & -00.2572 &                         & 1.61 & 19.4  &           & 9.9          &   \\
   & 10.32   & -00.26   & 32.15\tnote{6}   &                &       &           &              &   \\
C1 extension & 10.3551 & -00.1470 &               & 1.57  & 19.0  &           &  19.9       &  \\
\hline
\label{G10condensations}
\end{tabular}
\begin{tablenotes}
    \scriptsize    
\item[1] The coordinates (first line)and the column densities are those of the column density peaks measured on the column density map. The peak column densities have been corrected for a background emission ($\sim$0.2$\times$10$^{23}$~cm$^{-2}$, measured in the central region near $l$=10$\fdg$318 $b$=$-$0$\fdg$146). The coordinates (second line) are those of the ATLASGAL 870~$\mu$m emission peaks from Beuther et al. (\cite{beu11}).
\item[2] Mean temperature measured using the dust temperature map by integration in the aperture corresponding to the central region.
\item[3] Mass of the central region of the clumps obtained by integrating the column density in an aperture following the level at the N(H$_2$) peak's half intensity.
\item[4] Equivalent radius of the aperture following the level at the N(H$_2$) peak's half intensity (beam deconvolved).
\item[5] This clump is saturated at 250~$\mu$m; the coordinates and column densities have been measured on the column density map based on the 350~$\mu$m emission map; due to a smoothing by the beam (larger at 350~$\mu$m than at 250~$\mu$m), the peak column density is underestimated.
\item[6] V(NH$_3$) from Wienen et al. (\cite{wie12}); pointed observations in the direction of the ATLASGAL 870~$\mu$m intensity peaks; FWHM 40$\arcsec$.
\item[7] V(C$^{18}$O) from Beuther et al. (\cite{beu11}); C$^{18}$O spectra extracted from a velocity map towards the positions of the ATLASGAL 870~$\mu$m intensity peaks; FWHM 27$\farcs$5.
\end{tablenotes}
\end{threeparttable}
\end{table*}

\begin{figure*}[tb]
\centering
\includegraphics[width=18cm]{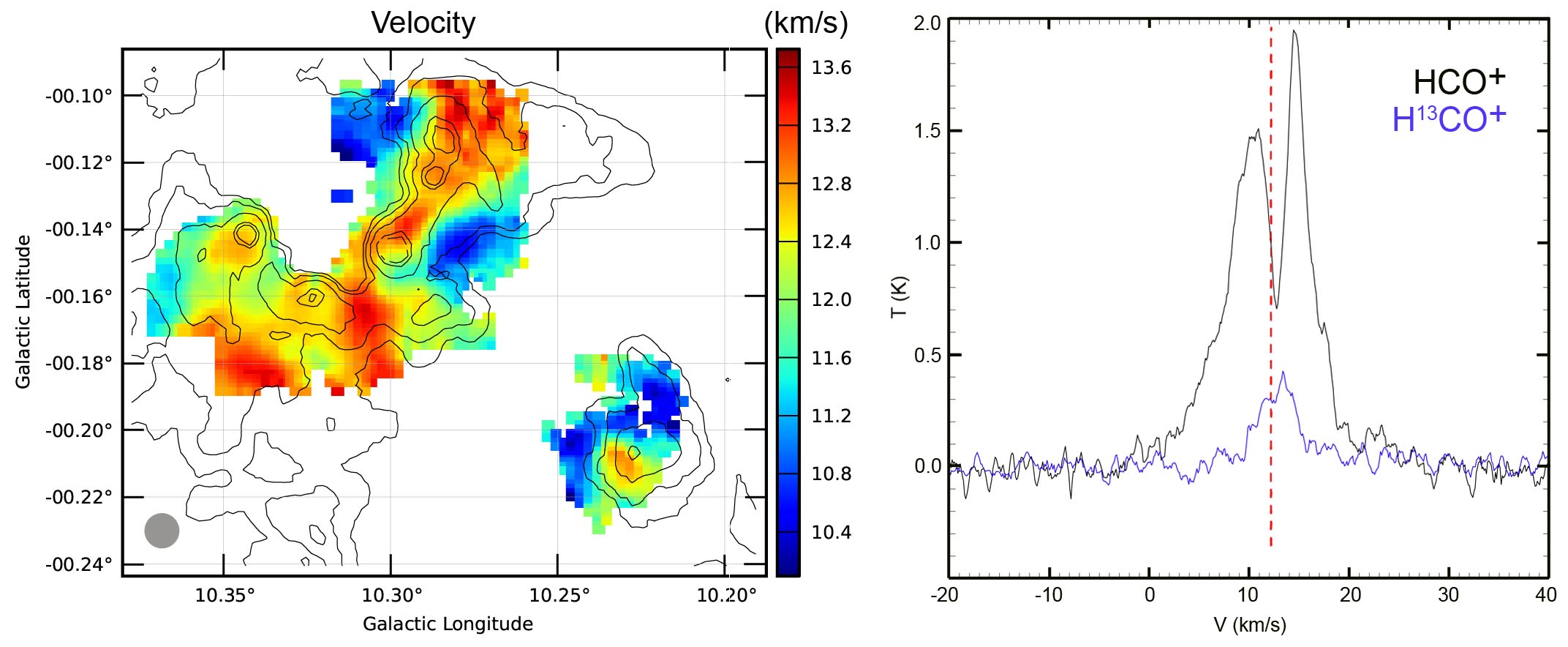}
  \caption{The main velocity component observed over the G010.32$-$00.15 field. {\it Left}: velocity field obtained with the HCO$^+$ (1-0) molecular line. The contours are for the column density (levels of 0.5, 0.75, 1, 1.5, and 2$\times$10$^{23}$~cm$^{-2}$). The 38$\arcsec$ beam is at the lower left. {\it Right}: velocity profile from a region of 36\arcsec$\times$36\arcsec centred on C3, at $l$=10\fdg299 $b$=$-$0\fdg147. The HCO$^+$ line shows self-absorption whereas the H$^{13}$CO$^+$ line is optically thin. The red dotted line shows the velocity obtained by fitting the splitted HCO$^+$ line with one gaussian. } 
\label{malt3}
\end{figure*}

\subsection{Velocity field of the molecular clumps}

Seven overlapping fields have been observed by MALT90 that cover this bright molecular complex.  Three main velocity components are present along the line of sight in different zones of the field.

$\bullet$ The first component, corresponding to velocities in the range 10.5~km~s$^{-1}$ to 13.5~km~s$^{-1}$ as shown in Fig.~\ref{malt3}, is observed in the direction of the clumps C1 to C5 (those at the waist of the bipolar nebula) and in the direction of C6. These velocities are  similar to that of the ionized gas, showing that these clumps and the central \HII\ region are parts of the same complex. It was already known, but indirectly, that C6 is in contact with the bipolar nebula, because the southern lobe is distorted at the border of C6. These velocities are also in good agreement with the velocities measured with the NH$_3$ line by Wienen et al. (\cite{wie12}) and with CO lines by Beuther et al. (\cite{beu11}, and references therein). Self-absorption is clearly observed in the direction of C3 (Fig.~\ref{malt3}~{\it Right}), the clump with the saturated 250~$\mu$m emission; these lines present a double-peaked profile over the C3 clump, the result of a central self-absorption. The H$^{13}$CO$^+$ line, which is optically thin, shows only one component centred on the self-absorption feature in the direction of C3 (Fig.~\ref{malt3}~{\it Right}).

\begin{figure}[h!]
\centering
\includegraphics[width=85mm]{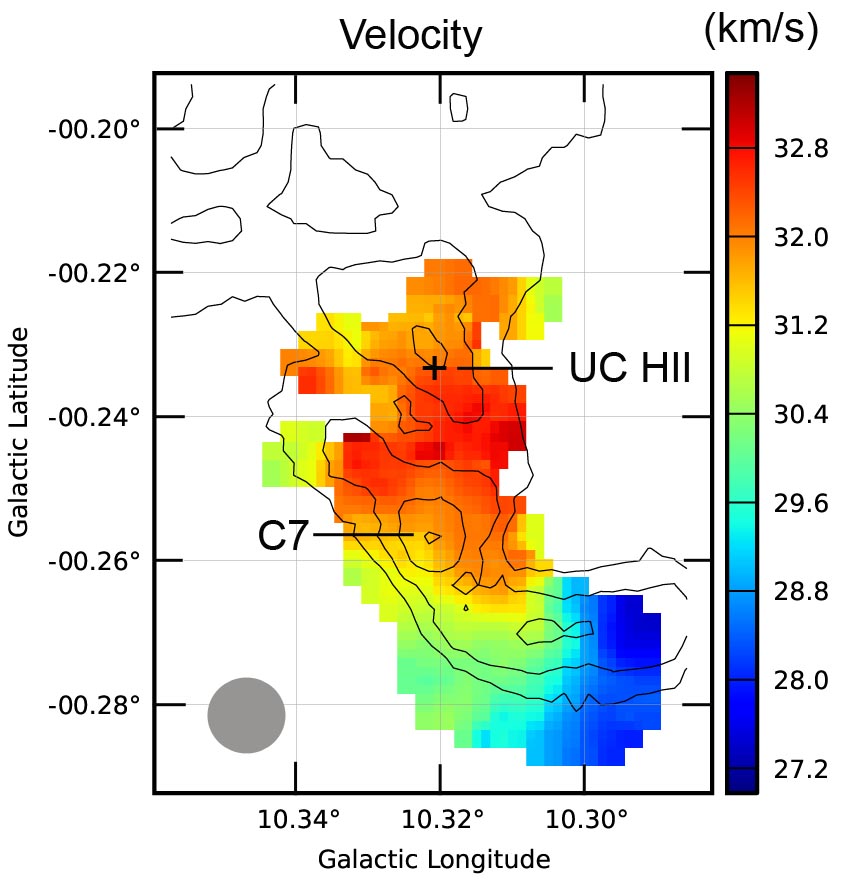}
  \caption{Second velocity component observed over the G010.32$-$00.15 field (results obtained with the HNC (1-0) molecular line). The contours are for the column density (levels of 0.4, 0.5, 0.75, and 1$\times$10$^{23}$~cm$^{-2}$). The black crosses indicate the position of the MAGPIS UC \HII\ region.}
\label{malt4}
\end{figure}

$\bullet$ The second component, corresponding to velocities in the range 28~km~s$^{-1}$ to 32.5~km~s$^{-1}$ as shown in Fig.~\ref{malt4}, is observed in the direction of C7 and vicinity. 

\begin{figure}[h!]
\centering
\includegraphics[width=85mm]{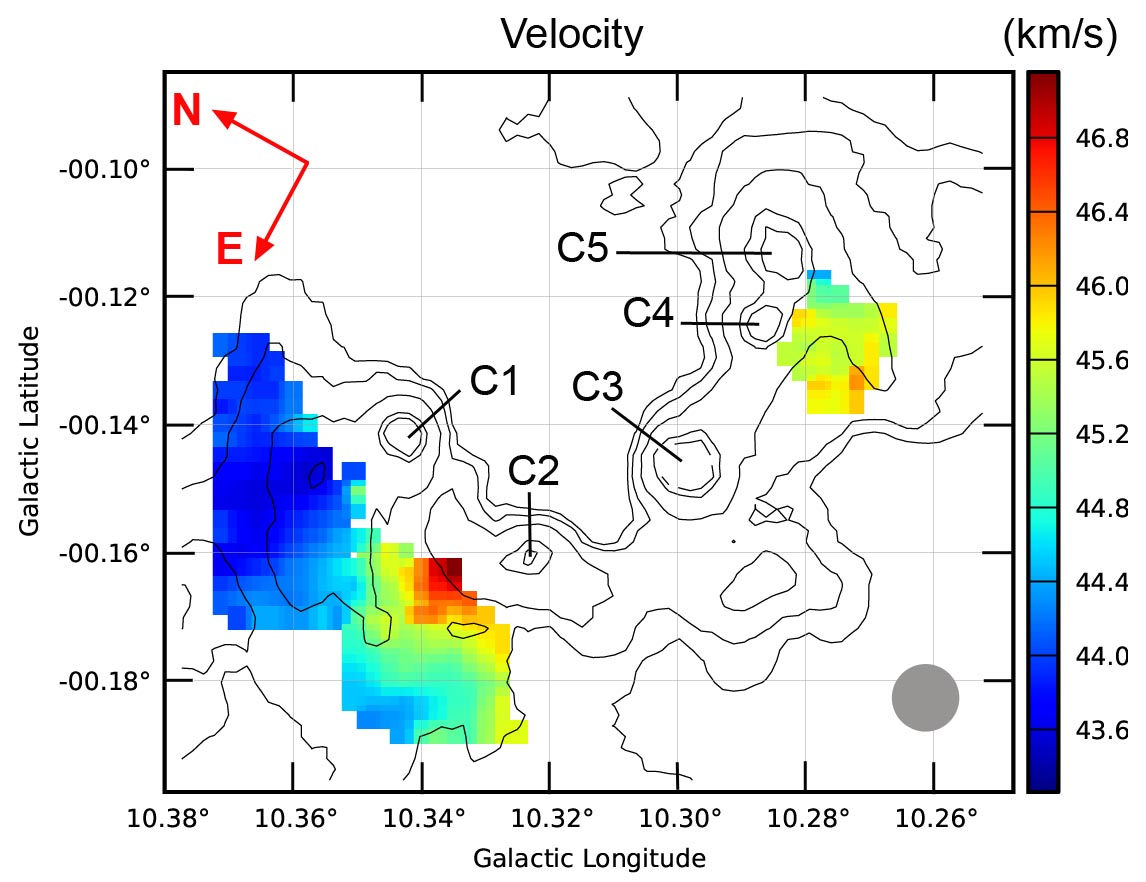}
  \caption{Third velocity component observed over the G010.32$-$00.15 field (results obtained with the HCO$^+$ (1-0) molecular line).} 
\label{malt5}
\end{figure}

$\bullet$ The third component, corresponding to velocities in the range 43.5~km~s$^{-1}$ to 47~km~s$^{-1}$ as shown in Fig.~\ref{malt5}, is observed east of C1 and C2 (especially in the direction of the C1 extension structure), and south of C4, in directions adjacent to these clumps.

The results presented in Figs.~\ref{malt3}, \ref{malt4}, and ~\ref{malt5} have been obtained with the H$^{12}$CO$^+$ (1-0) or HNC (1-0) molecular lines; similar results are obtained with these two lines. (The velocities agree to within 0.5~km~s$^{-1}$.) \\

These observations show that at least three molecular clouds are present in the field at velocities of the order of 12~km~s$^{-1}$, 30~km~s$^{-1}$, and 45~km~s$^{-1}$, thus presumably at different distances. These clouds are observed here via their emission, and   are probably the same clouds that are responsible for the H$_2$CO absorption of the radio-continuum emission of the central \HII\ region (Appendix~A.1).

\subsection{Young stellar objects}

\begin{figure}[!h]
\centering
\includegraphics[width=85mm]{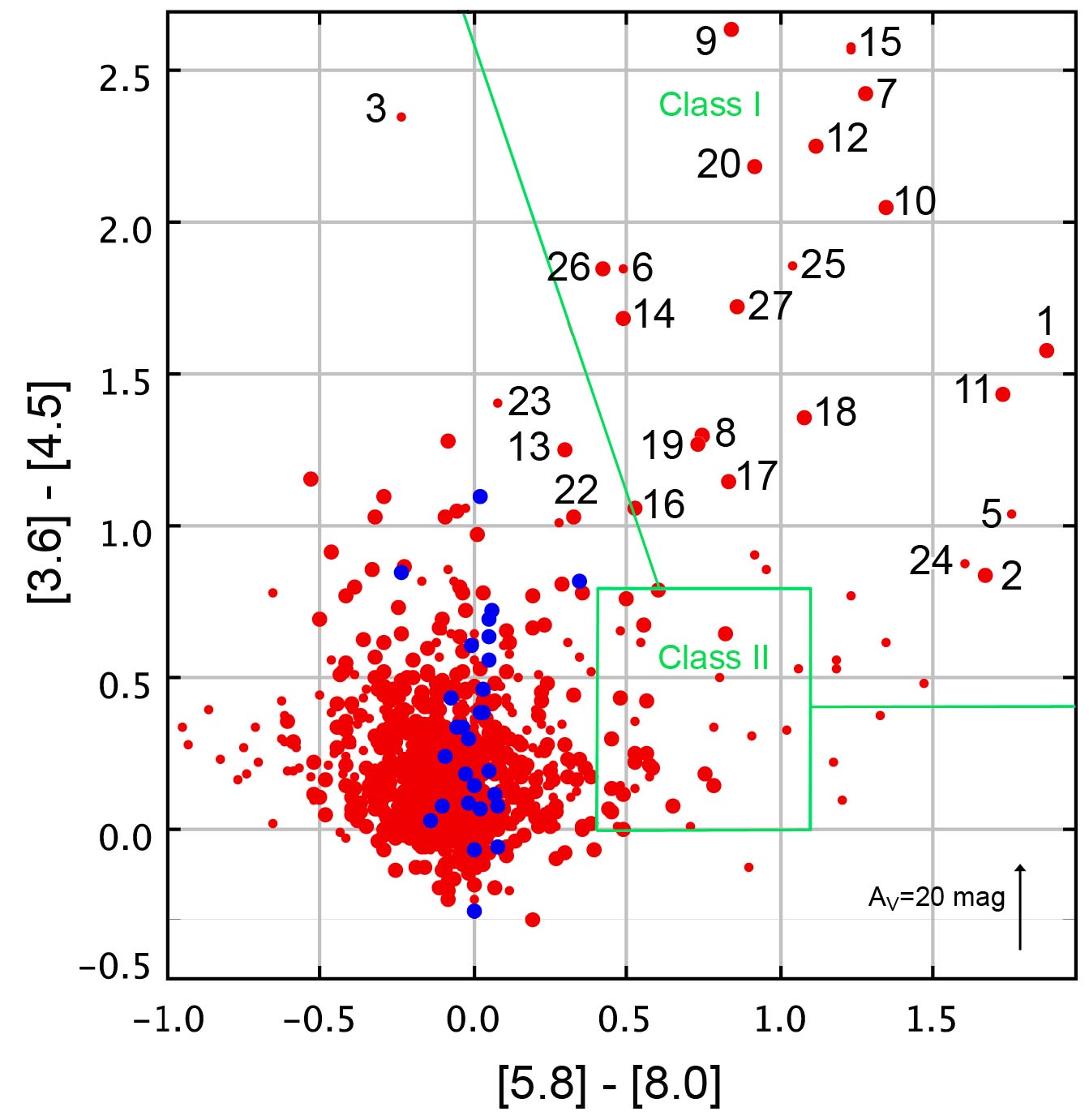}
  \caption{Colour-colour diagram, [3.6]$-$[4.5] versus [5.8]$-$[8.0], for all {\it Spitzer}-GLIMPSE sources with measurements in the four bands, situated at less than 7$\arcmin$ from the exciting star (\#1) of G010.32$-$00.15. The big red dots correspond to sources with measurements more accurate than 0.15~mag in all four bands, the small red dots for the less accurate sources. See Fig.~\ref{champ319f} for further information.}
\label{champ10g}
\end{figure}

In the following we try to detect candidate Class~0/I YSOs and to confirm their evolutionary status. Figure~\ref{champ10g} presents the {\it Spitzer} [3.6]$-$[4.5] versus [5.8]$-$[8.0] diagram: the red dots are for sources located less than 7$\arcmin$ from the exciting star~\#1, thus in a field covering the bipolar nebula, lobes included, its associated molecular clumps, and the parental filament. The blue dots correspond to candidate xAGB stars. G010.32$-$00.15 lies in the direction of the galactic bulge, so it is not surprising to find many candidate AGB stars in this field. We have identified in Fig.~\ref{champ10g} all the sources that we consider as candidate Class~I YSOs or that we  discuss in the following. Their parameters, name (column 1), coordinates (columns 2 \& 3),  magnitudes from the near-IR to 70~$\mu$m (columns 4 to 12), luminosity (column 13), spectral index $\alpha$ between $K$ and 24~$\mu$m (column 14), and colour [8]$-$[24]\footnote{We note a discrepancy between the 24~$\mu$m magnitudes in the Robitaille et al. (\cite{rob08}) and the MIPSGAL catalogues (see Table~\ref{G10_YSOs}). The consequence is that some sources classified (according to their [8]$-$[24] colour) as flat-spectrum sources using the Robitaille et al. magnitudes are classified as Class~II sources when using the MIPSGAL magnitudes.} (column 15) are given in Table~\ref{G10_YSOs}. YSOs \#16 to \#26 are located far from the waist of the bipolar nebula, and many of them are possibly not associated with it. Several candidate Class~I YSOs are located in the vicinity of the nebula, especially in the direction of the clumps and filaments present at the waist of the nebula. We discuss the content of each of the clumps in the following. (In this section the column densities are given without background subtraction.)

\begin{figure*}[tb]
\sidecaption
\includegraphics[width=12cm]{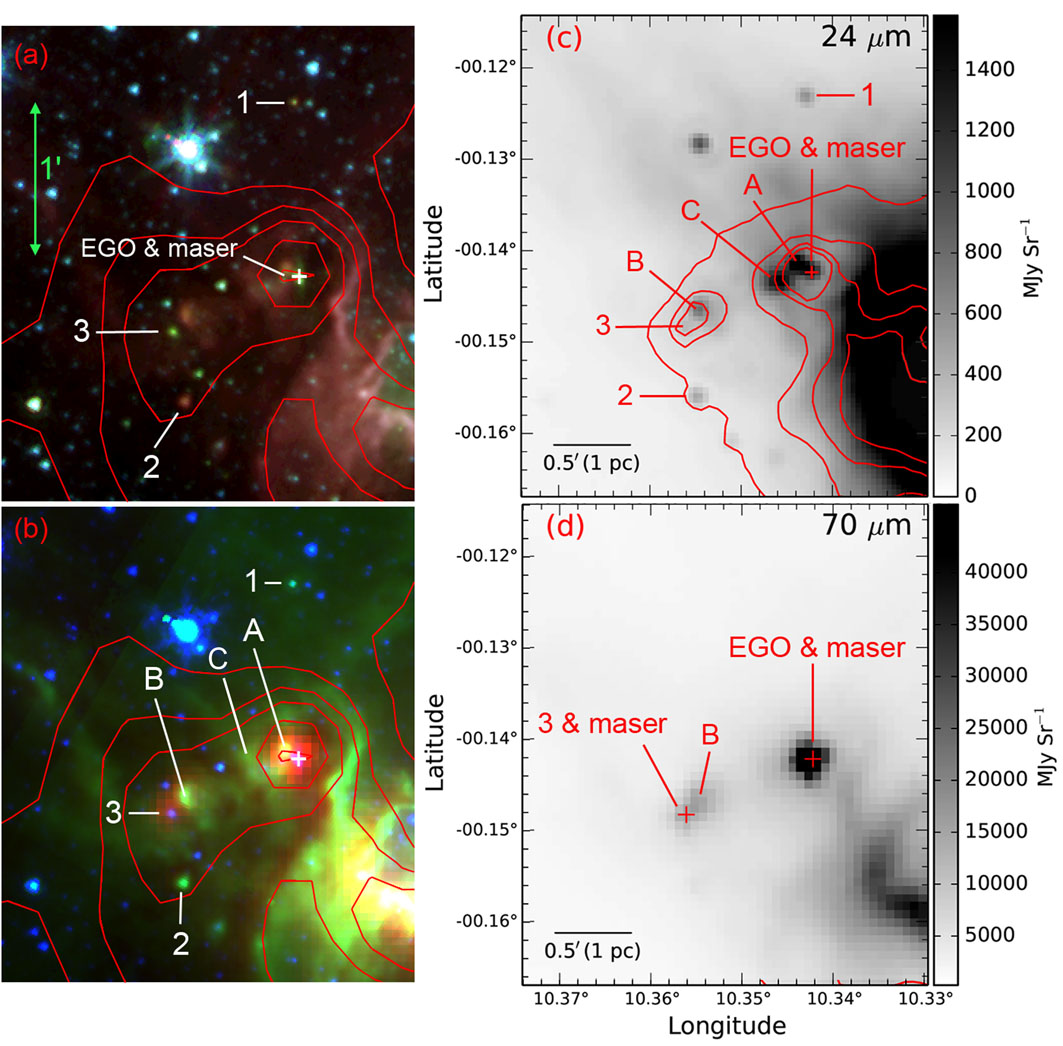}
 \caption{C1 clump and surroundings. The sources discussed in the text are identified. {\it (a)} Colour image with the 5.8~$\mu$m, 4.5~$\mu$m, and 3.6~$\mu$m emissions in red, green, and blue, respectively (logarithmic units).  The red contours refer to the column density (levels of 0.3, 0.5, 0.75, 1.0, 1.5, and 2~10$^{23}$~cm$^{-2}$). {\it (b)} Colour image with the 70~$\mu$m, 8.0~$\mu$m, and 3.6~$\mu$m emissions in red, green, and blue, respectively (linear units). {\it (c)} MIPSGAL image  at 24~$\mu$m; the contours are for the 70~$\mu$m emission (levels at 5000, 10000, 15000, and 20000~MJy/sr). A, B, and C are slightly extended sources of 8.0~$\mu$m, 24~$\mu$m, and 70~$\mu$m emission. {\it (d)} {\it Herschel} 70~$\mu$m image.} 
\label{champ10hbis}
\end{figure*}

$\bullet$ C1: Figure~\ref{champ10hbis} shows clump C1 and its vicinity.  YSO~\#1, with [8]$-$[24] in the range 4.16~mag to 4.49~mag, is a Class~I YSO. It is located outside of C1 but on the upper side of the waist of the bipolar nebula, in a region of intermediate column density (3.3$\times$10$^{22}$~cm$^{-2}$), and has a faint 70~$\mu$m counterpart (indicating a luminosity $\sim$12$\lsol$).

Figure~\ref{champ10hbis}~{\it (b)} \& {\it (d)} show that several sources lie at the very centre of C1 (at less than 3$\arcsec$ - 0.025~pc - from the column density peak): 1) an EGO (Cyganowski et al. \cite{cyg08}, \cite{cyg09}), which  appears as a rather bright green source in Fig.~\ref{champ10hbis}~{\it (a)}; it presents a 24~$\mu$m counterpart; 2) a class~II methanol maser (Walsh et al. \cite{wal98}, Green et al. \cite{gre10}), almost coincident with the EGO. Its velocity, 15.4~km~s$^{-1}$ (Green et al. \cite{gre10}), indicates that it is associated with C1.  A class~I methanol maser is also associated with the EGO and the class~II methanol maser (Cyganowski et al \cite{cyg09}). A bright and compact 70~$\mu$m source lies in the very centre of C1. It is the third-brightest source of the field at this wavelength with a flux density of 197$\pm$20~Jy. Its luminosity, estimated from its flux at 70~$\mu$m, is $\sim$1500~$\lsol$. It lies in the exact direction of the EGO and the methanol maser and has no radio counterpart. We suggest that C1 contains a MYSO in an early evolutionary stage because it has no near-IR counterpart (thus a massive Class 0/I YSO). Two slightly extended 8.0~$\mu$m and 24~$\mu$m sources (sources A \& C in Fig.~\ref{champ10hbis}~{\it (c)} are also present in the direction of C1. The bright source A lies $\sim$3$\farcs$5 - 0.3~pc - from the centre of C1. But the 70~$\mu$m emission peaks in the direction of the EGO \& methanol  masers rather than in the direction of source A. Sources A \& C have no detectable radio counterpart (flux $\leq$5~mJy on the MAGPIS map at 20~cm).
  
C1 has an extension of high column density. In this direction there are two YSOs: YSO~\#2, with a colour [8]$-$[24]=3.55~mag and a faint 70~$\mu$m counterpart is a flat-spectrum YSO. YSO~\#3, with a colour [8]$-$[24]=5.03~mag, and a bright 70~$\mu$m counterpart is clearly a Class~I YSO. A 6.7~GHz methanol maser has been detected in its direction by Green et al (\cite{gre10}), which lies at less than 1$\arcsec$ of YSO \#3. YSO \#3 is slightly extended at 4.5~$\mu$m, suggesting that it can be associated with an EGO. However YSO \#3 is most probably not at the same distance as the bipolar G010.32$-$0.15 \HII\ region. Its associated methanol maser has a velocity of 49.9~km~s$^{-1}$ showing that YSO \#3 is associated with the molecular material of the third velocity component (near 44~km~s$^{-1}$) that overlaps with the C1 extension (see Sect.~6.3.1). A slightly extended source, source B in Fig.~\ref{champ10hbis}~{\it (b),~(c),~(d)} is also present in the direction  of the C1 extension. (It lies at 3$\farcs$2 from the column density peak.) It is seen at 8.0~$\mu$m, 24~$\mu$m, and 70~$\mu$m and has a very faint radio counterpart on the MAGPIS map at 20~cm, so it is probably a faint UC \HII\ region. We estimate its flux at 20~cm to be $\sim$8~mJy, and it is unresolved. Its flux density at 70~$\mu$m is $\sim$38~Jy.  We do not know if it is associated with the bipolar nebula or with the molecular component at a velocity $\sim$44~km~s$^{-1}$ (same situation for YSOs~\#2).

$\bullet$ C2: C2 faces the exciting cluster, Fig.~\ref{champ10i}~{\it (a)}. It is bordered by a bright emission region on the side facing the cluster. It is bright at radio wavelengths (MAGPIS image; Fig.~\ref{champ10i}~{\it (c)}), where we see the emission of the dense ionized layer bordering the molecular clump, and it is bright at 8.0~$\mu$m or 5.8~$\mu$m (Fig.~\ref{champ10i}~{\it (b)}), since we see the emission of the PAHs in the PDR, excited by the radiation leaking from the \HII\ region.

A 6.67~GHz methanol maser lies at the very centre of C2 (Walsh et al. \cite{wal98}; Green et al. \cite{gre10}). It lies at  2$\farcs$4 -- 0.2~pc -- from the column density peak. Its velocity, 11.5~km~s$^{-1}$ (Green et al. \cite{gre10}), indicates that it is associated with the clump. In the same direction is YSO~\#4, a very bright near- and mid-IR unresolved source. YSO~\#4 is not part of the GLIMPSE catalog, probably because it lies in a region of bright nebular emission and is saturated in several GLIMPSE bands. (This region is also saturated at 24~$\mu$m.) YSO~\#4 is the second brightest 70~$\mu$m source of the field (after the UC \HII\ region in C3). It is point-like, with a flux of 630$\pm$50~Jy, indicating a luminosity $\sim$4600~$\lsol$. It has no detectable radio counterpart on the MAGPIS image (flux $\leq$5~mJy at 20-cm). Source \#4 lies in a massive condensation ($\sim$400~$\msol$) and shows methanol maser emission at 6.7~GHz. All this indicates that source \#4 is a massive YSO (MYSO), which  possibly is accreting material too strongly to form a detectable \HII\ region (to explain the lack of an associated UC \HII\ region; Walmsley \cite{wal95}). It has a near-IR counterpart, so it is probably a massive Class I and not a Class~0 YSO.

\begin{figure}[h!]
\centering
\includegraphics[width=75mm]{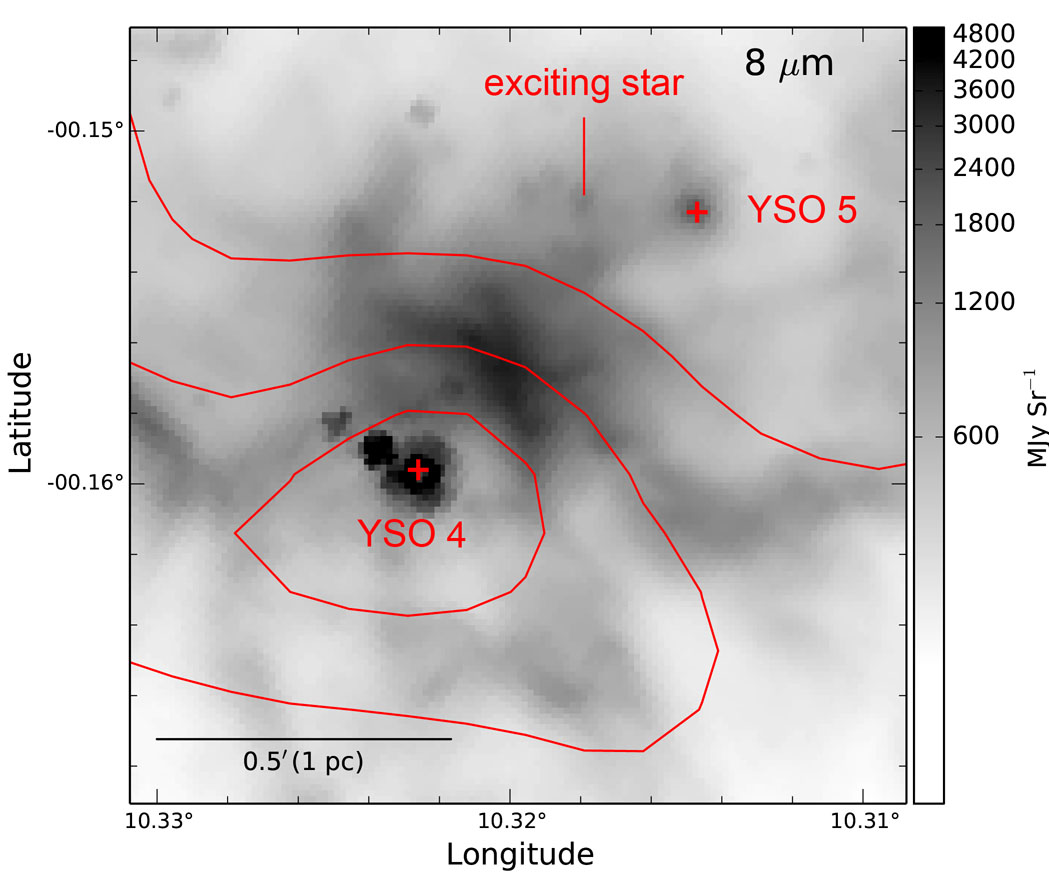}
 \caption{Identification of YSO \#5. The underlying grey image is the {\it Spitzer} 8.0~$\mu$m image. The contours are for the column density (levels of 0.5, 1.0, and 1.5$\times$10$^{23}$~cm$^{-2}$); they show the centre of clump C2. YSO~\#4 is also identified (saturated with two glints). } 
\label{YSO5}
\end{figure}

\begin{figure}[h!]
\centering
\includegraphics[width=85mm]{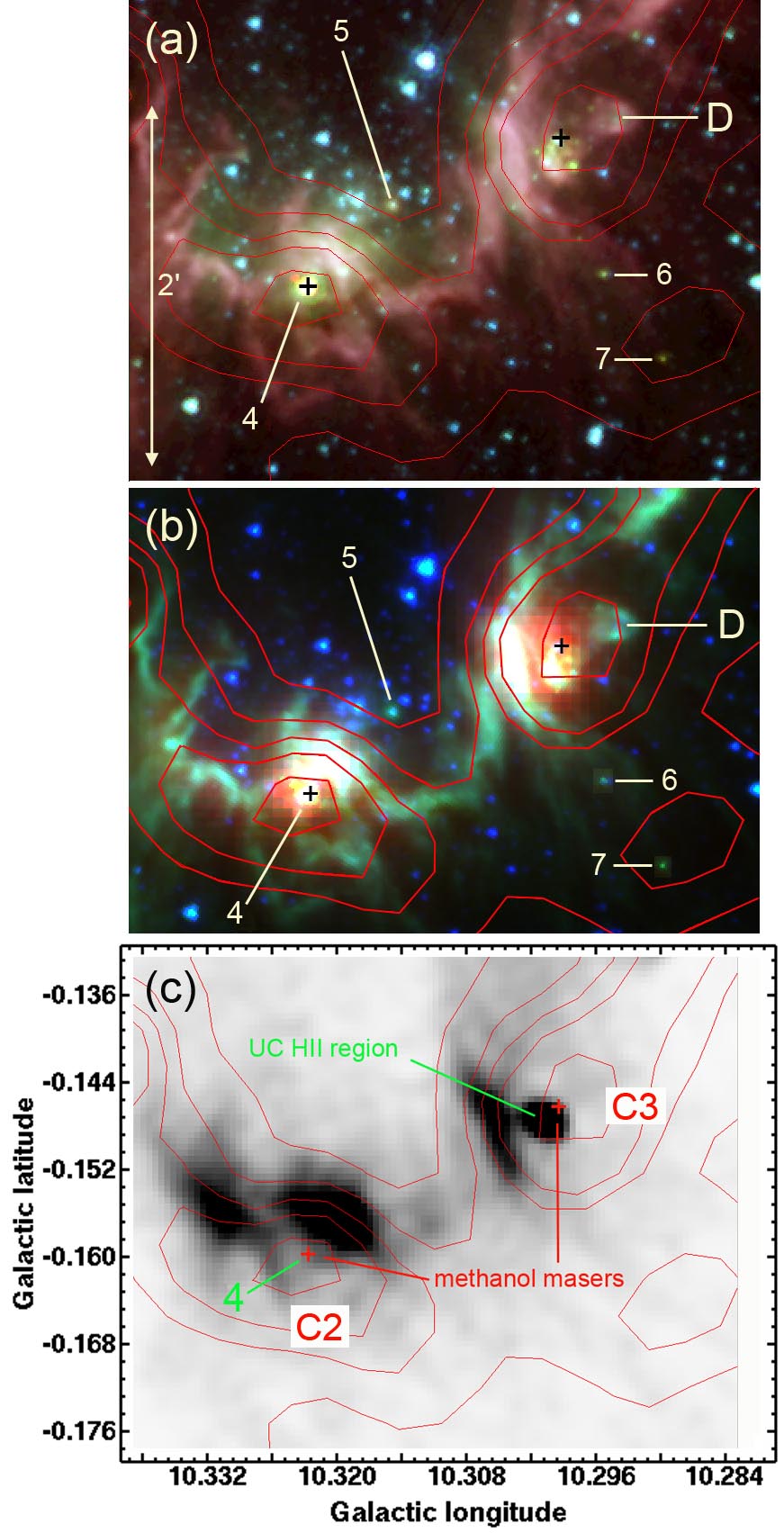}
  \caption{Clumps C2 and C3. {\it (a)} Colour image with the {\it Spitzer} 5.8, 4.5, and 3.6~$\mu$m emissions in red, green, and blue, respectively (logarithmic units). The red contours are the column density contours (same levels as in Fig.~\ref{champ10hbis}~{\it (b)}). The candidate Class~I YSOs \#4, \#5, \#6, and \#7 are identified. The positions of the methanol masers are indicated by plusses. {\it (b)} Colour image with the 70~$\mu$m, 5.8~$\mu$m, and 3.6~$\mu$m respectively in red, green and blue (linear units). {\it (c)} The red column density contours  are superimposed on the MAGPIS radio-continuum image. The UC \HII\ region is identified.}
\label{champ10i}
\end{figure}

The nature of YSO~\#5 is not clear. It lies in the direction of the central exciting cluster, 12$\arcsec$ from star \#1  (Fig.~\ref{YSO5}). Its $JHK$ magnitudes correspond to a B3V star with about 16.8~mag of visual extinction, thus possibly part of the exciting cluster. Its 8.0~$\mu$m counterpart is bright and slightly extended.  YSO \#5 lies in the saturated 24~$\mu$m zone. As shown by Whitney et al. (\cite{whi13}), an early B star illuminating an optically thin dust cloud can appear to have an IR excess and resemble an YSO. It is possibly the case for source \#5, superimposed on a relatively bright background at 8.0~$\mu$m. 

$\bullet$ C3: C3 is bordered by a bright rim, which is especially bright on its east side facing the exciting cluster (Fig.~\ref{champ10i}~{\it (a)}). Bright radio-continuum emission is observed in the direction of this bright rim. Figure~\ref{champ10i}~{\it (c)} also shows an UC \HII\ region in the direction of the centre of C3. We suggest that it is a separate, independent, \HII\ region embedded inside C3 or lying on the border of C3. Its size is 0.06~pc (FWHM, beam deconvolved).  Its flux is given in Table~\ref{UCH} (Appendix~A). Its ionizing photon flux (log(N$_{\rm Lyc}$)$\sim$47.2) corresponds, according to Smith et al. (\cite{smi02}), to a main exciting star (if single) of spectral type B0V -- B0.5V. A small cluster is present in this direction, visible on the UKIDSS $K$ image and on the {\it Spitzer} images up to 5.8~$\mu$m, which is probably the exciting cluster. But we have not been able to identify the exciting star of this UC \HII\ region.  This UC \HII\ region has a mid-IR counterpart, a small extended region of similar size emitting at all GLIMPSE wavelengths. (This region is saturated at 24~$\mu$m.)  It is the brightest 70~$\mu$m compact source of the field with a flux of 890$\pm$30~Jy.  A faint class~II methanol maser (Walsh et al. \cite{wal98}; Green et al. \cite{gre10}) has been detected in the direction of C3. It lies at 1.8$\farcs$4 - 0.016~pc - from the column density peak and at 6$\farcs$5 from the centre of the UC \HII\ region. It is probably associated with C3 since its velocity is 19.9~km~s$^{-1}$. 

Another small region of  extended GLIMPSE emission lies in the direction of C3 (region D, Fig.~\ref{champ10i}). This one has no detectable  radio-continuum counterpart (flux $\leq$5~mJy), but it has a saturated 24~$\mu$m counterpart and a faint extended 70~$\mu$m counterpart. Again a small cluster is present in its direction. It contains a star that could be a late B star with a visual extinction of about 22~mag (UKIDSS catalog: $J$=17.089~mag, $H$=14.795~mag, $K$=13.354~mag; coordinates $l$=10\fdg2941 $b$=$-$0\fdg1443). This region is probably located in front of C3 because, in its direction, the column density is of the order of 2$\times$10$^{23}$~cm$^{-2}$, which corresponds to a visual extinction of about 215~mag, so much higher than that of the star. 

Two faint candidate Class~I YSOs (\#6 and \#7) lie outside of C3 but still in a region of high column density and along a 8.0~$\mu$m filament originating in the waist of the bipolar nebula. The column densities in the direction of \#6 and \#7 are  5.2$\times$10$^{22}$~cm$^{-2}$ and 9.7$\times$10$^{22}$~cm$^{-2}$, respectively. YSO~\#6 has no measurable counterparts at 24~$\mu$m (due to saturation) and 70~$\mu$m; its nature is uncertain. YSO~\#7 has a faint 24~$\mu$m counterpart and no detectable one at 70~$\mu$m. Its nature is uncertain (a low-luminosity Class~II/flat-spectrum YSO).

\begin{figure}[h!]
\centering
\includegraphics[width=85mm]{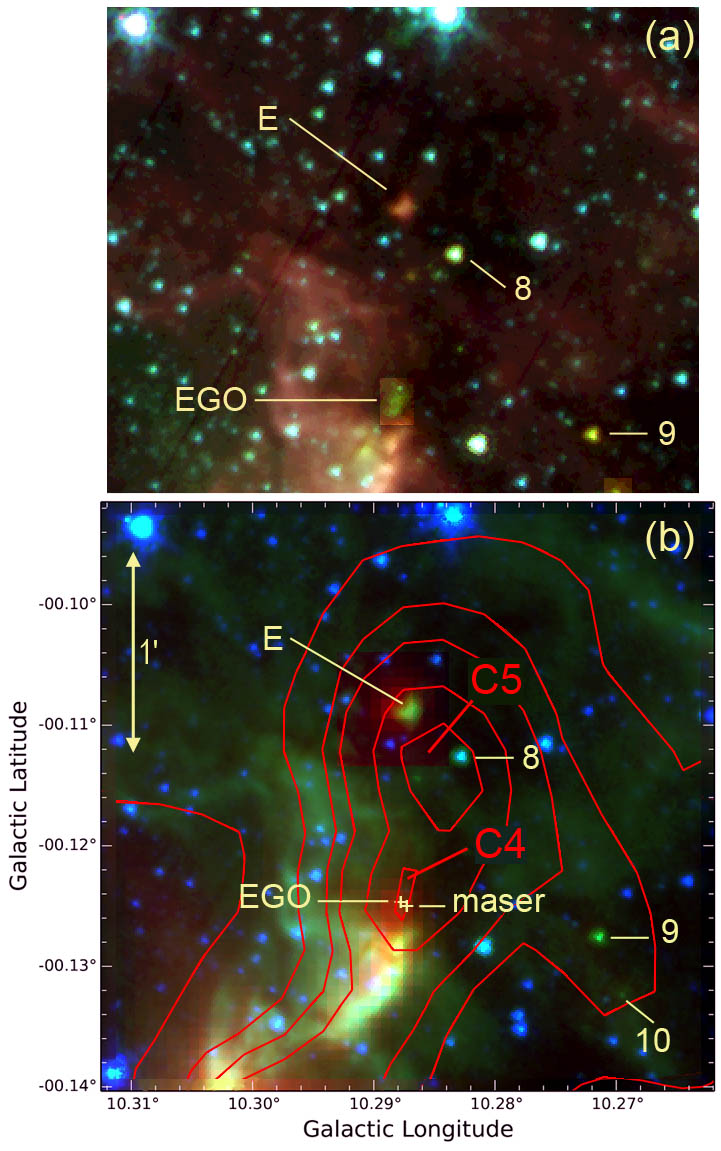}
  \caption{C4 and C5 clumps and vicinity. {\it (a)} Same colours and units as in Fig.~\ref{champ10hbis}~{\it (a)}. Sources \#8, \#9, and \#10 are candidate Class~I YSOs. The brightness has been enhanced in the vicinity of the EGO. The slightly extended GLIMPSE source E is identified. {\it (b)} Same colours, units, and contours  as in Fig.~\ref{champ10hbis}~{\it (b)}. The positions of the EGO and of the methanol maser are indicated.}
\label{champ10j}
\end{figure}

$\bullet$ C4 and C5: Fig.~\ref{champ10j} shows clumps C4, C5,  and vicinity. In the very centre of C4 (at the peak of column density) lies an EGO (Cyganowski et al. \cite{cyg08}), which is much fainter than the one in C1, and a class~II methanol maser  brighter than the one in C1 (Walsh et al. \cite{wal98}; Green et al. \cite{gre10}). The velocity of the maser, 4.6~km~s$^{-1}$, indicates that it is associated with C4. It lies at 1$\farcs$8 - 0.015~pc - from the column density peak.  The EGO and the class~II methanol maser have no {\it Spitzer} counterpart (except at 4.5~$\mu$m; but no 24~$\mu$m emission), but they have a faint 70~$\mu$m counterpart, as shown by Fig.~\ref{champ10j}~{\it (b)}. C4 with its EGO and methanol maser, and with no near-IR and radio counterparts, resembles C1.  However, its 70~$\mu$m emission is not as bright as the C1 one. (Its flux of 30$\pm$10~Jy indicates a luminosity $\sim$260~$\lsol$.) It is a less luminous Class~0/I YSO than the one in C1.  

C5 lies west of C4. Two sources lie in the direction of C5: 1) YSO~\#8, a flat-spectrum or a Class~II YSO with [8]$-$[24] in the range 2.95~mag -- 3.29~mag, which has no detectable 70~$\mu$m counterpart; 2) a small source of extended emission, which we call E, visible at all GLIMPSE wavelengths and at 24~$\mu$m and 70~$\mu$m (Fig.~\ref{champ10j}). Source E resembles source B, with a  70~$\mu$m flux $\sim$24~Jy, but it has no detectable 20-cm counterpart (flux $\leq$5~mJy). We suggest that it is a small region of PAH and thermal emission around a late B star, but this star is not detected in UKIDSS or {\it Spitzer}-GLIMPSE images. The column density in the direction of YSO~\#8 and of region E is in the range 1.8 -- 2.7$\times$10$^{23}$~cm$^{-2}$.

A faint 8.0~$\mu$m filament extends to the south of C4 and C5, originating in the waist of the bipolar nebula.  Two YSOs, \#9 and \#10, are located along this filament. The column density in the direction of this filament and  of YSOs \#9 and \#10 is in the range 9 -- 10$\times$10$^{22}$~cm$^{-2}$. YSO~\#9, with [8]$-$[24] in the range 2.62~mag -- 3.36~mag is a Class~II or a flat-spectrum YSO. It has a faint 70~$\mu$m counterpart, indicating a luminosity $\sim$50$\lsol$. YSO~\#10 is faint at all {\it Spitzer} wavelengths and has no detectable 70~$\mu$m counterpart, so it is possibly a low-luminosity Class~II YSO. Two velocity components, near 13~km~s$^{-1}$ and 46~km~s$^{-1}$, are present in the direction of YSO~\#9 and \#10, making their association with the bipolar nebula uncertain.

\begin{figure}[h!]
\centering
\includegraphics[width=85mm]{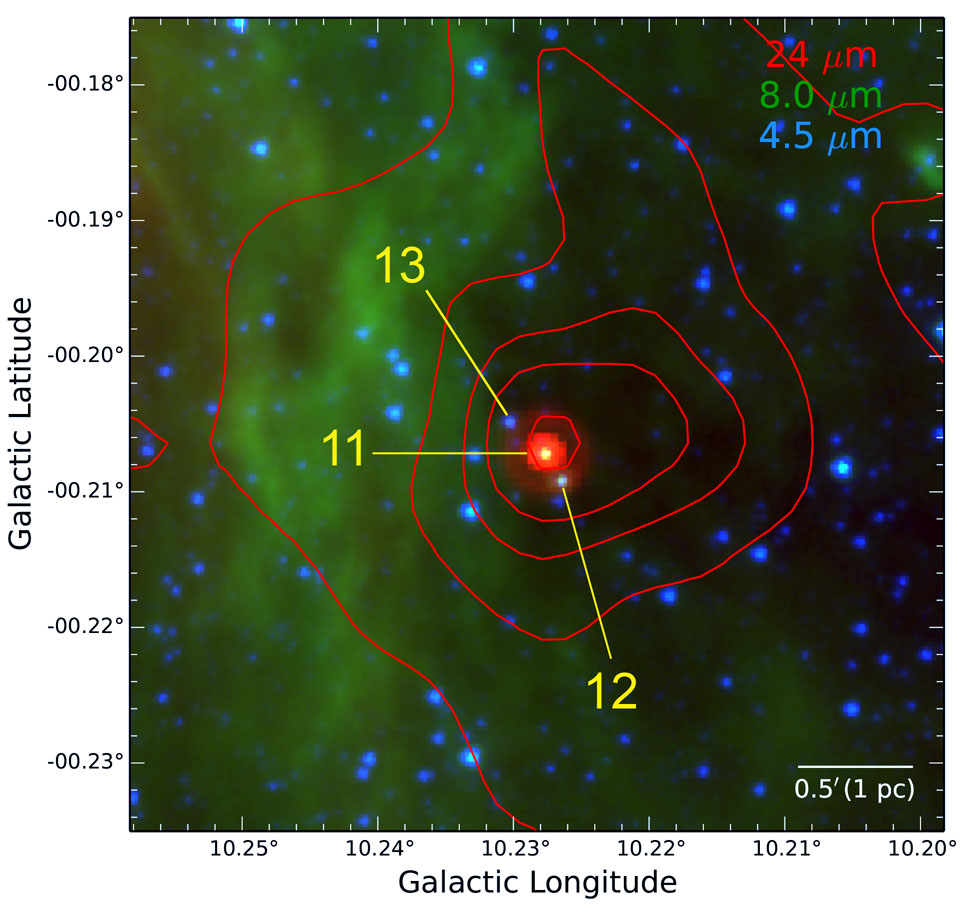}
\caption{C6 dust clump and vicinity. Colour composite image with the 24~$\mu$m, 8.0~$\mu$m, and 4.5~$\mu$m emissions in red, green, and blue, respectively (linear units). The three candidate Class~I YSOs are identified. The red contours are for the column density (same levels as  in Fig.~\ref{champ10hbis}~{\it (b)}).}
\label{champ10k}
\end{figure}

Two more clumps lie close to the bipolar nebula at the border of the southern lobe:

$\bullet$ C6: it  is a cold clump that belongs to the complex. Its velocity is similar to the velocities of the ionized gas and of clumps C1 to C5.  C6 is similar to the clumps C1 to C5 in terms of column density, mass, and density. C6 contains three  candidate Class~I YSOs, \#11 in its very centre (at 2$\farcs$0 - 0.017~pc - from the column density peak), and \#12 and \#13 nearby (Fig.~\ref{champ10k}). YSO~\#11 is bright at 24~$\mu$m (with [8.0]$-$[24]=6.06~mag) and bright at 70~$\mu$m, so it is clearly a Class~I YSO. Its 70~$\mu$m flux $\sim$30~Jy indicates a luminosity $\sim$260$\lsol$. YSO~\#12 is almost as bright as YSO~\#11 at GLIMPSE wavelengths. YSO~\#13 is brighter than \#11 and \#12 from $K$ to 4.5~$\mu$m, but much fainter at 8.0~$\mu$m. Both YSO~\#12 and \#13 lie in the direction of the diffraction ring of YSO~\#11, so their 24~$\mu$m magnitude is very uncertain. YSO~\#12 is possibly a flat-spectrum source. The nature of YSO~\#13 is unknown. No radio emission is detected on the MAGPIS radio map in the direction of C6 (flux $\leq$5~mJy at 20-cm). 

C6 was probably a pre-existing cloud, reached by the IF during the expansion of the southern lobe (as C6 protudes inside the southern lobe).  It interacts with the southern lobe and distorts its edge. C6 appears to be compressed on its north-east border by the ionized gas. Figure~\ref{champ10k} shows that the molecular material is not symmetrically distributed around YSO~\#11: we measure a column density gradient $\sim$6.3$\times$10$^{23}$~cm$^{-2}$ per parsec on the side of C6 turned towards the ionized gas and a column density gradient of $\sim$2.6$\times$10$^{23}$~cm$^{-2}$ per parsec on its side turned towards the outside.

$\bullet$ C7: this clump does not belong to the G010.32$-$00.15 complex because its velocity is too different from that of the ionized gas of the bipolar \HII\ region. The candidate Class~I YSOs found in the direction of C7 and the UC \HII\ regions present there are discussed in Appendix~A.2.\\

\begin{figure}[h!]
\centering
\includegraphics[width=80mm]{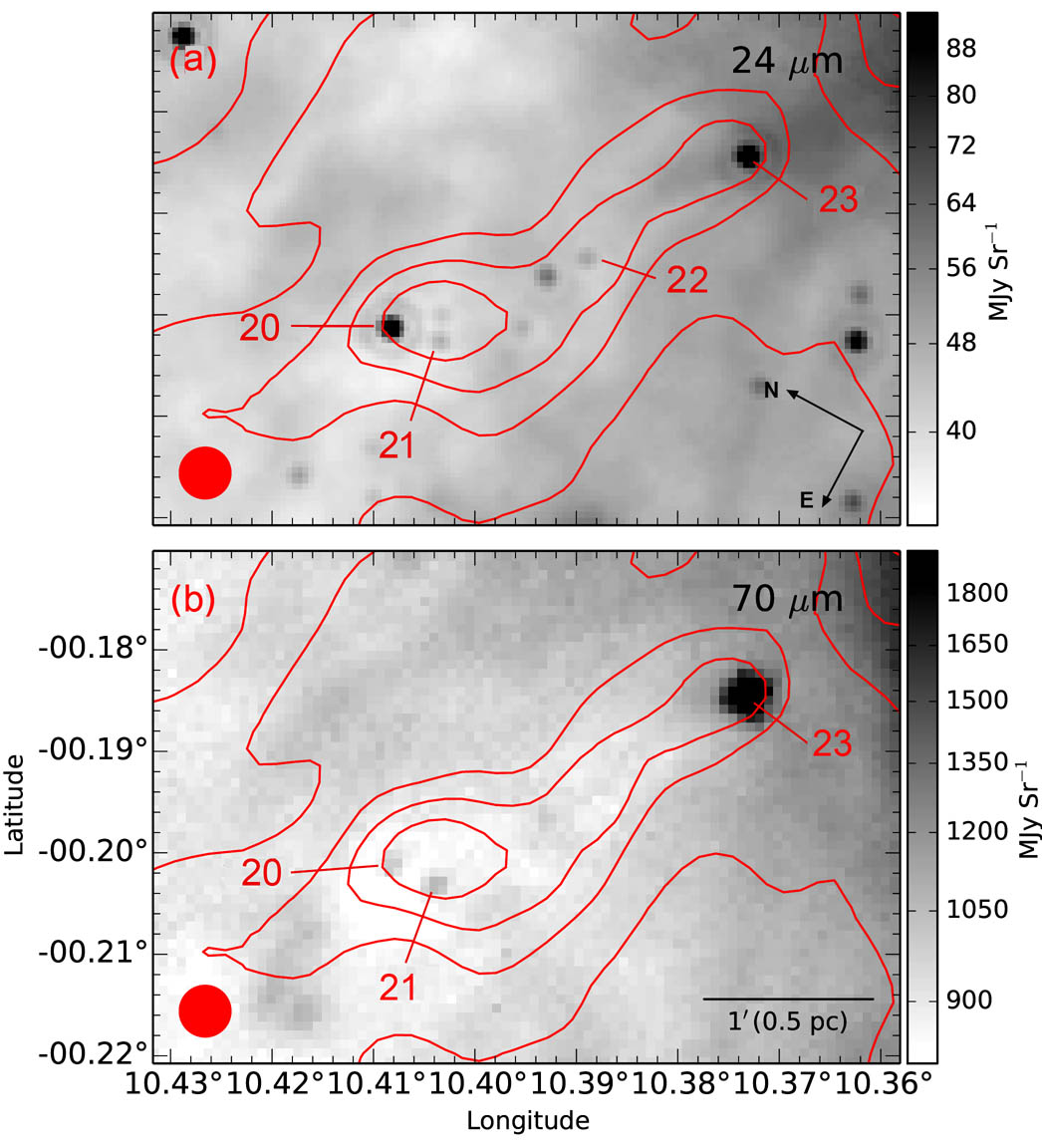}
  \caption{YSOs in the direction of the eastern parental filament. The red contours correspond to column densities of 3, 4, 5, 6, and 7~10$^{22}$~cm$^{-2}$. They are superimposed on grey images (logarithmic units): {\it (a)} the 24~$\mu$m image; {\it (b)} the 70~$\mu$m Hi-GAL image. The 18$\arcsec$ beam of the column density map (also that of the 250~$\mu$m image) is at the lower left.}
\label{fil}
\end{figure}

Several other candidate Class~I YSOs lie in the vicinity of G010.32$-$00.15. They are identified in Appendix~A,  Fig.~\ref{autresYSO}. Four of them lie in the direction of the parental filament east of the bipolar nebula.  They are most probably part of the complex. The column density is high in their direction (Fig.~\ref{fil}) with N(H$_2$)$\geq$7$\times$10$^{22}$~cm$^{-2}$. YSO~\#20 has a bright 24~$\mu$m counterpart ([8]$-$[24] in the range 2.87~mag to 3.36~mag) and a faint 70~$\mu$m one. It is a low-luminosity Class~II or flat-spectrum YSO (L$\sim$6$\lsol$).  At 70~$\mu$m, YSO~\#20 has a ``twin'' of similar brightness, YSO~\#21, that is  very faint at 24~$\mu$m.  These two sources are separated by about 16$\arcsec$. YSO~\#21 is possibly a Class~0 YSO of low-luminosity ($\sim$9~$\lsol$). But it is not an extreme Class~0 (a PBR according to Stutz et al. (\cite{stu13}). YSO~\#22 has a faint 24~$\mu$m counterpart and no detectable 70~$\mu$m one. Its colour, [8]$-$[24]=2.36~mag, indicates that it is a Class~II YSO. YSO~\#23 is a bright 24~$\mu$m and 70~$\mu$m source. Its colour, [8]$-$[24] in the range 4.50~mag to 4.83~mag, and its 70~$\mu$m flux density of 10.5~Jy indicate a Class~I of intermediate luminosity ($\sim$100~$\lsol$). Furthermore, YSO~\#23 has bright 160~$\mu$m and 250~$\mu$m point-like counterparts (with fluxes of   20.1~Jy and 18.5~Jy, respectively); we estimate the mass of its envelope to be 17~$\msol$ (for a temperature of 18.2~K taken from the temperature map). The YSOs \#20, \#21, \#22, and \#23 are most probably associated with the bipolar nebula. But they lie far from the IF limiting the \HII\ region (the closest, YSO~\#23, is located in projection some 1.4~pc away from the IF), so their formation has probably not been triggered by the central nebula.

The other candidate Class~I YSOs are not observed in the direction of structures linked to the nebula, and their velocity is unknown. Their association with G010.32$-$00.15 is therefore uncertain. They are discussed in Appendix~A.2.

To summarize, among the sources selected as candidate Class~I YSOs on the basis of their {\it Spitzer}-IRAC colours and clearly associated with the bipolar \HII\ region, we find four confirmed Class~I YSOs (\#1, \#4, \#11, and \#23), three flat-spectrum  or Class~II sources (thus of uncertain nature between Class~I and Class~II; \#8, \#12, and \#20), and one Class~II YSO (\#22). Furthermore, YSO \#21, in the direction of the parental filament, is detected as a Class~0, and the clumps C1 and C4 contain Class~0 sources detected thanks to their associated outflows (via EGOs and class~II methanol masers). All the high-luminosity sources (bright  at 70~$\mu$m) are observed in the direction of the massive clumps forming the waist of the nebula.\\

Several small-size diffuse regions, regions A to F, are present in the vicinity of the clumps, generally in the direction of rather high column density regions. They are detected at {\it Spitzer}-GLIMPSE wavelengths, especially at 8.0~$\mu$m. They have counterparts at 24~$\mu$m and 70~$\mu$m. We suggest that they are associated with B stars, mostly stars later than B2 because they have no detectable radio-continuum emission (region B is an exception), but stars that are hot enough to heat the very small dust grains that emit at 24~$\mu$m and possibly at 70~$\mu$m.

\begin{table*}
\caption{Candidate Class~I YSOs and other sources discussed in the field of G010.32$-$00.15.}
\resizebox{18cm}{!}{
\begin{tabular}{rrrrrrrrrrrrrrr}
\hline\hline
Name & l         & b         & $J$         & $H$              & $K$              & [3.6]             & [4.5]            & [5.8]          & [8.0]            & [24]               & S(70$\mu$m) & L & $\alpha$ & [8]$-$[24]  \\
    & ($\degr$) & ($\degr$) & (mag)        & (mag)            & (mag)            & (mag)            & (mag)            & (mag)          & (mag)            & (mag)              & (Jy) & ($\lsol$) &   & (mag)  \\
\hline
\multicolumn{4}{l}{C1} \\
1   &  10.34286 & -0.12324 &                & 16.906$\pm$0.044 & 14.980$\pm$0.016 & 12.330$\pm$0.070 & 10.764$\pm$0.050 & 9.066$\pm$0.052 & 7.199$\pm$0.024 & 3.03 (2.71 ROB)   & 1.1$\pm$0.3$^*$ & 11.6 & 1.84 & 4.17 (4.49 ROB) \\
2   &  10.35481 & -0.15608 &                &                  & 15.932$\pm$0.038 & 11.921$\pm$0.116 & 11.096$\pm$0.125 & 8.136$\pm$0.062 & 6.469$\pm$0.114 & 2.92              & 6.2$\pm$2.0$^*$ &  & 2.21 & 3.55 \\
3   & 10.35612 & -0.14847 &              &                     &                  & 11.930$\pm$0.087 &  9.595$\pm$0.079 & 8.721$\pm$0.044 & 8.962$\pm$0.162 & 3.93$\pm$0.10$^*$   & 41$\pm$10$^*$  &     &   & 5.03  \\
EGO &  10.3423 & -0.1422 &  &    &  &    &   &    &   & 1.81$\pm$0.20$^*$   & &  &  & \\
methanol maser & 10.3422 & -0.1422  &       &       &       &        &        &        &        &         &      &   & & \\
methanol maser & 10.3561 & -0.1483  &       &       &       &        &        &        &        &         &      &   & & \\
A &  10.3439 & -0.1415  &       &       &       &        &        &        &        &         &                &  & &  \\
B & 10.3547 & -0.1463  &       &       &       &        &        &        &        &         &   38$\pm$10$^*$ &  & &  \\
C & 10.3460 & -0.1432  &       &       &       &        &        &        &        &         &                &  & &  \\
\multicolumn{4}{l}{C2} \\
4   & 10.3226 & -0.1596 &                & 16.55$^*$           & 13.08$^*$          & 7.86$^*$           & saturated        & saturated      & saturated      & saturated             & 630$\pm$50$^*$ & 4600 &  &  \\
5   & 10.31474 & -0.15225  & 14.220$\pm$0.002 & 12.447$\pm$0.001 & 11.424$\pm$0.001 & 10.429$\pm$0.077 & 9.403$\pm$0.145 & 7.599$\pm$0.113 & 5.846$\pm$0.217 &  saturated    &   &    &    & \\
methanol maser & 10.3226 & -0.1597 &       &       &       &        &        &        &        &         &     &   &   & \\
\multicolumn{4}{l}{C3} \\
6   &  10.29520 & -0.15858 &              &                   &                  & 12.125$\pm$0.147  & 10.293$\pm$0.081 & 8.945$\pm$0.075 & 8.461$\pm$0.174   &       & $\leq$1.0$^*$ & $\leq$10.6 &   & \\
7   &  10.28971 & -0.16641 &              &                    & 17.500$\pm$0.159 & 13.440$\pm$0.096  & 11.022$\pm$0.05 & 9.169$\pm$0.051 & 7.894$\pm$0.032 &  4.5$\pm$0.5$^*$  & $\leq$0.5$^*$ & $\leq$5.5 & & 3.4 \\
methanol maser & 10.2993 & -0.1461 &       &       &       &        &        &        &        &         &    &   &  & \\
UC HII region  & 10.3008 & -0.1471 &       &       &       &        &        &        &        &  & 890$\pm$30$^*$ & & & \\
D & 10.2941 & -0.1443  &       &       &       &        &        &        &        &         &                &  & & \\
\multicolumn{4}{l}{C4 and C5} \\
8   & 10.28264 & -0.11271 &              &                    & 14.772$\pm$0.013   & 9.864$\pm$0.039  & 8.573$\pm$0.038 & 7.357$\pm$0.03 & 6.609$\pm$0.026 &  3.66 (3.32 ROB)  & $\leq$1.0$^*$  & $\leq$10.6 & 1.52 & 2.95 (3.29 ROB) \\
9   & 10.27116 & -0.12776 &              &                    &                   & 12.221$\pm$0.061 & 9.599$\pm$0.054 & 7.759$\pm$0.039 & 6.920$\pm$0.033 & 4.30 (3.56 ROB)  & 5.1$\pm$1.0$^*$ & 49 &  & 2.62 (3.36 ROB) \\
10  & 10.26919 & -0.13257 &              &                     &                 & 14.014$\pm$0.137 & 11.980$\pm$0.074 & 10.252$\pm$0.084 & 8.910$\pm$0.055 &  6.0$\pm$0.5$^*$  & $\leq$4.5$^*$ &  &  & 2.9 \\
EGO & 10.2874 & -0.1247 &       &       &       &        &        &        &        &         &      &   &  &  \\
methanol maser & 10.2872 & -0.1252 &       &       &       &        &        &        &        &         &   &   &   & \\
E   & 10.2873 & -0.1088 &       &       &       &        &        &        &        &         &  24$\pm$5$^*$  & &  & \\
\multicolumn{4}{l}{C6} \\
11  & 10.22781 & -0.20707 &                 &                  & 15.225$\pm$0.020 & 12.261$\pm$0.067 & 10.841$\pm$0.079 & 8.744$\pm$0.034 & 7.021$\pm$0.022 & 0.96             & 30.0$\pm$5.0$^*$ & 262 & 2.64 & 6.06  \\
12  & 10.22664 & -0.20910   &              &                  & 16.572$\pm$0.067 & 12.473$\pm$0.069 & 10.234$\pm$0.052 & 8.799$\pm$0.041 & 7.684$\pm$0.035 & 4.2$\pm$1.0$^*$ & & & 1.65 & 3.5 \\
13  & 10.23044  &  -0.20471 &  &  17.006$\pm$0.049 & 13.151$\pm$0.003 & 10.678$\pm$0.049 & 9.443$\pm$0.042 & 8.865$\pm$0.039 & 8.569$\pm$0.063 &  & & &  &  \\
\multicolumn{4}{l}{C7} \\
14  & 10.32254 & -0.22372 &                   &                 &                & 13.702$\pm$0.122 & 12.032$\pm$0.141 & 10.558$\pm$0.088 & 10.070$\pm$0.070 &  $\geq$6.7$^*$  & $\leq$0.4$^*$ &  &  & $\leq$3.4 \\
UC HII region  & 10.3202 & -00.2331 &       &       &       &        &        &        &        &     & 130$\pm$10$^*$   &  & & \\
15  & 10.32035 & -0.25867  &              &                      &                & 11.412$\pm$0.119 & 8.849$\pm$0.113 & 7.231$\pm$0.090 & 6.002$\pm$0.296 & 1.71 & 113$\pm$30$^*$ & &  & 4.30  \\
methanol maser & 10.3204 & -0.2585  &       &       &       &        &        &        &        &         &      &   & &  \\
HII region & 10.3218 & -0.2587 &       &       &       &        &        &        &        &         &  & &  & \\ 
F   & 10.3151 & -0.2547 &       &       &       &        &        &        &        &         &  14$\pm$3$^*$  & &  & \\
\multicolumn{4}{l}{External} \\
16  & 10.25432 & -0.24416 & 18.969$\pm$0.121 & 16.088$\pm$0.021 & 13.083$\pm$0.003 & 10.163$\pm$0.070 & 9.114$\pm$0.073 & 8.234$\pm$0.038 & 7.706$\pm$0.036 &   4.56$\pm$0.10$^*$   & 1.6$\pm$0.2$^*$ & & 0.39 & 3.15 \\
17  & 10.27684  & -0.23287 & 19.278$\pm$0.161 &  & 15.234$\pm$0.020 & 12.800$\pm$0.069 & 11.641$\pm$0.074 & 10.697$\pm$0.092 & 9.865$\pm$0.057 & $\geq$7.5$^*$ & $\leq$0.3$^*$ & &  & $\leq$2.4 \\
18  & 10.36044 & -0.23713   &             &                    & 15.595$\pm$0.027 & 11.595$\pm$0.060 & 10.249$\pm$0.047 & 9.217$\pm$0.045 & 8.137$\pm$0.032 & 5.61 (5.31 ROB)   & 0.2$\pm$0.1$^*$ &  & 1.07 & 2.53 (2.83 ROB) \\
19  & 10.35452 & -0.22285  & 18.489$\pm$0.079 & 16.901$\pm$0.045 & 15.479$\pm$0.025 & 12.443$\pm$0.062 & 11.182$\pm$0.076 & 10.188$\pm$0.055 & 9.453$\pm$0.051 & 6.44 ROB      &  $\leq$0.3$^*$  &  & 0.59 & 3.01 ROB \\
20 & 10.40812 & -0.20145  &              &                     & 16.939$\pm$0.095 & 12.015$\pm$0.086 & 9.839$\pm$0.058 & 8.549$\pm$0.039 & 7.633$\pm$0.027 & 4.76 (4.27 ROB)     & 0.6$\pm$0.2$^*$  & 6.5 & 1.99 & 2.87 (3.36 ROB) \\
21 & 10.4038  & -0.2028   &         &        &        &       &       &    &            &  7.47 ROB  & 0.8$\pm$0.3$^*$ & 8.6 & &  \\
22 & 10.38881 & -0.19460 & & 16.443$\pm$0.029  &  13.718$\pm$0.005 & 11.327$\pm$0.051 & 10.315$\pm$0.050  & 9.590$\pm$0.048  & 9.261$\pm$0.074  & 6.9$\pm$0.5$^*$ & $\leq$0.3$^*$ & $\leq$3 & -0.26 & 2.36 \\
23 & 10.37284 & -0.18482  &               &                   & 14.631$\pm$0.012 & 11.630$\pm$0.089 & 10.236$\pm$0.154 & 9.255$\pm$0.062 & 9.180$\pm$0.057 &  4.68 (4.35 ROB)  & 10.5$\pm$2.0$^*$  & 97 & 1.07  & 4.50 (4.83 ROB) \\
24  & 10.33769  & -0.08369 & & 17.941$\pm$0.114  & 15.592$\pm$0.027 & 13.362$\pm$0.090 & 12.502$\pm$0.116  & 11.014$\pm$0.144 & 9.411$\pm$0.217  & 5.06 & 0.9$\pm$0.3$^*$ & & 1.19 & 4.35 \\
25  & 10.36273 & -0.05901 &  & 17.617$\pm$0.085 &  16.249$\pm$0.049 & 14.468$\pm$0.177 &  12.621$\pm$0.129 & 11.510$\pm$0.133 & 10.473$\pm$0.104  & 6.5$\pm$0.5$^*$ & $\leq$0.3$^*$ &  & 0.87 & 3.97 \\
26  & 10.23508 & -0.12661  &                 &                   &               & 13.425$\pm$0.115 & 11.594$\pm$0.128 & 10.272$\pm$0.087 & 9.854$\pm$0.104 &  6.50$\pm$0.20$^*$ & 0.4$\pm$0.2$^*$ & &  & 3.35 \\
27  & 10.25199 & -0.07062 & 18.642$\pm$0.089 & 18.236$\pm$0.149 & 16.330$\pm$0.053 & 11.443$\pm$0.039 & 9.736$\pm$0.045 & 8.199$\pm$0.032 & 7.340$\pm$0.038 &  4.00$\pm$0.10$^*$  & $\leq$1.0$^*$ &  & 1.86 & 3.34 \\
\\
\hline
\label{G10_YSOs}
\end{tabular}\\
}
\tablefoot{
The 24~$\mu$m magnitudes are from the MIPSGAL catalogue or from Robitaille et al. (\cite{rob08}; magnitude in brackets followed by ROB), or have been measured by us. A $^*$ indicates our own measurements. The luminosity of the sources (associated with G010.32$-$00.15) is estimated from their 70~$\mu$m flux, using the correlation established by Duham et al. (\cite{duh08}) and Ragan et al. (\cite{rag12}).The positions of the 6.7~Ghz methanol masers are from Green et al. (\cite{gre10}; positional accuracy $\sim$0$\farcs$4), the positions of the EGOs from Cyganowski et al. (\cite{cyg09}; positional accuracy $\sim$0$\farcs$1).
}
\end{table*}

\section{Discussion}

\subsection{IRDCs and star formation}

IRDCs were first identified as regions of low brightness at mid-IR wavelengths, this property being attributed to absorption of the background emission by dense dusty clouds (Perault et al. \cite{per96}; Egan et al. \cite{ega98}; Carey et al. \cite{car98}). Infrared dark clouds (IRDCs) are often considered as the birthplace of massive stars or clusters (for example, Battersby et al. \cite{bat10}). Thousands of IRDCs have been detected in the Galactic plane (Simon et al. \cite{sim06}; Peretto \& Fuller \cite{per09}).

When considering 3171 IRDCs that were ``{\it Spitzer}-dark regions'', Wilcock et al. (\cite{wil12}) have shown that only 1205 of them were ``{\it Herschel}-bright clouds'', hence dense structures seen in emission at SPIRE wavelengths and not simply ``holes'' in the sky, zones devoid of material. This mis-identification of IRDCs occurs in our regions; for example, four IRDCs are catalogued in the G319.88$+$00.79 field, and three of them do not correspond to dense structures, as shown by Fig.~C.1 (Appendix C).  They correspond to the centres of the ionized bubbles devoid of 8.0~$\mu$m emission (SDC~G319.882$+$0.810 at the centre of S98 and SDC~G319.886$+$0.753 at the centre of S97), or to a zone of low 8.0~$\mu$m emission between two filaments (SDC~G319.859$+$0.818). Only one of them corresponds to a zone of real absorption owing to the presence of a dense structure along the line of sight (SDC~G319.910$+$0.796; on the left border of C1). Seven IRDCs are listed in the vicinity of G010.32$-$00.15, as shown by Fig.~C.2 (Appendix~C). Five of them are observed in the direction of high column density structures, especially SDC~G010.215$-$0.206 in the direction of the C6 clump and SDC~G010.358$-$0.166 + SDC~G010.405$-$0.200 in the direction of the parental filament. But two of them are not at all associated with dense structures: SDC~G010.315$-$0.133 which lies in the direction of the central hole in column density (a zone occupied by the ionized region and devoid of 8.0~$\mu$m emission); SDC~G010.291$-$0.187 which lies in the direction of the centre of the bottom lobe (also a zone occupied by ionized gas).

The very dense clumps that lie at the waist of G010.32$-$00.15 and contain massive sources presently forming or just formed  are not detected as IRDCs. This situation is frequently observed (see also the C2 clump in G319.88+00.79, or the massive-star forming condensation~1 in RCW120, Deharveng et al. \cite{deh09}, Zavagno et al. \cite{zav10}). These massive clumps, not detected as IRDCs and containing MYSOs, are adjacent to the PDRs of \HII\ regions, and the absorption they induce is hidden at near or mid-IR wavelengths by the bright emission of the nearby PDRs.

\subsection{The global morphology of G319.88$+$00.79 and G010.32$-$00.15 - Filaments and sheets}

Many studies are devoted to filaments and to star formation along filaments. It is often forgotten that 2D structures - sheets, viewed  edge-on, can be mistaken as filaments. Also perfect geometrical structures (cylinders or flat planes) are never met in nature where density gradients and wavy structures are everywhere present. What have we learned about the overall morphology of  the G319.88$+$00.79 and G010.32$-$00.15 bipolar nebulae and their parental clouds?\\

\begin{figure}[h!]
\centering
\includegraphics[width=85mm]{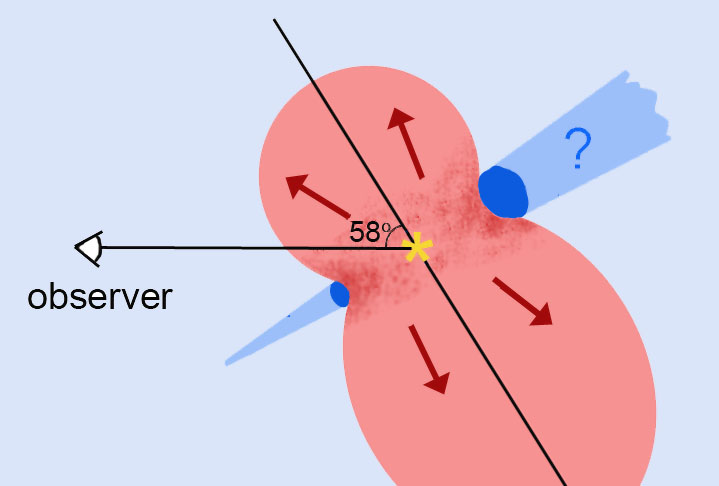}
  \caption{Morphology of G319.88$+$00.79. The ionized material appears in pink, the neutral one in blue. We do not know the extent of the molecular material at the back of the nebula.}
\label{schema1}
\end{figure}

That we are seeing a ring (or torus) of dense absorbing material at the waist of the G319.88$+$00.79 bipolar \HII\ region, centred on the exciting star(s) (Sect.5; Fig.~\ref{G319cc}), indicates two things: 1) the exciting star(s) formed in a 2D dense structure, a sheet rather than a cylindrical filament; 2) this parental sheet is not seen edge-on; the line of sight makes an angle of about 58$\degr$ with the symmetry axis of the bipolar nebula (assuming a circular ring; the observed size of the elliptic ring is $\sim 1.95\arcmin \times 1.03\arcmin$, thus 1.47~pc $\times$ 0.78~pc for a distance of 2.6~kpc).  Figure~\ref{schema1} shows our view of the morphology of this region. Only the bottom part of the ring is clearly seen in absorption, which implies that this absorption is due to material located in front of the \HII\ region. We do not detect {\it Herschel}-SPIRE emission associated with this structure, which suggests that there is little material.  The bright rims observed left of the ring are associated with the C1 clump, and the bright rims observed right of the ring are associated with C2. We suggest that C2 lies slightly at the back of the \HII\ region and that it is composed of several small-size dense cores (each bordered by the bright rims observed, for example, at 8.0~$\mu$m, Fig.~\ref{G319cc}), not resolved by {\it Herschel}-SPIRE. Thus dense structures are located at the back of the \HII\ region,  but we do not know the extent of the molecular material there.

The mean velocity of the ionized region is $-$38~km~s$^{-1}$ (Caswell \& Haynes \cite{cas87}), and the velocity of the molecular material is in the range $-$44~km~s$^{-1}$ to $-$42~km~s$^{-1}$ (C1 to C3 clumps). If the morphology illustrated by Fig.~\ref{schema1} is correct, then the ionized bottom lobe is dominant at radio wavelengths. (In the lobes the ionized gas flows away from the dense central region.) Observations of the velocity field of the ionized region is necessary to confirm the proposed morphology.\\

\begin{figure}[h!]
\centering
\includegraphics[width=85mm]{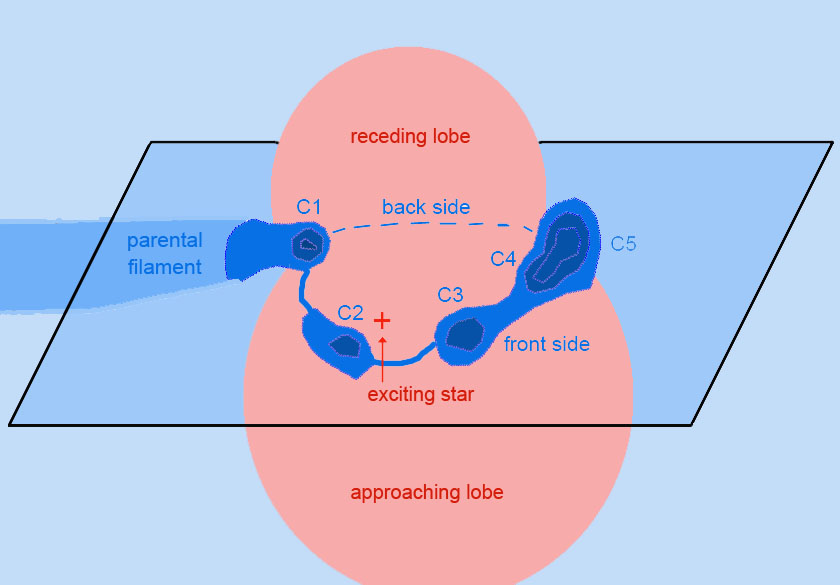}
  \caption{Morphology of the G010.32$-$00.15 complex. The plane represented here contains the parental filament, the clumps at the waist of the nebula, and the exciting star. The line of sight makes a small angle with this plane.}
\label{schema2}
\end{figure}

In the case of the G010.32$-$00.15 bipolar \HII\ region we know a little more about the velocity field of the ionized gas. The velocity of the ionized gas has been measured by Araya et al. (\cite{ara07}) at three positions with a beam of 2.5$\arcmin$. In the  direction of the central region ($l$=10\fdg313, $b$=$-$0\fdg152), the velocity is 9.4~km~s$^{-1}$; the velocity is 12.6~km~s$^{-1}$ 2.5$\arcmin$ north in the direction of the upper lobe; the velocity is 5.6~km~s$^{-1}$ 2.5$\arcmin$ south of the central region in the direction of the bottom lobe. This velocity gradient indicates an approching  bottom lobe with respect to the central region, and a receding upper one (thus Fig.~\ref{schema2}).

The parental filament (left of the \HII\ region) is seen in absorption, and in line with it, the C1 and the C4 + C5 external clumps are also seen in absorption at {\it Spitzer}-IRAC wavelengths.  Thus these clumps are slightly in front of the \HII\ region. The C2 and C3 clumps are not clearly seen in absorption, but they are bordered by bright rims (PAH emission) on their side turned towards the exciting star (Fig.~\ref{champ10f}); thus they lie in front of the ionized region and its exciting star\footnote{If they were lying behind the exciting star, their whole area would show a bright emission at 8~$\mu$m. Here we see only their top border turned towards the exciting star.}. This  suggests that the C1 to C5 clumps form a single structure, half a torus of dense material, encircling the waist of the bipolar \HII\ region and located in front of this region. 

The \HII\ region is not open on its back side (it is ionization-bounded), as we see a faint {\it Spitzer}-IRAC PAHs emission closing the clumpy  torus on its back side. The parental structure is much more dense on the front side of the \HII\ region than on the back one. We suggest that the parental structure was a flat 2D cloud presenting a strong density gradient or containing a dense pre-existing filament.

\subsection{Triggered star formation}

 
Several processes can trigger star formation in the vicinity of \HII\ regions, as illustrated by Deharveng et al. (\cite{deh10}; their figure 4): small-scale gravitational instabilities of the IF, large-scale gravitational instabilities of the material accumulated in a shell behind the IF (the so-called ``collect and collapse process''), and  radiation-driven implosion (``RDI process'') of a pre-existing dense clump compressed by the ionized gas. The process proposed by Fukuda \& Hanawa (\cite{fuk00}; Sect.~3) is somewhat different, involving the gravitational collapse of clumps, formed by the compression of a pre-existing dense filament due to an adjacent \HII\ region.  It is often difficult to establish which triggering process is at work. And, as shown by the simulations of Dale et al. (\cite{dal12}; \cite{dal15}), it is often difficult to deduce which objects have been induced to form and which formed spontaneously.\\

We propose the G010.3$2$-00.15 region as an example of massive-star formation triggered  by an \HII\ region and involving different processes.  Concerning the C1 to C5 clumps, the signatures of triggering follow.

$\bullet$ An incomplete shell of dense molecular material surrounds the ionized region at its waist. It is fragmented, and the five fragments are massive (several hundreds of solar masses). The shape of the PDR shows the interaction of the high-pressure ionized gas with the surrounding molecular material: the ionized gas tries to escape from the central \HII\ region via low density zones present between the dense fragments (Fig.~\ref{champ10f}).

$\bullet$ Each of the fragments contains at least one embedded young stellar object, most often of high mass. C1 contains a MYSO in an early stage of evolution (probably a Class~0), C2 contains a high-luminosity YSO (probably a Class~I) with a high accretion rate, C3 contains an UC \HII\ region, and C4 contains a YSO probably of Class~0.   All these massive YSOs have methanol maser emission at 6.7~GHz. According to Breen et al. (\cite{bre10}), this maser emission indicates an age $\leq$5$\times$10$^4$~yr for the YSOs. We cannot determine the dynamical age of the bipolar \HII\ region  accurately because appropriate models are missing, but we can estimate a lower limit for its age. The waist of the \HII\ region has a diameter of 1.44~pc (maximum size in the parental structure). An \HII\ region of radius 0.72~pc, which is excited by an O5V -- O6V star (N$_{\rm Lyc}$=10$^{49}$), and is evolving in a homogeneous medium of density 10$^4$~H$_2$~cm$^{-3}$ (a reasonable density for the parental structure) has an age $\sim$170\,000~yr (according to the evolution of an \HII\ region described by Dyson \& Williams \cite{dys97}). A bipolar \HII\ region should expand more slowly in the plane of the parental structure than a spherical \HII\ region. A fraction of the ionized gas flows away from the dense central region towards the lobes, reducing the pressure of the ionized gas in the central region. The bipolar \HII\ region is most probably older than 170\,000~yr, hence much older than the Class~0/I YSOs located in the clumps at its waist. 

$\bullet$ The  C1 to C5 clumps form a single structure: a dense half shell encircling the waist of the nebula. The probability that these clumps were pre-existing structures distributed along a ring and simultaneously reached by the expanding IF is very low.  We argue that their formation results from the interaction of the \HII\ region with the dense surrounding material.  Under the assumption that triggering is indeed taking place, there are two alternative solutions for the triggering process.  The first option is the equivalent of the collect and collapse process, but taking place in a 2D rather than in a 3D cloud.  The material of the parental sheet is collected during the expansion of the central \HII\ region, forming a dense torus. Later, gravitational instabilities lead to the fragmentation of the torus, followed by star formation.  However, at present, no theoretical model or simulation can replicate this scenario.  The second option is that, according to Fukuda \& Hanawa's similations, the \HII\ region compressed a nearby pre-existing filament, leading to the formation of the C2 and C3 central cores that contain the more evolved and massive YSOs (YSO~\#4 and an UC \HII\ region). The external clumps, C1 and C4 + C5, contain the less evolved YSOs, which are detected via the presence of EGOs, methanol masers, and 70~$\mu$m emission.  The present data do not allow us to determine that C1 and C4 + C5 are first-generation cores, or if only C2 and C3 are first-generation cores while C1 and C4 + C5 are second-generation ones.  However, some observational details do not agree with the simulations.  In Fukuda \& Hanawa's simulations, the filament is not distorted by the \HII\ region. (It does not encircle the waist of the \HII\ region.) In the simulations, the \HII\ region is density bounded on the side opposite to the filament, whereas our \HII\ regions are ionization bounded in the plane of their parental structure.

The C6 clump suggests that another triggering process may be at work. It is clearly a massive pre-existing structure that has been reached by the IF and probably compressed by the high-pressure ionized gas.  An intermediate-mass Class~I YSO possibly results from this compression, which is located at the column density peak of the clump. We estimate that its luminosity is $\sim$260$\lsol$.  We suggest that its formation results from  the radiation driven implosion of the pre-existing dense cloud, as described and simulated by numerous authors (Bisbas et al. \cite{bis09}, \cite{bis11}, and references therein). Bisbas et al. (\cite{bis11}) have shown that the incident ionizing flux is the critical parameter determining a cloud evolution. The surface of C6, located in projection at about 2.9~pc from the O5 -- O6 exciting star, receives a ionizing photon flux $\sim$10$^9$~cm$^{-2}$~s$^{-1}$, which is a case for low photon flux. The shock front progressing in the cloud is weak, and it sweeps up matter slowly. According to the simulations, star formation occurs in a dense filament, well behind the IF, towards the centre of the cloud. The observations of C6 fit this picture well; however, the resolution of the {\it Herschel} observations does not allow access to the small-scale structure of the clump. Furthermore, the clouds involved in the simulations of Bisbas et al. have much lower mass than the C6 clump ($\sim$600$\msol$, obtained by integration in a circular aperture of radius 50\arcsec covering the whole clump). Thus triggered star formation in C6 via the RDI process also needs confirmation.\\
 
Star formation is less active around G319.88$+$00.79 than around G010.32$-$00.15 (Sect.~5.4). We suggest that the formation of YSOs \#4 and \#5, located at the waist of the nebula, near the ionization front, could - based on their location - have been triggered by the central \HII\ region. It is especially the case of YSO~\#5, which is located in the direction of a bright rim (part of the C2 clump), thus close to the IF. All the other YSOs in this region are located along the parental filament, outside the C1 and C2 clumps. The Class~0 YSO A is located about 0.6~pc (in projection) from the IF at the right waist of the nebula. The same situation is observed in the G010.32$-$00.15 field; for example, YSOs~\#21 and \#23 are located along the parental filament, too far away from the IF for triggering. (The Class~I YSO~\#23 is located about 1.4~pc away - in projection - from the IF at the left waist of the nebula; the Class~0 YSO~\#21 lies even farther away.) The formation of these Class~0/I YSOs has most likely not been triggered by the expansion of the bipolar nebulae. Their formation along a filament probably  results from some other process, such as the filament's fragmentation via gravitational instabilities (Jackson et al. \cite{jac10}; Andr\'e et al. \cite{and10} and references therein).

\section{Conclusions}

Using the {\it Spitzer} GLIMPSE and MIPSGAL surveys and the {\it Herschel} Hi-GAL survey we have identified 16 candidate bipolar nebulae along the Galactic plane, between $+$60$\degr$ and $-$60$\degr$ in Galactic longitude, $+$1$\degr$ and $-$1$\degr$ in latitude. This first paper focuses on two of them, G319.88$+$00.79 and G010.32$-$00.15, which are especially interesting for illustrating the morphology of these regions and the formation of stars in their vicinity. Our conclusions follow.

$\bullet$ Concerning their morphology both regions present similar signatures.

- The parental structure, whether filament or sheet, is cold (13~K--17~K), and dense ($\sim$10$^{4}$~cm$^{-3}$ or more in G319.88$+$00.79). 

- The brightest parts of the \HII\ region and its exciting cluster are centred on the parental structure. 

- Two ionized lobes, closed or open, extend perpendicular to the parental structure. Warm dust (20~K--25~K) is present in the PDRs bordering the ionized regions, including the lobes.

- High column density clumps are present at the waist of the nebulae, adjacent to the ionized regions. MALT90 observations show that they are associated with the ionized regions, because they present similar velocities. Their mass is a few hundred solar masses. High densities are measured in their central regions (several 10$^4$~cm$^{-3}$ in G319.88$+$00.79, several 10$^5$~cm$^{-3}$ in G010.32$-$00.15).

$\bullet$ We identified the exciting stars of the two \HII\ regions. This has been especially useful for estimating the distance of G010.32$-$00.15. The distance of this region, as adopted by different authors, is in the range 2~kpc to 19~kpc. We have estimated a spectrophotometric distance of 1.75$\pm$0.25~kpc, which is compatible with the near kinematic distance.

$\bullet$ We used colour-colour diagrams created from the {\it Spitzer}-GLIMPSE observations to identify candidate Class~I YSOs in the vicinity of the \HII\ regions. Some of these have been confirmed, using their 24~$\mu$m and Hi-GAL 70~$\mu$m emissions. Class~0 YSO were detected thanks to their 70~$\mu$m emission.

Near G319.88$+$00.79, eight candidate Class~I YSOs were identified by their {\it Spitzer}-GLIMPSE colours. Thanks to their [8]$-$[24] colour, three of them (YSO~\#2, \#4, and \#9) are confirmed Class~I, three  are flat-spectrum sources (thus of uncertain evolutionary status between Classes~I and II; YSOs \#5, \#7, \#8), and two are probable Class~II (YSOs \#1 and \#6). A Class~0 is detected (YSO~A), which has a luminosity $\sim$80$\lsol$ and probably a high infall rate. Only two YSOs (\#4 and \#5) are located in a clump at the waist of the nebula. The other YSOs, except YSO~\#2, are located along the parental filament.

Near G010.32$-$00.15 twenty-six sources are candidate Class~I YSOs, according to their {\it Spitzer}-GLIMPSE colours; among them,  eight are confirmed Class~I YSOs (\#1, \#3, \#4, \#11, \#15, \#23, \#24, and \#25), five are Class~II YSOs, and ten have uncertain natures, Class~II or flat-spectrum sources, due to uncertain 24~$\mu$m magnitudes. Three Class~0 are present (2 of them, in C1 and C4, with associated EGOs and methanol masers and without near-IR or {\it Spitzer}-IRAC  counterparts, except at 4.5~$\mu$m; and YSO \#21). Several of these YSOs are possibly not associated with the bipolar nebula. This region  demonstrates the importance of velocity measurements to associate a clump or a YSO with a nearby \HII\ region. All the high-luminosity sources, bright  at 70~$\mu$m, are observed in the direction of the massive clumps forming the waist of the nebula. The two brightest and possibly most evolved sources (the UC \HII\ region in C3, and the Class~I YSO~\#4 in C2, of luminosity $\sim$4600$\lsol$) are embedded in the central C2 and C3 clumps. The less luminous sources, also possibly the least evolved, Class~0 YSOs (associated with EGOs and class~II methanol masers), are located in the outside C1 and C4 clumps.

$\bullet$ The G010.32$-$00.15 bipolar \HII\ region has triggered the formation of stars in the surrounding medium via various processes. We suggest that the C6 massive clump was a pre-existing cloud reached by the ionization front limiting the southern lobe and compressed by the high-pressure ionized gas. An intermediate luminosity Class~I YSO ($\sim$260$\lsol$) is located at its centre, probably formed by triggering via the RDI process. A cluster is possibly forming there because two other YSOs are located nearby.

On the contrary, we suggest that the C1 to C5 massive clumps were not pre-existing clouds, since they form a single circular structure, but that their formation results from the interaction of the \HII\ region with the surrounding dense molecular material. This triggering can result from the collect of the dense material of a flat cloud (thus the formation of a dense torus at the waist of the \HII\ region) or via  the compression of a pre-existing dense filament as suggested by Fukuda \& Hanawa (\cite{fuk00}). However, no model or simulation is presently available to confirm one or the other process.

$\bullet$ G010.32$-$00.15 is more extreme than G319.88$+$00.79. Its exciting star is more massive (O5V--O6V versus O8V--O9V), and  the condensations present at its waist are more numerous (5 against 2), more compact (equivalent radius in the range 0.05~pc -- 0.18~pc against $\sim$0.30~pc), and denser (in the range 3$\times$10$^5$~cm$^{-3}$ -- 5$\times$10$^6$~cm$^{-3}$ against less than 10$^5$~cm$^{-3}$).  More Class~0/I YSOs are present in the vicinity of G010.32-00.14 than near G319.88$+$00.79, and several of them are massive. The high density of the molecular material into which the exciting star of G010.32$-$00.15 formed and evolved\footnote{A higher density in the vicinity of G010.32$-$00.15 than in the vicinity of G319.88$+$00.79 is inferred from the comparison of the column densities of the parental filaments (which differ by a factor $\sim$8).} could be the reason for such a situation, but only a statistical approach can confirm or disprove this assumption. \\

After this work was completed, we became aware of the paper by Dewangan et al. (\cite{dew15}), a study of star formation around the G010.3$-$0.1 \HII\ region. These authors base their estimate of the evolutionary status of the sources on their near-IR and {\it Spitzer}-GLIMPSE magnitudes. They do not mention that there are several velocity components in some directions, hence the uncertain association of many sources with the central \HII\ region. Furthermore, and this is an important point, they mention the bipolar nature of the central \HII\ region, but in their discussion of (triggered) star formation they do not take the very specific morphology of this star-forming complex into account.


\begin{acknowledgements}

 We thank the referee for a careful reading of the manuscript: his comments and suggestions have been very helpful for clarifying  the text and improving the paper.  We also thank N. Fukuda for answering our questions about the simulations and D. Elia for his explanations concerning CuTEx.
 
This research made use of the SIMBAD database operated at the CDS, Strasbourg, France, and of the interactive sky atlas Aladin (Bonnarel et al. \cite{bon00}). 

The authors would first like to thank the Herschel Hi-GAL team for their continuing work on the survey. 

Observations were obtained with the {\it Herschel}-PACS and {\it Herschel}-SPIRE photometers. PACS was developed by a consortium of institutes led by MPE (Germany), including UVIE (Austria); KU Leuven, CSL, IMEC (Belgium); CEA, LAM (France); MPIA (Germany); INAF-IFSI/OAA/OAP/OAT, LENS, SISSA (Italy); IAC (Spain). This development was supported by the funding agencies BMVIT (Austria), ESA-PRODEX (Belgium), CEA/CNES (France), DLR (Germany), ASI/INAF (Italy), and CICYT/MCYT (Spain). SPIRE has been developed by a consortium of institutes led by Cardiff Univ. (UK) and including Univ. Lethbridge (Canada); NAOC (China); CEA, LAM (France); IFSI, Univ. Padua (Italy); IAC (Spain); Stockholm Observatory (Sweden); Imperial College London, RAL, UCL-MSSL, UKATC, Univ. Sussex (UK); Caltech, JPL, NHSC, Univ. Colorado (USA). This development was supported by national funding agencies: CSA (Canada); NAOC (China); CEA, CNES, CNRS (France); ASI (Italy); MCINN (Spain); SNSB (Sweden); STFC and UKSA (UK); and NASA (USA). We thank the French Space Agency (CNES) for financial support.

During the MALT90 observations used herein, the Mopra telescope was part of the Australia Telescope National Facility and was funded by the Commonwealth of Australia for operation as a National Facility managed by CSIRO. The University of New South Wales Mopra Spectrometer Digital Filter Bank used for the observations with the Mopra telescope was provided with support from the Australian Research Council, together with the University of New South Wales, University of Sydney, Monash University, and the CSIRO. The authors also thank the staff of the Paul Wild Observatory for their assistance during these observations.
 
This work is also based on observations made with the Spitzer Space Telescope, which is operated by the Jet Propulsion Laboratory, California Institute of Technology, under contract with NASA. We made use of the NASA/IPAC Infrared Science Archive to obtain data products from the 2MASS, {\it Spitzer}-GLIMPSE, and {\it Spitzer}-MIPSGAL surveys.

Some of the data reported in this paper were obtained as part of the United Kingdom Infrared Telescope (UKIRT) Service Programme.  The UKIDSS project is defined in Lawrence et al. (2007).

MS thanks the CNES for financial support through a post-doctoral fellowship. ADC acknowledges funding from the European Research Council for the FP7 ERC starting grant project LOCALSTAR.

\end{acknowledgements}



\clearpage
\begin{appendices}
\appendix

\section{More about G010.32$-$00.15}

\subsection{Distance of G010.32$-$00.15}

The distance of this region is very uncertain, and different indicators (and authors) give distances covering the range 2~kpc to 19~kpc.

$\bullet$ Radio recombination lines give the velocity of the ionized gas: V$_{\rm LSR}$(H109$\alpha$)=9.7~km~s$^{-1}$ (Reifenstein et al.~\cite{rei70}),  V(H110$\alpha$)=12.0~km~s$^{-1}$ (Downes et al.~\cite{dow80}), V(H110$\alpha$)=9.4~km~s$^{-1}$ in the central region (Araya et al.~\cite{ara07}). The molecular condensations present at the waist of the nebula have similar velocities in the range 11.3 -- 13.5~km~s$^{-1}$ (NH$_3$ emission lines; Wienen et al. \cite{wie12}; see Table~\ref{G10condensations}); see also the CO maps in Beuther et al.~(\cite{beu11}; their figures 7 and 8). These velocities lead to a kinematic near distance of $\sim$2~kpc or a far distance $\sim$15~kpc.

$\bullet$ H$_2$CO absorption lines at 6~cm are often used to resolve the distance ambiguity. For distances within the Sun's orbit,  the radial velocity increases with distance from the Sun in the first quadrant, reaches a maximum at the tangent point and then decreases to zero at the Galactocentric radius of the Sun. If the thermal continuum of the \HII\ region is absorbed by the H$_2$CO molecules, the H$_2$CO-absorbing cloud must be in front of the \HII\ region. H$_2$CO absorption lines in the direction of G10.31$-$0.15 are observed at velocities of 11.1~km~s$^{-1}$, 22.3~km~s$^{-1}$, and 38.4~km~s$^{-1}$, with faint components at 3.9~km~s$^{-1}$, 22.3~km~s$^{-1}$, and 47.3~km~s$^{-1}$ (Sewilo et al. \cite{sew04}; HPBW 2.56$\arcmin$). Similar velocities were measured for the H$_2$CO absorption lines by Wilson  (\cite{wil74}) and by Downes et al. (\cite{dow80}). That absorption lines are observed at velocities higher than for the \HII\ region points to the far kinematic distance. A far distance was adopted by Georgelin \& Georgelin (\cite{geo76}; 18.7~kpc), Sewilo et al. (\cite{sew04}; 15.0~kpc), and Du et al. (\cite{du11}; 15.12~kpc).

$\bullet$ The velocity at the tangent point is $\sim$150~km~s$^{-1}$, as indicated by the \HI\ emission observed in this direction (Kalberla et al. \cite{kal82}), but the H$_2$CO lines show no absorption near 150~km~s$^{-1}$ (Sewilo et al. \cite{sew04}). If G010.32$-$00.15 was at the far distance, we should observe absorption up to the maximum velocity in this direction. The H$_2$CO absorption only extends up to 47~km~s$^{-1}$, which could be due to a lack of H$_2$CO material near the tangent point. But the same situation is observed with the atomic hydrogen: the \HI\ absorption extends only to 45~km~s$^{-1}$ in the direction of G010.3$-$00.1 (and of the two nearby \HII\ regions G010.2$-$00.3 and G010.6$-$00.4), whereas its emission extends to 152~km~s$^{-1}$ (Kalberla et al. \cite{kal82}). Thus the \HII\ region G010.32$-$00.15 cannot be at the far distance. This situation has been discussed by Wilson (\cite{wil74}) and Downes et al. (\cite{dow80}). They adopt a distance of $\sim$5--6~kpc, corresponding to the highest H$_2$CO absorption velocity. A distance of 6~kpc also has been adopted by Beuther et al. (\cite{beu11}) in their study of W31.

$\bullet$ The nearby region G010.2$-$00.3 has velocities similar to our region, both for the ionized gas and for the associated molecular material (Wilson \cite{wil74}; Downes et al. \cite{dow80}; Araya et al. \cite{ara07}; Beuther et al. \cite{beu11}; Wienen et al. \cite{wie12}), so these two regions probably lie nearby. Near-IR photometry and $K$--band spectroscopy of the exciting stars of G010.2$-$00.3 by Blum et al.~(\cite{blu01}) indicate a distance of 3.4$\pm$0.3~kpc. \\

We identified the exciting cluster of the \HII\ region in Sect.~6.1 and its main O5V--O6V exciting star. The  spectrophotometric distance of this star is estimated to be 1.75~kpc, in good agreement with the near kinematic distance. And we have shown that this distance could not be greater than 2.5~kpc. 

How can we reconcile this maximum distance of 2.5~kpc for G010.32$-$00.15 with the H$_2$CO and \HI\ absorption spectra? We note that this situation is not unique to G010.32$-$00.15.  A parallax of a maser spot associated with G010.6$-$00.4 was measured by Sanna et al. (\cite{san14}), and they find a distance of 4.95$^{+0.51}_{-0.43}$\,pc.  Quireza et al. (\cite{qui06}) find a radio recombination line velocity of $-$1.1\,~km~s$^{-1}$ for G010.6$-$00.4, which leads to a kinematic distance of $\sim$18\,kpc (and no near kinematic distance).  Similarly, the \HII\ region G009.62+00.19 has a parallax distance of 5.15$^{+0.69}_{-0.55}$\,pc (Sanna et al., \cite{san09}) and near and far kinematic distances of $\sim$1 and $\sim$16\,kpc.  Hofner et al. (\cite{hof94}) found \HI\ absorption up to 55\,~km~s$^{-1}$ for G009.62$+$00.19.  All three sources are near a Galactic longitude of $10\degr$, have LSR velocities $\leq$10\,~km~s$^{-1}$, have H$_2$CO and \HI\ absorption up to $\sim$50\,~km~s$^{-1}$, and have large discrepancies between their distances estimated using different indicators.  The H$_2$CO and \HI\ clouds that are seen in absorption up to $\sim$50\,~km~s$^{-1}$ must lie in front of the three sources.  This implies that the velocity field cannot be described by approximately circular orbits about the Galactic centre.  The deviations from circular velocity here have been noticed by many previous authors and are probably the effects of the Galactic bar(s) (e.g., Rodriguez-Fernandez \& Combes, \cite{rod08}). Clearly, kinematic distances in this part of the Galaxy are not accurate.

\subsection{Other candidate Class~I YSOs in the vicinity of G010.32$-$00.15}

We discuss here the candidate Class~I YSOs present in the vicinity of G010.32$-$00.15,  which are not associated with it or whose association with the central \HII\ region is uncertain.

$\bullet$ YSOs in the direction of C7: this clump does not belong to the complex; its distance is unknown, preventing us from estimating its mass and density.

\begin{figure}[h!]
\centering
\includegraphics[width=80mm]{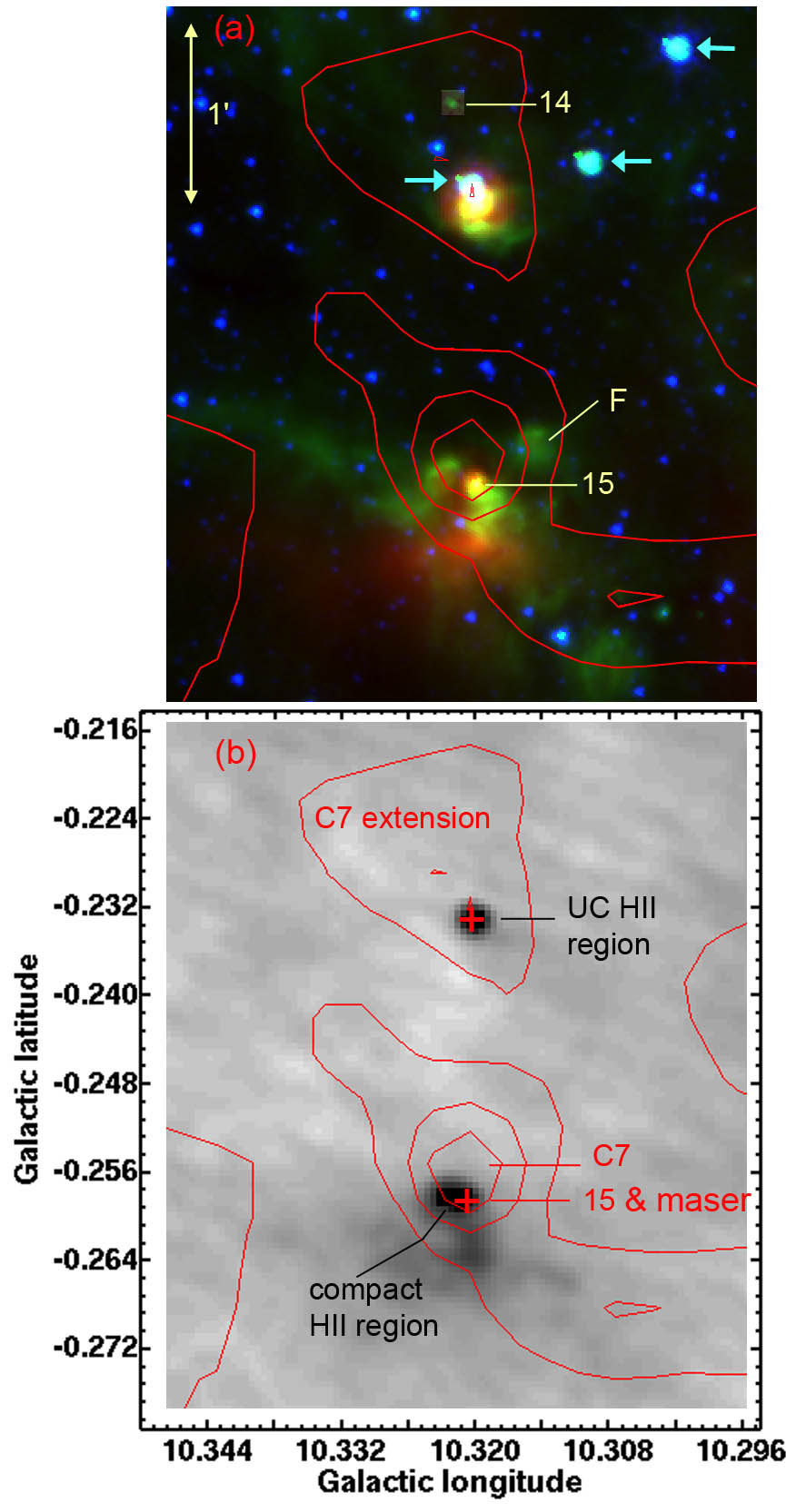}
  \caption{C7 clump. {\it (a)} Colour composite image with the 24~$\mu$m, the 8.0~$\mu$m, and the 3.6~$\mu$m emissions, respectively in red, green, and blue (linear units). Two candidate Class~I YSOs, \#14 and \#15, are identified; the brightness of YSO~\#14 has been enhanced. The three blue arrows point to candidate AGB stars. The red contours are for the column density (same levels as in Fig.~\ref{champ10hbis}~{\it (b)}). {\it (b)} The column density contours are superimposed on the  MAGPIS radio map at 20-cm. The UC and compact \HII\ regions are identified, which have 24~$\mu$m and 70~$\mu$m counterparts.}
\label{champ10l}
\end{figure}

The northern extension of C7 contains a small \HII\ region detected on the MAGPIS radio map and the faint candidate  Class~I YSO~\#14 (Fig.~\ref{champ10l}~{\it (a)}). YSO~\#14 has no detectable 24~$\mu$m and 70~$\mu$m counterparts, and its nature is uncertain (a low-luminosity Class~II or flat-spectrum source). The flux of the UC \HII\ region is given in Table~\ref{UCH}; its exciting star has not been identified. Its size is $\sim$3$\farcs$7 (beam deconvolved). The 70~$\mu$m source associated with this UC \HII\ region has a flux of 130$\pm$10~Jy. 

C7 contains a compact \HII\ region (see Table~\ref{UCH}) and the candidate Class~I YSO \#15 in a nearby direction. YSO \#15 is bright at 24~$\mu$m ([8.0]$-$[24] in the range 4.30~mag to 4.47~mag) and at 70~$\mu$m, but at this last wavelength it is difficult to separate it from the extended emission linked to the compact \HII\ region. At 70~$\mu$m we measured a flux of 113$\pm$30~Jy for this \HII\ region plus YSO~\#15. A class~II methanol maser was detected by Green et al. (\cite{gre10}) at less than 1$\arcsec$ from YSO~\#15. The velocity of the maser, 39.0~km~s$^{-1}$, indicates that it is associated with C7 and not with the G010.32$-$00.15 bipolar nebula. Thus YSO~\#15 is clearly a Class~I YSO embedded in C7.

A small {\it Spitzer} and {\it Herschel}-PACS diffuse region lies about 25$\arcsec$ west of YSO~\#15, source F in Fig.~\ref{champ10l}. It has no detectable radio-continuum counterpart on the MAGPIS map (flux at 20~cm $\leq$5~mJy). \\
 
\begin{figure}[h!]
\centering
\includegraphics[width=85mm]{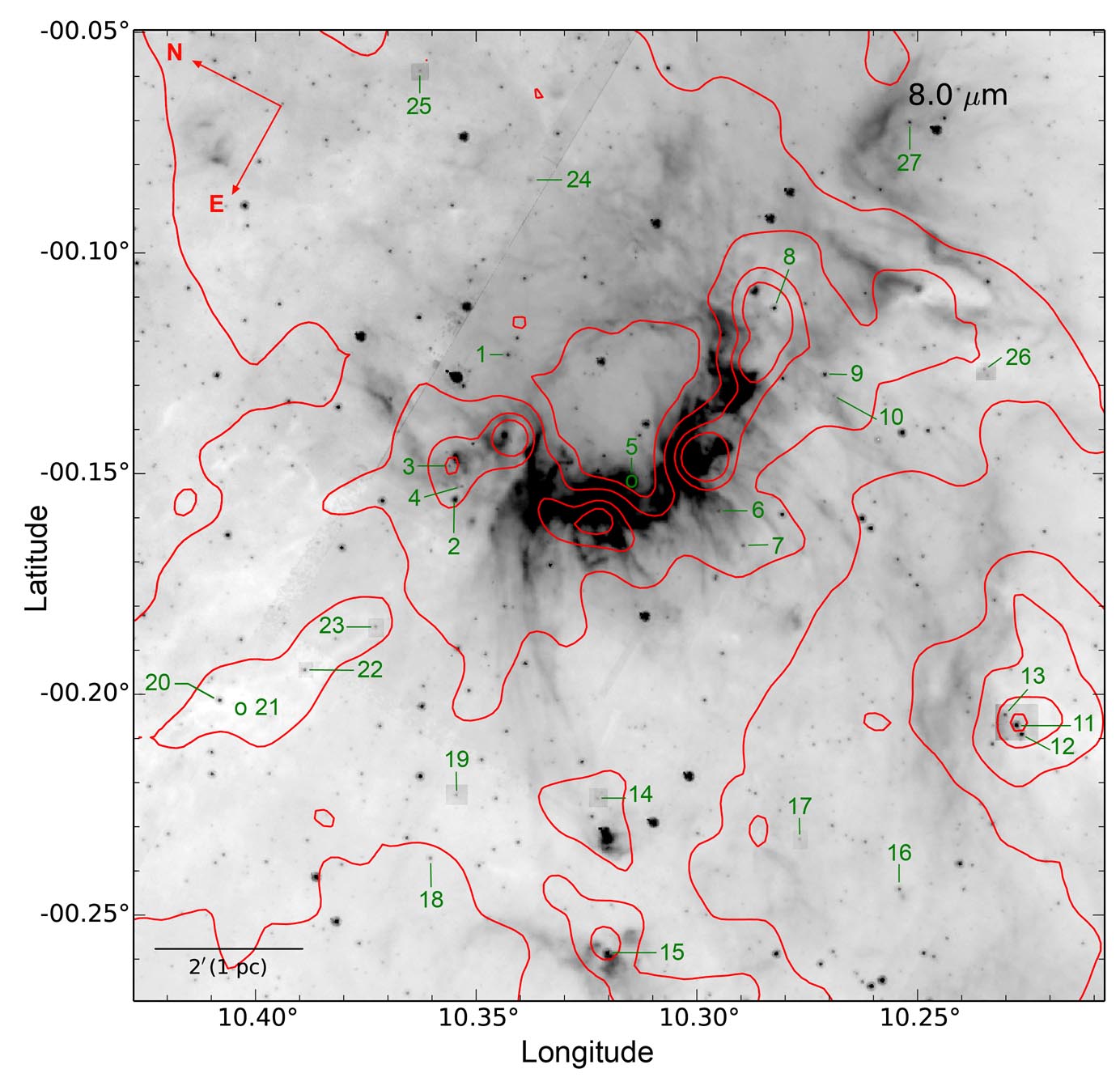}
  \caption{Identification of the candidates Class~I YSOs in the vicinity of G010.32$-$00.15. The underlying grey image is the {\it Spitzer} 8.0~$\mu$m image. The red contours are the column density contours (levels of 3, 5, 10, and 15~$\times$10$^{22}$~cm$^{-2}$. The intensity of some YSOs has been enhanced.) }
\label{autresYSO}
\end{figure}

$\bullet$ Sources at the periphery of the bipolar nebula: they are identified in Fig.~\ref{autresYSO}.

YSOs~\#16 and \#17 lie at the extremity of the bottom lobe. YSO~\#16 presents a 24~$\mu$m  and a faint 70~$\mu$m counterpart ([8]$-$[24]=3.15~mag, hence an uncertain nature, Class~II or flat-spectrum YSO). YSO~\#17 has no detectable 24~$\mu$m and 70~$\mu$m counterparts so it is probably a low-luminosity Class~II YSO. They are not associated with a condensation (the column density in their direction is $\sim$2.5$\times$10$^{22}$~cm$^{-2}$). YSOs \#18 and \#19 lie east of the southern lobe in a region where the column density is in the range 3 to 4$\times$10$^{22}$~cm$^{-2}$). They have a 24~$\mu$m counterpart but no detectable 70~$\mu$m one ([8]$-$[24] is in the range 2.53~mag to 2.83~mag for YSO~\#18 and $\sim$3.01~mag for YSO~\#19). They are probably low-luminosity Class~II YSOs.

YSOs~\#24 and \#25 lie in the direction of the northern lobe, in directions of intermediate column density (in the range 4 to 5$\times$10$^{22}$~cm$^{-2}$). They have 24~$\mu$m counterparts and colours typical of Class~I YSOs ([8]$-$[24]=4.35~mag for \#24 and 3.97~mag for \#25). YSO~\#24 has a faint 70~$\mu$m counterpart, and YSO~\#25 has none.   
 
YSO~\#27 lies in the direction of a nearby faint \HII\ region.  We argue that it is probably associated with this \HII\ region and not with the bipolar nebula. (There is no velocity measurement to indicate the distance of this nearby nebula.) The association is uncertain for YSO~\#26, which lies in the direction of a filament west of the bipolar nebula (column density 4.8$\times$10$^{22}$~cm$^{-2}$). The colour of these sources indicates that they are probably flat-spectrum YSOs ([8]$-$[24]$\sim$3.35~mag for both).

\subsection{Tables}

\begin{table*}[h!]
\begin{threeparttable}[h!]
\caption{Exciting cluster of G010.32$-$00.15. Data extracted from the UKIDSS catalogue except for star \#1.}
\begin{tabular}{rrrrrrl}
\hline\hline
N  &  $l$          &$ $b            & $J$      & $H$              & $K$              & $d$\tnote{1} \\
   & ($\degr$)   & ($\degr$)    & (mag)    & (mag)           & (mag)             & ($\arcsec$) \\
\hline
 1\tnote{2} & 10.31798 & -0.15185  & 11.631$\pm$0.043 &  9.694$\pm$0.055 &  8.613$\pm$0.039 & 0.00 \\
 1\tnote{3} &          &           & 11.9             &                  &  9.0             & 0.00 \\  
 2 &  10.31773  & -0.15134  & 13.662$\pm$0.002 & 11.915$\pm$0.001 & 10.875$\pm$0.001 &  2.16\\
 3 &  10.31872  & -0.15123  & 14.065$\pm$0.002 & 12.180$\pm$0.001 & 11.034$\pm$0.001 &  3.47\\
 4 &  10.31812  & -0.15058  & 17.175$\pm$0.024 & 14.874$\pm$0.007 & 13.578$\pm$0.005 &  4.79\\
 5 &  10.31686  & -0.15266  & 17.247$\pm$0.026 & 14.783$\pm$0.007 & 13.405$\pm$0.004 &  4.99\\
 6 &  10.31945  & -0.15219  & 14.286$\pm$0.002 & 12.392$\pm$0.001 & 11.316$\pm$0.001 &  5.61\\
 7 &  10.31958  & -0.15174  & 15.372$\pm$0.005 & 13.337$\pm$0.002 & 12.005$\pm$0.001 &  5.83\\
 8 &  10.31849  & -0.15348  & 14.196$\pm$0.002 & 12.079$\pm$0.001 & 10.789$\pm$0.001 &  6.53\\
 9 &  10.31614  & -0.15226  & 17.224$\pm$0.025 & 15.190$\pm$0.009 & 14.049$\pm$0.007 &  6.81\\
10 &  10.31641  & -0.15316  & 15.404$\pm$0.005 & 13.481$\pm$0.002 & 12.392$\pm$0.002 &  7.39\\
11 &  10.32038  & -0.15154  & 16.169$\pm$0.010 & 14.141$\pm$0.004 & 12.969$\pm$0.003 &  8.78\\
12 &  10.31986  & -0.15340  & 17.306$\pm$0.027 & 14.910$\pm$0.007 & 13.472$\pm$0.004 &  9.28\\
13 &  10.31880  & -0.14947  & 12.105$\pm$0.001 & 11.535$\pm$0.001 & 10.750$\pm$0.001 &  9.32\\
14 &  10.32000  & -0.15004  & 15.917$\pm$0.008 & 15.075$\pm$0.008 & 14.595$\pm$0.011 &  9.81\\
15 &  10.32061  & -0.15105  & 16.503$\pm$0.013 & 14.435$\pm$0.005 & 13.190$\pm$0.003 &  9.92\\
\hline
\label{G10cluster}
\end{tabular}
\begin{tablenotes}
\item[1] {\it d} is the projected distance from the exciting star.
\item[2] 2MASS photometric data.
\item[3] photometry from Bik et al.(\cite{bik05}).
\end{tablenotes}
\end{threeparttable}
\end{table*}

\begin{table*}[h!]
\caption{Ultracompact \HII\ regions near G010.32$-$00.15.}                               
\begin{tabular}{llll}                
\hline\hline                          
Name          & Radio flux, $\lambda$         &  Diameter   & Reference  \\
              &(mJy, cm)                      &  ($\arcsec$) &   \\
\hline
\multicolumn{3}{l}{C3} \\ 
G010.301$-$00.147 & 654.6, 6 & & Becker et al. (\cite{bec94}) \\
G010.30058$-$00.14734 & 426.2, 20 & & Helfand et al. (\cite{hel06}) \\
G010.3009$-$00.1477 & 631.4, 5 & 5.2 & Urquhart et al. (\cite{urq13}) \\
G010.321$-$00.147     & 320.7,20 & 4.9$\times$4.2 & White et al. (\cite{whi05}) \\
G010.321$-$00.147     & 628.0,6 & 5.5$\times$4.1  & White et al. (\cite{whi05}) \\
\hline
\multicolumn{3}{l}{C7 extension} \\
G010.321$-$00.233 & 67.9, 6 & &  Becker et al. (\cite{bec94}) \\
G010.32029$-$00.23323 & 56.7, 20 & &  Helfand et al. (\cite{hel06}) \\
                    &          & 3.7$^*$ & MAGPIS \\
\hline
\multicolumn{3}{l}{C1 extension} \\
Source B & 8$^*$, 20 &  unresolved$^*$  & MAGPIS \\
\hline 
\label{UCH}  
\end{tabular}
\tablefoot{
Our own measurements are indicated by an $^*$. The radio continuum source G10.321$-$0.147 is a UC \HII\ region as defined by Kurtz (\cite{kur05}): its size is $\sim$0.045~pc$\leq$0.1~pc, its emission measure is $\sim$10$^7$~cm$^{-6}$~pc, its density is 1.9$\times$10$^4$~cm~$^{-3}\geq$10$^4$~cm$^{-3}$, its mass is $\sim$2$\times$10$^{-2}$~$\msol$. The nature of the two other radio sources is uncertain because their distance is unknown, but they have small sizes.} 
\end{table*}

\section{Apertures used to estimate the mass of the clumps}

\begin{figure}[h!]
\centering
\includegraphics[width=75mm]{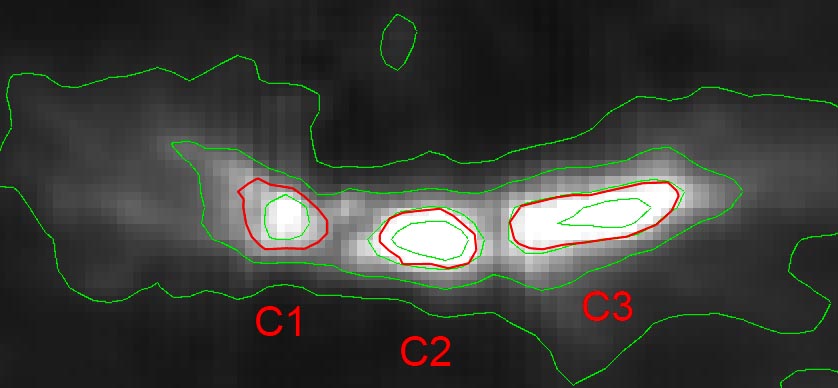}
  \caption{C1, C2, and C3 clumps in the field of G319.88$+$00.79. The underlying grey image is the column density map. The apertures used to integrate the column density and estimate the masses are indicated (red contours; these apertures follow the level at half peak-values, after correction for a background emission). The green contours correspond to column densities of 1, 2, 4, and 6$\times$10$^{22}$~cm$^{-2}$ (as in Fig.~\ref{dusttemperature1}~{\it (d)}).}
\label{G319apertures}
\end{figure}

\begin{figure}[h!]
\centering
\includegraphics[width=85mm]{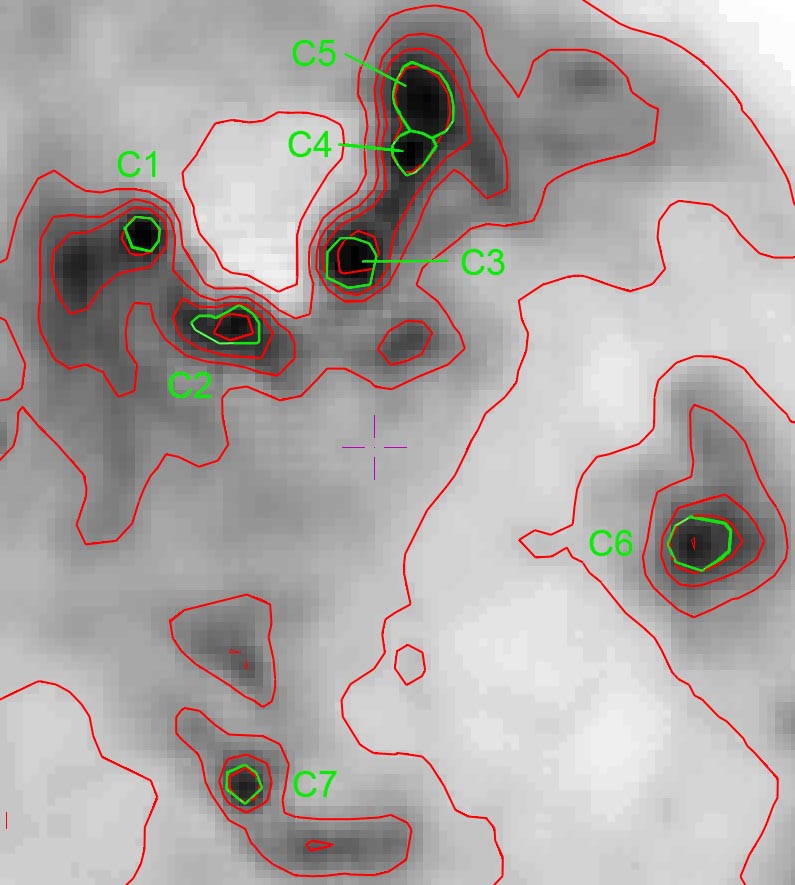}
  \caption{C1 to C7 clumps in the field of G010.32-00.14.  The underlying grey image is the column density map (logarithmic units). The apertures used to integrate the column density and estimate the masses are indicated (green contours; the polygonial apertures follow the level at half peak-values, after correction for the background emission, the same for all clumps). The red  contours correspond to column densities of 0.3, 0.5, 0.75, 1.0, and 1.5$\times$10$^{23}$~cm$^{-2}$ (as in Fig.~\ref{champ10e}~{\it (d)}).}
\label{G010apertures}
\end{figure}

\section{Infrared dark clouds in the direction of the bipolar \HII\ regions}

\begin{figure}[h!]
\centering
\includegraphics[width=80mm]{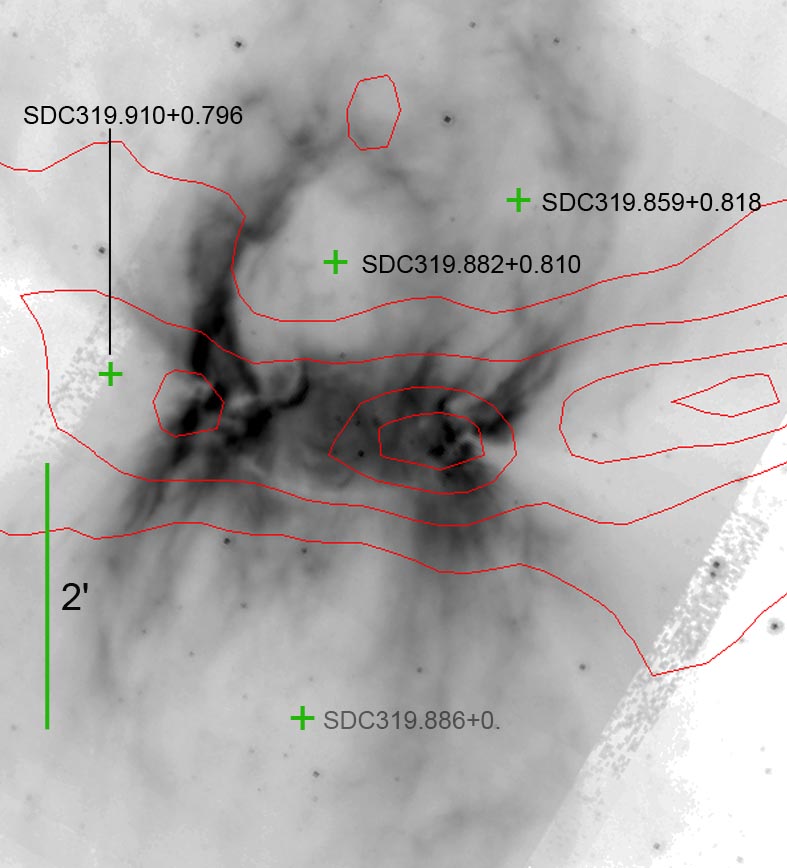}
  \caption{IRDCs detected by Peretto \& Fuller (\cite{per09}) in the vicinity of G319.88$+$00.79. The underlying grey image is the {\it Spitzer} 8.0~$\mu$m image. The red contours correspond to column densities of 1, 2, 4, and 6$\times$10$^{22}$~cm$^{-2}$.}
\label{G319IRDCs}
\end{figure}

\begin{figure}[h!]
\centering
\includegraphics[width=85mm]{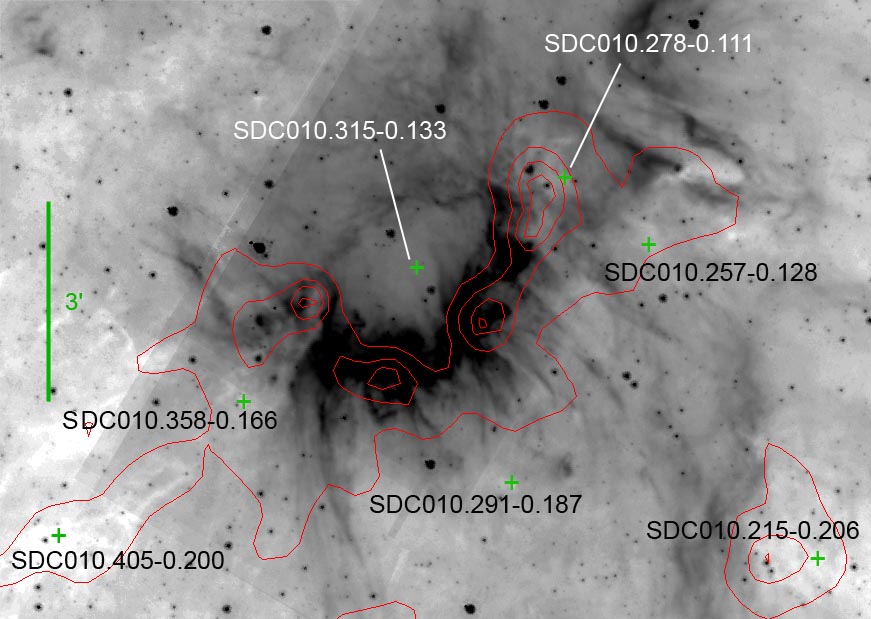}
  \caption{IRDCs detected by Peretto \& Fuller (\cite{per09}) in the vicinity of G010.32$-$00.15. The underlying grey image is the {\it Spitzer} 8.0~$\mu$m image. The red contours correspond to column densities of 0.5, 1.0, 1.5, and 2.0$\times$10$^{23}$~cm$^{-2}$.}
\label{G010IRDCs}
\end{figure}

\end{appendices}


\begin{thebibliography}{}

\bibitem[2004]{all04} Allen, L.E., Calvet, N., D'Alessio, P. et al. 2004, ApJSS, 154, 363
\bibitem[2011]{and11} Anderson, L.D., Bania, T.M., Balser,D.S., Rood, T. 2011, ApJSS, 194, 32 
\bibitem[2012]{and12} Anderson, L.D., Bania, T.M., Balser,D.S., Rood, T. 2012, ApJ, 754, 62
\bibitem[2000]{and00} Andr\'e, Ph., Ward-Thompson, D., Barsony, M. 2000, Protostars and Planets IV, eds Mannings, V., Boss, A.P., Russell, S. S., Tucson: University of Arizona Press, p.59 
\bibitem[2010]{and10} Andr\'e, Ph., Men'shchikov, A., Bontemps, S. et al. 2010, A\&A, 518, L102
\bibitem[2007]{ara07} Araya, E., Hofner, P., Goss, W.M. et al. 2007, ApJSS, 170, 152 
\bibitem[2011]{arz11} Arzoumanian, D., Andr\`e, Ph., Didelon, P. et al. 2011, A\&A, 529, L6
\bibitem[2013]{arz13} Arzoumanian, D., Andr\'e, Ph., Peretto, N., K\"onyves, V. 2013, A\&A, 553, 119 
\bibitem[2010]{ban10} Bania, T.M., Anderson, L.D., Balser, D.S., Rood, R.T. 2010, ApJ, 718, L106
\bibitem[2010]{bat10} Battersby, C., Bally, J., Jackson, J.M. et al. 2010, ApJ, 721, 222 
\bibitem[2010]{bea10} Beaumont, C.N., Williams, J.P. 2010, ApJ, 709, 791
\bibitem[1994]{bec94} Becker, R.H., White, R.L., Helfand, D.J., Zoonematkermani, S. 1994, ApJS, 91, 347
\bibitem[2003]{ben03} Benjamin, R.A., Churchwell, E., Babler, B.L., et al. 2003, PASP, 115, 953
\bibitem[2011]{beu11} Beuther, H., Linz, H., Henning, Th. et al. 2011, A\&A, 531, 26 
\bibitem[2010]{bie10} Bieging, J.H., Peters, W.L., Kang, M. 2010, ApJSS, 191, 232
\bibitem[2005]{bik05} Bik, A., Kaper, L., Hanson, M.M., Smits, M. 2005, A\&A, 440, 121
\bibitem[2009]{bis09} Bisbas, T.G., W\"{u}nsch, R., Whitworth, A.P., Hubber, D. A. 2009, A\&A, 497, 649
\bibitem[2011]{bis11} Bisbas, T.G., W\"{u}nsch, R., Whitworth, A.P. et al. 2011, ApJ, 736, 142
\bibitem[2001]{blu01} Blum, R.D., Damineli, A., Conti, P.S. 2001, AJ, 121, 3149
\bibitem[1999]{boc99} Bock, D.C.-J., Large, M.I., Sadler, E.M. 1999, AJ, 117, 1578
\bibitem[1979]{bod79} Bodenheimer, P., Tenorio-Tagle, G., Yorke, H.W. 1979, ApJ, 233, 85
\bibitem[2000]{bon00} Bonnarel, F., Fernique, P., Bienayme, O., et al. 2000, A\&ASS, 143, 33
\bibitem[1986]{bra86} Brand, J. 1986, PhDT 
\bibitem[2010]{bre10} Breen, S.L., Ellingsen, S.P., Caswell, J.L., Lewis, B.E. 2010, MNRAS, 401, 2219
\bibitem[1998]{car98} Carey, S.J., Clark, F.O., Egan, M.P. et al. 1998, ApJ, 508, 721
\bibitem[2009]{car09} Carey, S.J., Noriega-Crespo, A., Mizuno, D.R., Shenoy, S., Paladini, R. et al. 2009, PASP, 121, 76
\bibitem[1987]{cas87} Caswell, J.L., Haynes, R.F. 1987, A\&A, 171, 261
\bibitem[2006]{chu06} Churchwell, E., Povich, M.S., Allen, D. et al. 2006, ApJ, 649, 759
\bibitem[2007]{chu07} Churchwell, E., Watson, D.F., Povich, M.S. et al. 2007, ApJ, 670, 428
\bibitem[1998]{con98} Condon, J.J., Cotton, W.D., Greisen, E.W. et al. 1998, AJ, 115, 1693
\bibitem[2008]{cyg08} Cyganowski, C.J., Whitney, B.A., Holden, E. et al. 2008, AJ, 136, 2391
\bibitem[2009]{cyg09} Cyganowski, C.J., Brogan, C.L., Hunter, T.R., Churchwell, E. 2009, ApJ, 702, 1615
\bibitem[2012]{dal12} Dale, J.E., Ercolano, B., Bonnell, I.A. 2012, MNRAS, 427, 2852
\bibitem[2015]{dal15} Dale, J.E., Haworth, T.H., Bressert, E. 2015, arXiv:1502.05865
\bibitem[2009]{deh09} Deharveng, L., Zavagno, A., Schuller, F. et al. 2009, A\&A, 496, 177
\bibitem[2010]{deh10} Deharveng, L., Schuller, F., Anderson, L.D. et al. 2010, A\&A, 523, 6
\bibitem[2012]{deh12} Deharveng, L., Zavagno, A., Anderson, L.D. et al. 2012, A\&A, 546, 74
\bibitem[2015]{dew15} Dewangan, L.K., Ojha, D.K., Grave, J.M.C., Mallick, K.K. 2015, MNRAS, 446, 2640 
\bibitem[1980]{dow80} Downes, D., Wilson, T.L., Bieging, J., Wink, J. 1980, A\&ASS, 40, 379
\bibitem[2011]{du11}  Du, Z.M., Zhou, J.J., Esimbeck, J. et al. 2011, A\&A, 532, 127
\bibitem[2008]{duh08} Dunham, M.M., Crapsi, A., Evans II, N.J. et al. 2008, ApJSS, 179, 249  
\bibitem[2003]{dut03} Dutra, C.M., Bica, E., Soares, J., Barbuy, B. 2003, A\&A, 400,533
\bibitem[1997]{dys97} Dyson, J.E., \& Williams, D.A. 1997, The physics of the 
interstellar medium, 2nd ed., ed. R J Tayler \& M Elvis (Institute of Physics Publishing, Bristol and Philadelphia)
\bibitem[1998]{ega98} Egan, M.P., Shipman, R.F., Price, S.D. et al. 1998, ApJ, 494, 199
\bibitem[2012]{eli13} Elia, D., Molinari, S., Fukui, Y.  et al. 2013, ApJ 772, 45
\bibitem[2011]{fos11} Foster, J.B., Jackson, J.M., Barnes, P.B. et al. 2011, ApJSS, 197, 25
\bibitem[2013]{fos13} Foster, J.B., Rathborne, J.M., Sanhueza, P. et al. 2013, PASA, 30, 38
\bibitem[2000]{fuk00} Fukuda, N., Hanawa, T. 2000, ApJ, 533, 911
\bibitem[1976]{geo76} Georgelin, Y.M., georgelin, Y.P. 1976, A\&A, 49, 57
\bibitem[1994]{gre94} Greene, T.P., Wilking, B.A., Andr\'e, Ph. et al. 1994, ApJ, 434, 614
\bibitem[2010]{gre10} Green, J.A., Caswell, J.L., Fuller, G.A. et al. 2010, MNRAS, 409, 913 
\bibitem[2010]{gva10} Gvaramadze, V.V., Kniazev, A.Y., Fabrika, S. 2010, MNRAS, 405 ,1047
\bibitem[2006]{hel06} Helfand, D.J., Becker, R.H., White, R.L. et al. 2006, AJ, 131, 2525
\bibitem[1983]{hil83} Hildebrand, R.H. 1983 Q.Jl. R. astr. Soc., 24, 267
\bibitem[2011]{hil11} Hill, T., Motte, F., Didelon, P. et al. 2011, A\&A, 533, A94
\bibitem[2012]{hoa12} Hoare, M.G., Purcell, C.R., Churchwell, E.B. et al. 2012, PASP, 124, 939
\bibitem[1994]{hof94} Hofner, P., Kurtz, S., Churchwell, E. et al. 1994, ApJ, 429, 85 
\bibitem[2005]{ind05} Indebetouw, R., Mathis, J.S., Babler, B.L. et al. 2005, ApJ, 619, 931
\bibitem[2010]{jac10} Jackson, J.M., Finn, S.C., Chambers, E.T. et al. 2010, ApJ, 719, L185
\bibitem[2013]{jac13} Jackson, J.M., Rathborne, J.M., Foster, J.B. et al. 2013, PASA, 30, 57
\bibitem[1982]{kal82} Kalberla, P.M.W., Goss, W.M., Wilson, T.L. 1982, A\&A, 106, 167
\bibitem[2001]{kim01} Kim, K.T., Koo, B.C. 2001, ApJ, 549, 979
\bibitem[2002]{kim02} Kim, K.T., Koo, B.C. 2002, ApJ 575, 327
\bibitem[1997]{kuc97} Kuchar, T.A., Clark, F.O. 1997, ApJ, 488, 224
\bibitem[2005]{kur05} Kurtz, S. 2005, IAUS, 227, 111
\bibitem[1987]{lad87} Lada ,C.J. 1987, IAUS, 115, 1 
\bibitem[2007]{law07} Lawrence, A., Warren, S.J., Almaini, O. et al. 2007, MNRAS, 379, 1599
\bibitem[2005]{mar05} Martins, F., Schaerer, D., Hillier, D.J. 2005, A\&A, 436, 1049
\bibitem[2006]{mar06} Martins, F., Plez, B. 2006, A\&A, 457, 637
\bibitem[2004]{meg04} Megeath, S.T., Allen, L.E., Gutermuth, R.A. et al. 2004, ApJS, 154, 367
\bibitem[2010]{min10} Minniti, D., Lucas, P.W., Emerson, J.P. et al. 2010, New Astronomy, 15, 433
\bibitem[2010a]{mol10a} Molinari, S., Swinyard, B., Bally, J. et al. 2010a, PASP, 122, 314 
\bibitem[2010b]{mol10b} Molinari, S., Swinyard, B., Bally, J. et al. 2010b, A\&A, 518, L100
\bibitem[2011]{mol11} Molinari, S., Schisano, E., Faustini, F.  et al. 2011, A\&A, 530, 133
\bibitem[1994]{oss94} Ossenkopf, V., Henning, Th 1994, A\&A, 291, 943
\bibitem[2014]{par14} Paradis, D., M\'eny, C., Noriega-Crespo, A. et al. 2014, arXiv1409.6892 
\bibitem[2005]{par05} Parker, Q.A., Phillipps, S., Pierce, M.J. et al. 2005, MNRAS, 362, 689
\bibitem[2013]{pav13} Pavlyuchenkov, Ya.N., Kirsanova, M.S., Wiebe, D.S. 2013, ARep., 57, 573
\bibitem[1996]{per96} Perault, M., Omont, A., Simon, G. et al. 1996, A\&A, 315, 165
\bibitem[2009]{per09} Peretto, N., Fuller, G.A. 2009, A\&A, 505, 405
\bibitem[2013]{per13} Peretto, N., Fuller, G.A., Duarte-Cabral, A. et al. 2013, A\&A, 555, 112
\bibitem[2007]{pov07} Povich, M.S., Stone,J.M., Churchwell, E. et al. 2007, ApJ, 660, 346 
\bibitem[2013]{pur13} Purcell, C.R., Hoare, M.G., Cotton, W.D. et al. 2013, ApJS, 201,1 
\bibitem[2006]{qui06} Quireza, C., Rood, R.T., Balser, D.S., Bania, T.M. 2006, ApJS, 165, 338
\bibitem[2012]{rag12} Ragan, S., Henning, Th., Krause, O. et al. 2012, A\&A, 547, 49 
\bibitem[1970]{rei70} Reifenstein, E.C., Wilson, T.L., Burke, B.F. et al. 1970, A\&A, 4, 357 
\bibitem[1985]{rie85} Rieke, G.H., Lebofsky, M.J. 1985, ApJ, 228, 618
\bibitem[2006]{rob06} Robitaille, T.P., Whitney, B.A., Indebetouw, R. et al. 2006, ApJSS, 167, 256
\bibitem[2008]{rob08} Robitaille, T.P., Meade, M.R., Babler, B.L. et al. 2008, AJ, 136, 2413
\bibitem[2008]{rod08} Rodriguez-Fernandez, N.J., Combes, F. 2008, A\&A, 489, 115 
\bibitem[2013]{sad13} Sadavoy, S.I., Di Francesco, J., Johnstone, D. et al. 2013, ApJ, 767, 126
\bibitem[2009]{san09} Sanna, A., Reid, M.J., Moscadelli, L. et al. 2009, ApJ, 706, 464
\bibitem[2014]{san14} Sanna, A., Reid, M.J., Menten, K.M. et al. 2014, ApJ, 781, 108
\bibitem[2009]{sch09} Schuller, F., Menten, K.M., Contreras, Y. et al. 2009, A\&A, 504, 415
\bibitem[2004]{sew04} Sewilo, M., Watson, C., Araya, E., et al. 2004, ApJS, 154, 553
\bibitem[2006]{sim06} Simon, R., Jackson, J.M., Rathborne, J.M., Chambers, E.T. 2006, ApJ, 639, 227
\bibitem[1990]{sim90} Simpson, J.P., Rubin, R.H. 1990, ApJ, 354, 165
\bibitem[2012]{sim12} Simpson, R.J., Povich, M.S., Kendrew, S. et al. 2012, MNRAS, 424, 2442
\bibitem[2006]{skr06} Skrutskie, M.F., Cutri, R.M., Stiening, R. et al. 2006, AJ, 131, 1163
\bibitem[2002]{smi02} Smith, L.J., Norris, R.P.F., Crowther, P.A. 2002, MNRAS, 337, 1309
\bibitem[1987]{ste87} Stetson, P.B. 1987, PASP, 99, 191
\bibitem[2013]{stu13} Stutz, A.M., Tobin, J.J., Stanke. T. et al. 2013, ApJ, 767, 36
\bibitem[2011]{tra11} Traficante, A., Calzoletti, L., Veneziani, M. et al. 2011, MNRAS, 416, 2932
\bibitem[2013]{urq13} Urquhart,J.S, Thompson,M.A., Moore, T.J.T. et al. 2013, MNRAS, 435, 400 
\bibitem[1995]{wal95} Walmsley, M. 1995, RevMexAA, 1, 137
\bibitem[1998]{wal98} Walsh, A.J., Burton, M.G., Hyland, A.R., Robinson, G. 1998, MNRAS, 301, 640
\bibitem[2005]{whi05} White, R.L., Becker, R.H., Helfand, D.J. 2005, AJ, 130, 586
\bibitem[2013]{whi13} Whitney, B.A., Robitaille, T.P., Bjorkman, J.E. et al. 2013, ApJS, 207, 30
\bibitem[2012]{wie12} Wienen, M., Wyrowsky, F., Schuller, F., et al. 2012, A\&A, 544, 146
\bibitem[2012]{wil12} Wilcock, L.A., Ward-Thompson, D., Kirk, J.M. et al. 2012, MNRAS, 422, 1071
\bibitem[1974]{wil74} Wilson, T.L. 1974, A\&A, 31, 83
\bibitem[2010]{wri10} Wright, E.L., Eisenhardt, P.R., Mainzer, A.K. et al. 2010, AJ, 140, 1868
\bibitem[2010]{zav10} Zavagno, A., Russeil, D., Motte, F. et al. 2010, A\&A, 518, 81  
\end{thebibliography}
\end{document}